\documentstyle[11pt,psfig,epsfig]{article}
\newcommand{\be}{\begin{equation}}
\newcommand{\bn}{\begin{eqnarray}}
\newcommand{\ee}{\end{equation}}
\newcommand{\en}{\end{eqnarray}}
\newcommand{\nn}{\nonumber \\}
\newcommand{\HS}{$m_h^{SM}$}

\begin{document}
\vspace*{-10mm}

\noindent\hbox to\hsize{February 1997 \hfill BROWN-HET-1076}\\
\noindent\hbox to\hsize{hep-ph/9702323 \hfill BROWN-TA-549}\\
\baselineskip18pt
\vspace{1.0cm}
\pagestyle{plain}
\begin{center}
{\bf FIELD THEORY FOR THE STANDARD MODEL
\footnote{Based on the invited lectures given by  one of us(KK) at the 
15th Symposium on Theoretical
Physics, Seoul National University, Seoul, Korea, 22-28 August 1996.}}
\vglue 15mm
\normalsize
{\bf Kyungsik Kang } \\
\vglue 2mm
{\it Department of Physics, Brown University, Providence, RI 02912, USA}\\
\vglue 5mm
{\it and }\\
\vglue 5mm
{\bf Sin Kyu Kang
\footnote{Korea Science and Engineering Foundation
Post-doctoral Fellow.} } \\
\vglue 2mm
{\it Department of Physics, Brown University, Providence, RI 02912, USA}\\
\vglue 10mm
{\bf ABSTRACT}\\
\vglue 8mm
\begin{minipage}{14cm}
{\normalsize 
 We review the origin of the standard electroweak model and discuss
in great detail the on-shell renormalization scheme of the standard model
as a field theory.
One-loop radiative corrections are calculated by the dimensional
regularization and the dominant higher-order corrections are also
discussed.
We then show how well the standard model withstands the precision test against
the LEP and SLD data along with its prediction on the indirect bound of the
Higgs boson mass $m_H$. The uncertainty in $m_H$ and in
$Z-$decay parameters due to the experimental uncertainties in $m_t, M_W$
and QED and QCD couplings is also discussed.
}
\end{minipage}
\end{center}
\newpage
\tableofcontents
\newpage
\section{ Introduction}
\setcounter{equation}{0}
In modern terms, a gauge theory is a renormalizable quantum
field theory$^{[1]}$  with a {\bf local gauge symmetry} requirement, i.e. a choice of a
{\bf gauge group}. The quantum electrodynamics (Q.E.D.) of the electrons and
phtons is a well-known renormalizable gauge theory with the Abelian gauge
group U(1). The pure Lagrangian of the matter (electron) and gauge field
(photon) is modified to use the Covariant Derivatives in places of ordinary
derivatives. This procedure induces the minimal coupling of the electrons and
photons and allows us to avoid the divergence problem in the higher order
diagrams. The gauge principle can also be extended to a non-Abelian symmetry,
as in the case of the original Yang-Mills paper$^{[2]}$
 that attempted to explain the
origin of the isotopic spin conservation in the strong interaction. It
turns out that the Yang-Mills idea is more appropriate for describing the
electroweak intercations of the leptons and quarks and the color symmetry of
the quarks, the modern version of the strong interaction. In other words, such
Yang-Mills theories abound in Nature and we now have the standard
model for  the elementary particle interactions based on the gauge invariance
principle. We believe that the standard model may be
just a combination (actually a direct product) of three simple Lie
groups corresponding to the three fundamental forces: electromagnetism,
weak interactions and strong interactions. To each of these forces
correspond roughly the (Special) Unitary gauge groups $U(1), SU(2)$ and $SU(3)$.
The standard model based on the $SU(3) \times SU(2) \times U(1)$ gauge theory
gives us a nearly complete understanding of all the interactions of
elementary particles.
The modern perspective is to promote all gauge symmetries from global
to local, thus turning a conserved quantum number (such as baryon
number $B$ or lepton number $L$) into a quantity
which may now be violated to a certain degree by the dynamical exchange
of a massive gauge boson. For example, we can even build a typical
Grand Unified Theory (GUT) or some generalized GUT, which
would only conserve $B-L$ and hence predict the baryon number non-conserving
processes such as  the proton decay and also the lepton number violation !
Throughout our journey in and beyond the standard model,
we will have to deal almost exclusively
with two distinct types of fields: fermionic matter fields
denoted generically by $\psi$ and gauge bosons (a.k.a. force fields) denoted
by $A_\mu$. Leptons (such as electrons) and quarks are examples of fermions
while the photon, weak intermediate vector bosons and gluons exhaust the list
of gauge bosons in the standard model.
These fields will reside in certain representations of the
appropriate gauge groups, with fermions usually in some fundamental
representation and gauge bosons in the adjoint representation.
For quantum electrodynamics (QED) the existence of a unique photon ($\gamma$)
means that the local gauge symmetry is just an abelian $U(1)$. Not only is this
conceptually the easiest but it is also remarkably predictive with
complete agreement between theory and experiment up to 7 significant figures
in $(g-2)$ experiments.
We thus introduce$^{[1]}$ the basic concepts of the gauge theory
in the context of QED.

Let us consider a general Lagrangian density
${\cal L} (\psi, \partial_\mu \psi, A_\mu, F_{\mu\nu})$
where $F_{\mu\nu}$ is the electromagnetic energy-momentum tensor
formed from the four-potential $A_\mu$ and $\partial_\mu$ is shorthand for
$\partial/\partial x^\mu$. This Lagrangian density must be invariant under
an $U(1)$ gauge transformation that locally changes fermion fields by
\begin{equation}
\psi \rightarrow \psi'= e^{ie\Lambda(x)} \psi = U \psi \approx
(1+ie \Lambda) \psi 
\end{equation}
and simultaneously changes gauge fields by
\begin{equation}
A_\mu \rightarrow A_\mu' \approx A_\mu -\partial_\mu \Lambda .
\end{equation}
Infinitesimally therefore we have
\begin{equation}
\delta \psi=ie\Lambda\psi~~~~~~~~\delta (\partial_\mu \psi)=\partial_\mu
(\delta \psi)=ie\Lambda\partial_\mu \psi +ie (\partial_\mu\Lambda) \psi .
\end{equation}
We will find it convenient to introduce the Covariant Derivative
\begin{equation}
\medskip
 D_\mu = \partial_\mu +ieA_\mu       
\end{equation}
since $D_\mu \psi$ then transforms just like the fermion fields themselves, i.e.
using (1.3) we obtain the desired result
\begin{equation}
 \delta(D_\mu\psi) = 
ie\Lambda (D_\mu \psi).~~~~~~                                     
\end{equation}
The promotion of $\partial_\mu$ to $D_\mu$ is often called the minimal
substitution rule (or minimal coupling).
Exponentiating we have that
\begin{equation}
D_\mu\psi \rightarrow D'_\mu \psi'=U(\partial_\mu \psi + ieA_\mu \psi)
\end{equation} 
but since we also have that
\begin{equation}
D'_\mu \psi'= (\partial_\mu+ieA'_\mu) U\psi= (\partial_\mu U)\psi
+U\partial_\mu \psi +ieA'_\mu U\psi               
\end{equation}
the gauge field must transform as
\be
A'_\mu=UA_\mu U^\dagger +{i\over{e}} (\partial_\mu U) U^\dagger .
\ee
The generalization for non-Abelian gauge groups$^{[2]}$ such as $SU(n)$ for
which the generators of the group $t_\alpha$ do not commute is simply done
by noting that in these cases $\Lambda$ of (1.1) becomes
a sum $\Lambda=\sum_\alpha \Lambda_\alpha t_\alpha$ such that
\bn
\psi \rightarrow \psi' & = &U\psi \approx (1+ig \sum_\alpha \Lambda_\alpha
t_\alpha) \psi \nn
A_\mu \rightarrow A'_\mu & = &UA_\mu U^\dagger +{i\over g} (\partial_\mu U)
U^\dagger                                         
\en 
with the shorthand notation $A_\mu=\sum_\alpha A^\alpha_\mu t_\alpha$.
This shows that there is a gauge boson for each generator of the group
and the group generators satisfy the commutation relations
\be
 \left[ t_\alpha,t_\beta\right]=i f_{\alpha\beta\gamma}t_\gamma 
\ee
with structure constants $f_{\alpha\beta\gamma}$ which will allow
complex vertices among the gauge bosons.
We adopt the usual notation that indices
at the start of the greek alphabet are related to the gauge
group while those in the middle of the alphabet are Lorentz indices.
By looking at small changes in $A_\mu$ according to (1.9), i.e. by expanding
$U$ about the identity matrix, one can easily show that
\be
\delta A_\mu=ig\left[\Lambda,A_\mu\right] -\partial_\mu \Lambda .
\ee
The introduction of the Covariant Derivative by
\be
 D_\mu = \partial_\mu +igA_\mu =\partial_\mu +ig\sum_\alpha A_\mu^\alpha
t_\alpha,                                            
\ee
then again insures that $D_\mu \psi$ transforms like the fermion
field
\be
D_\mu \psi \rightarrow D'_\mu \psi' = U(D_\mu \psi),
\ee
and that ${\cal L} (\psi, \partial_\mu \psi, A_\mu, F_{\mu\nu})$
using such a Covariant Derivative
must be gauge invariant. This insures that a free field in $\cal L$
with the standard kinetic energy term will automatically introduce
interactive couplings between the fields through the Covariant Derivative
while maintaining its gauge invariance.
It is understood now that the energy-momentum tensor
of the gauge field is formed with
\be
F_{\mu\nu}=D_\mu A_\nu - D_\nu A_\mu = \partial_\mu A_\nu - \partial_\nu A_\mu
+ig \left[A_\mu,A_\nu\right] =\sum_\alpha t_\alpha F^\alpha_{\mu \nu}
\ee
with
\be
F_{\mu\nu}^\alpha = \partial_\mu A^\alpha_\nu-\partial_\nu A^\alpha_\mu
-g f_{\alpha\beta\gamma}A^\beta_\mu A^\gamma_\nu .
\ee
One can explicitly check that $F'_{\mu\nu}=UF_{\mu\nu}U^\dagger$ as it should.
We now explicitly write down the Lagrangian density in QED
\be
{\cal L} = \bar\psi i\gamma^\mu D_\mu \psi -{1\over 4} F_{\mu\nu}F^{\mu\nu}
\ee
with $D_\mu$ as in (1.4).
Under $U(1)$ gauge transformations we have seen that the different terms
transform as
\bn
\psi \rightarrow \psi'&=& e^{ie\Lambda(x)} \psi \approx (1+ie\Lambda) \psi \nn
 A_\mu \rightarrow A_\mu'&=& UA_\mu U^\dagger +{i\over e}
(\partial_\mu U)U^\dagger \approx A_\mu -\partial_\mu \Lambda \nn
F_{\mu\nu} \rightarrow F'_{\mu\nu} &=& UF_{\mu\nu}U^\dagger = F_{\mu\nu}
\approx \partial_\mu A_\nu -\partial_\nu A_\mu      
\en
thus leaving $\cal L$ invariant.
The generalization to an $SU(n)$ gauge transformation with $\Lambda= \sum_a
t_\alpha \Lambda_a$ is achieved by using the covariant derivative
(1.12) and gives
\be
 A^\alpha_\mu \rightarrow A^{'\alpha}_\mu \approx A^\alpha_\mu
-\partial_\mu\Lambda_\alpha-gf_{\alpha\beta\gamma} \Lambda_\beta A^\gamma_\mu .
\ee
Since $\delta {\cal L}(\psi,\cdots)={\cal L}(\psi',\cdots)-
{\cal L}(\psi,\cdots) =0$ under a gauge transformation (1.9), there should be a
conserved current associated with the gauge symmetry, if $\cal L$ depends only
on $\psi$ and $\partial_\mu\psi$, according to Noether's theorem,
\be
\delta {\cal L} = g \partial_\mu\left( \sum_\alpha \Lambda_\alpha
J^\mu_\alpha \right) = 0                           
\ee
where
\be
J_\alpha^\mu=i{{\delta {\cal L}}\over{\delta (\partial_\mu\psi)}} t_\alpha\psi
= i\sum_{j,k} {{\delta {\cal L}}\over{\delta (\partial_\mu\psi_j)}}
(t_\alpha)_{jk} \psi_k .                                  
\ee
Note that if $\delta {\cal L}$ does not contain the gauge function $\Lambda_
\alpha$ explicitly, a conserved current arises,$^{[3]}$
\be
\partial_\mu J_\alpha^\mu={{\partial (\delta {\cal L})}\over{g \partial
\Lambda_\alpha}} =0 ,                                     
\ee
while
\be
 J^\mu_\alpha = {1\over g} {{\partial (\delta {\cal L})}\over{\partial
(\partial_\mu\Lambda_\alpha)}}.
\ee
The fermionic fields $\psi$ satisfy equal time anti-commutation relations
from the quantization condition at $x_0=t$,
\bn
\left[\psi_j(\vec x, x_0),\Pi_k(\vec y,x_0)\right]_+ &=&
i\delta_{jk}\delta^{(3)} (\vec x-\vec y) \nn
\left[\psi_j(\vec x, x_0),\psi_k(\vec y,x_0)\right]_+ &=&
\left[\Pi_j(\vec x, x_0),\Pi_k(\vec y,x_0)\right]_+ =0,
\en
where $\Pi_i = {{\delta {\cal L}}\over{\delta(\partial_0\psi_i)}}$
is the conjugate momentum to $\psi_i$. The time components
of the currents (1.20) are related to the charge operators
\be
Q_\alpha=\int d^3x J^0_\alpha.
\ee
One can also show that the charge operators satisfy the same algebraic relation
(1.10) and also
\be
\left[Q_\alpha,\psi_j(\vec x,t)\right]_+ = \sum_m \left(t_\alpha\right)_{jm}
\psi_m(\vec x,t).                                       
\ee
\section{ Electroweak theory: the $SU(2)_L\times U(1)$
 Weinberg-Salam Model} 
\setcounter{equation}{0}
We start with the observation that the low-energy four-point Fermi
interaction amongst leptons relevant for muon decay $\mu^- \rightarrow
e^-\bar \nu_e \nu_\mu$
can be thought to originate from the exchange of a massive negatively charged
intermediate vector boson (IVB) of mass $M_B$ with coupling strength $g$
between the muonic and leptonic branches
if the couplings satisfy $G_F/\sqrt 2 \propto g^2/M_B^2$.
The low-energy pointlike nature of the coupling then implies that $M_B \approx
300$ GeV. After historical forays into Scalar, Pseudoscalar and Tensor
forms for the weak current, experiments finally established
the Vector and Axial-vector nature of the charged currents and the universality
of the weak interactions of all particles.
In fact the purely $V-A$ structure in the lepton sector allows us to form two
weak isospin doublets,
\bn
\left( \begin{array}{c}
        \nu_{\mu} \\
         \mu^- \end{array} \right)_L,
  ~~~ \qquad
 \left( \begin{array}{c}
          \nu_e \\
          e^- \end{array} \right)_L
\en
for the $\mu$-decay which can be interpreted as
a transition between upper and lower
components mediated by a $\Delta Q=-1$ IVB. Mathematically, we define
left (right) projection operators
\be
P_\pm ={{1\pm \gamma_5}\over 2}                      
\ee
such that
\be
\psi_{L,R}=P_\pm \psi
\ee
(note that $V-A$ involves $P_+$ in our conventions).
It is a simple exercise to show that $\bar \psi_{L,R}= \bar \psi P_{\mp}$
(recall that $\bar \psi = \psi^\dagger \gamma^0$)
implies that a mass term must be of the form
\be
\bar\psi \psi  = \bar\psi_R \psi_L +\bar \psi_L \psi_R .
\ee
The Fermi interaction for $\mu$-decay can be written as
${{G_F}\over{\sqrt 2}} J_\mu^\dagger(\mu_L)J^\mu(e_L) + h.c.$ where
$J^\mu(l_L)=\bar l_L\gamma^\mu \left({{\tau_+}\over 2}\right) l_L$, $\tau_+$
being the charge-raising operator $\tau_1+i\tau_2$ and $h.c.$ denoting the
Hermitean conjugate of what precedes it.
At this point, we may as well complete the list of known weak left-handed
doublets both in the leptonic and quark sectors
\bn
l_L &=& \left( \begin{array}{c}
             \nu_e \\
              e^-\end{array}\right)_L
   ~~~~~~~\qquad
        \left(  \begin{array}{c}
          \nu_\mu \\
         \mu^- \end{array} \right)_L
   ~~~~~~~\qquad
        \left( \begin{array}{c}
          \nu_\tau \\
         \tau^- \end{array} \right)_L \nn
q_L &=& \left( \begin{array}{c}
              u \\
              d \end{array}\right)_L
   ~~~~~~~\qquad
        \left( \begin{array}{c}
            c \\
            s  \end{array} \right)_L
   ~~~~~~~\qquad
        \left( \begin{array}{c}
            t \\
            b \end{array} \right)_L 
\en
where we have stopped at three generations, the first of each consisting
of the elementary entities
needed for everyday life: the electron, its neutrino and the
up and down quarks which help form the proton and the neutron.
As we shall discuss later, recent
measurements at the CERN LEP Collider
of the total decay width of the neutral intermediate vector
boson $Z_0$ showed that there is room for only three neutrino species implying
the existence of three family generations of leptons and quarks
associated with the three neutrino species.
The long anticipated discovery of the top quark was announced by CDF
and D$\O$ groups at Fermilab$^{[4]}$, 
thus completing the three generations of the
leptons and quarks.
For the moment$^{[5]}$, we stress that the weak
interaction is governed by the $SU(2)$ gauge symmetry and the fermions
participating in weak interaction processes transform as the left-handed
doublets under $SU(2)$.
Since the electromagnetic current obtained by using Noether's theorem
on the Lagrangian density (1.16) is just $j^\mu=-\bar \psi \gamma^\mu\psi =
-(\bar\psi_L\gamma^\mu\psi_L+\bar\psi_R\gamma^\mu\psi_R)$,
in order to include the electromagnetic interaction we conclude that
charged particles must have right handed assignments as weak singlets
under the weak interaction gauge symmetry $SU(2)$:
\be
l_R= e_R^-,~~\mu_R^-~~~~\tau_R^-;~~~~~~~ q_R=u_R,~~d_R,~~s_R,~~c_R,~~b_R,~~
t_R.
\ee
Absence of the right-handed neutrinos insures their masslessness at this stage.
We shall denote the 3 gauge bosons associated with $SU(2)$
by $A^\mu_i$, with $i=1,2,3$ and the unique boson of $U(1)$ by $B^\mu$.
We note in passing that the number of generators in the adjoint representation
of $SU(n)$, thus the number of associated gauge bosons in the
$SU(n)$ gauge theory, is just
$(n^2-1)$. If $T_i$ and $Y/2$ are the generators of $SU(2)_L$ and $U(1)$
we will find it useful to introduce the shorthand notation
\be
{\cal A}^\mu=\sum_{i=1}^3 A_i^\mu T_i = \vec A^\mu \cdot \vec T
~~~~~~~~B'_\mu ={Y\over 2} B_\mu                  
\ee
and we will denote the
respective coupling constants by $g$ and $g'$.
A useful representation for the generators is in terms of the Pauli
matrices $\vec \tau$ such that $T_i=\tau_i/2$. The structure constants
for SU(2) form the Levi-Cevita tensor, hence
\be
[T_i,T_j]=i\epsilon_{ijk}T_k.
\ee
In order to incorporate QED in the theory, we shall have to make sure that
some linear combination of the generators is the electric charge $Q$ and that
its accompanying boson remain massless (the photon). Clearly $T_3$ is related
to $Q$ since $\Delta T_3=\Delta Q$ in each doublet. The obvious relationship
is then
\be
Q=T_3+{Y\over 2}                                 
\ee
with $[T_3,Y]=0$ since they belong to different groups.
The assignments for the left-handed doublets can be obtained by concentrating
on the upper component, for example, while those for the right-handed singlets
is straightforward with the use of $Y(q_R)=2Q(q_R)$. The results are
\be
Y(l_L)=-1,~~Y(l_R) =-2,~~Y(q_L)={1\over 3}, ~~Y(u_R)=4/3,~~Y(d_R)=-2/3
\cdots                                            
\ee
We shall later need to introduce doublets of scalar fields with
charge assignments
\bn
  \left( \begin{array}{c}
          \phi^+ \\
          \phi^0 \end{array} \right)
\en
resulting in a value $Y_\phi=1$.

Again local gauge invariance of $\cal L$ is guaranteed by the use of the
Covariant Derivative, but since the gauge group is now semi-simple i.e.
is a direct product of simple Lie groups, we have to write
\be
D_\mu=\partial_\mu+igA_\mu+ig'B'_\mu =\partial_\mu+ig\vec A_\mu \cdot \vec T
+ig'{Y\over 2} B_\mu.
\ee
When introduced in the Lagrangian density for matter fields (with
the understanding to sum over the fermion doublets (2.5) and singlets (2.6)
for repeated symbols $l_{L,R}$ and $q_{L,R}$),
\be
{\cal L}_{fermion}=\bar l_L i\gamma^\mu D_\mu l_L
+ \bar l_R i\gamma^\mu D_\mu l_R
+\bar q_L i\gamma^\mu D_\mu q_L + \bar q_R i\gamma^\mu D_\mu q_R 
\ee
this gives explicitly for each lepton sector
\be
{\cal L}_{lepton} = \bar l_L i \gamma^\mu \left(\partial_\mu
+{i\over 2}g\vec \tau \cdot \vec A_\mu-{i\over 2}g'B_\mu \right)l_L+
\bar l_R i\gamma^\mu\left(\partial_\mu-ig'B_\mu \right)l_R+\cdots
\ee
and for each quark sector
\bn
{\cal L}_{quark} &=& \bar q_L i \gamma^\mu \left(\partial_\mu
+{i\over 2}g\vec \tau \cdot \vec A_\mu+{i\over 6}g'B_\mu\right)q_L \nn
& &+\bar u_R i\gamma^\mu\left(\partial_\mu+{{2i}\over 3}g'B_\mu \right)u_R+
\bar d_R i\gamma^\mu\left(\partial_\mu-{i\over 3}g'B_\mu \right)d_R+\cdots
\en
The Lagrangian density for the gauge fields is simply
\be
{\cal L}_{gauge}= -{1\over 4}  {\cal A}_{\mu\nu}  {\cal A}^{\mu\nu}
-{1\over 4} B'_{\mu\nu}B'^{\mu\nu}               
\ee
where, for example
\begin{eqnarray}
{\cal A}_{\mu\nu}&=& D_\mu {\cal A}_\nu -D_\nu {\cal A}_\mu  \nn
 &=& (\partial_\mu+ig {\cal A}_\mu) {\cal A}_\nu
-(\partial_\nu+ig {\cal A}_\nu) {\cal A}_\mu \nn
 &=& \partial_\mu {\cal A}_\nu -\partial_\nu {\cal A}_\mu
+ig [{\cal A}_\mu,{\cal A}_\nu] \\
&=& \partial_\mu {\cal A}_\nu -\partial_\nu {\cal A}_\mu
+ig  A_{i,\mu} A_{j,\nu} [T_i,T_j]\nn
&=& \partial_\mu {\cal A}_\nu -\partial_\nu {\cal A}_\mu
-g \epsilon_{ijk} A_{i,\mu} A_{j,\nu}
 T_k \nonumber
\end{eqnarray}
with summation implied over repeated latin indices $i,j = 1,2,3$.

Let us concentrate on the charged current coupling for one
generic family generation
\bn
{\cal L}^{\Delta Q \neq 0} = i^2 g \left\lbrace
\bar l_L\gamma^\mu \left( {{\tau_1}\over 2}A_{1\mu}+{{\tau_2}\over 2}A_{2\mu}
\right)l_L+
\bar q_L\gamma^\mu \left( {{\tau_1}\over 2}A_{1\mu}+{{\tau_2}\over 2}A_{2\mu}
\right)q_L \right\rbrace \nn
=-{g\over{\sqrt 2}} \left\lbrace
\bar \nu_L \gamma^\mu W^+_\mu e_L + \bar e_L \gamma^\mu W^-_\mu \nu_L +
\bar u_L \gamma^\mu W^+_\mu d_L + \bar d_L \gamma^\mu W^-_\mu u_L \right\rbrace
\en
where the fields
\be
W^\pm = {1\over{\sqrt 2}} \left( A_{1\mu} \mp i A_{2\mu}\right) 
\ee
annihilate the gauge bosons $W^\pm$ or create $W^\mp$.
At this stage all fermions and gauge bosons are massless, i.e. the gauge
symmetry alone does not allow the fermion mass terms $\bar e_L e_R$ or
$W_\mu^{\dagger}W^\mu$. Adding the latter would ruin the renormalizability of
the theory. More compactly, we can rewrite (2.18) as
\be
{\cal L}^{\Delta Q \neq 0} = - {g\over{\sqrt 2}} \sum_i \bar \psi_i
\gamma^\mu P_+ \left( T^+W^-_\mu+T^-W^+_\mu\right) \psi_i      
\ee
where
\be
 T^\pm = {\tau^\pm\over 2} = {{\tau_1\pm i\tau_2}\over 2}  
\ee
and $\psi_i (i=1,2,\cdots,6)$ represents the fermion doublet of the leptons
$l$ and quarks $q$.

We now turn to the neutral couplings where new effects will be seen to arise.
We explicitly have
\bn
{\cal L}^{\Delta Q =0} &=& -g \left[
\bar l_L\gamma^\mu {{\tau_3}\over 2} A_{3\mu} l_L +
\bar q_L\gamma^\mu {{\tau_3}\over 2} A_{3\mu} q_L \right]  \nn
& &+ {{g'}\over 2}\left[
\bar l_L\gamma^\mu B_\mu l_L + 2 \bar l_R\gamma^\mu B_\mu l_R \right] \nn
& &-{{g'}\over 2}\left[{1\over 3}
\bar q_L\gamma^\mu B_\mu q_L + {4\over 3} \bar u_R\gamma^\mu B_\mu u_R -
{2\over 3} \bar d_R\gamma^\mu B_\mu d_R \right].
\en
The lepton couplings themselves can be regrouped to give
\bn
{\cal L}^{\Delta Q =0}_{leptons} &=&  \bar \nu_L\gamma_\mu \nu_L
\left( -{g\over 2}A_{3\mu} +{g'\over 2} B_\mu\right) \nn
& &+ \bar e_L\gamma_\mu e_L \left( {g\over 2}A_{3\mu} +
{g'\over 2} B_\mu\right) + g' \bar e_R\gamma_\mu e_R B_\mu .
\en
The first term implies the existence of a new type of neutral current
interaction while the other terms are familiar from the electromagnetic
interaction.
The first term would cause a new kind of neutrino induced reactions which
originate from the weak interactions mediated by the neutral current.
This is perhaps the most striking prediction of the $SU(2)\times U(1)$ gauge
theory and has since been established by various experiments as we
shall see later.

At this point, we impose that a linear combination of $A_{3\mu}$
and $B_\mu$ must represent the photon field $A_\mu$ coupling to the charge
generator $Q$. We therefore perform a rotation from the above basis to
the physical basis of $A_\mu$ and $Z_\mu$, where $Z_\mu$ is a new field
orthogonal
to the photon $\gamma$ and is the familiar neutral gauge boson
$Z_0$. The discovery of the $W^\pm$ and $Z_0$ at the
S$p\bar p$S of CERN by two groups of the Underground
Area experiments UA1 and UA2 promoted this theory as the standard model of
electroweak interactions. However, at this stage, these gauge bosons
are massless and we will return to the mass generating mechanism and
the question of renormalizability later.
We introduce the weak-mixing angle $\theta_w$
(the so-called Weinberg angle)
\bn
\left(\begin{array}{c}
    A_\mu  \\
    Z_\mu \end{array} \right)
 =\left(\begin{array}{cc}
    \cos \theta_w & \sin \theta_w  \\
    -\sin \theta_w & \cos \theta_w \end{array} \right)
\left(\begin{array}{c}
    B_\mu   \\
    A_{3\mu} \end{array} \right)
\en
in order to separate the electromagnetic interaction from the remaining
neutral current weak interaction.
Then, since $\nu$ does not couple to the photon, we must have that
$g\sin\theta_w-g'\cos\theta_w=0$
and thus a relationship between the coupling strengths
\be
{g'\over g}=\tan\theta_w.
\ee
Using this relation in the electron coupling to the photon we can
obtain the further constraint that the electronic charge takes the form
\be
 e=g'\cos \theta_w=g\sin\theta_w                        
\ee
which can be rewritten conveniently as
\be
\frac{1}{e^2} = \frac{1}{g^2} + \frac{1}{g^{\prime 2}}.
\ee
Note that if there are more U(1) factors with coupling strengths
$g'_i$, as is fashionable
in some superstring-inspired low-energy phenomenology, the second term of the
RHS of (2.27) just contains a sum of the inverse coupling constants squared,
i.e. $1/g'^2 \rightarrow \sum_i 1/g'^2_i$.
Some straightforward algebra leads to the $\nu$ coupling
purely left-handedly to $Z_\mu$
\be
 -{g\over {2\cos\theta_w}} \bar \nu_L \gamma^\mu \nu_L Z_\mu 
\ee
while the electron's coupling contains both left- and right-handed components
summing up to
\be
{g\over{2\cos\theta_w}} \bar e_L \gamma^\mu e_L Z_\mu -
g' \sin\theta_w \bar e \gamma^\mu e Z_\mu.
\ee
Using (2.25) both factors can be combined to yield the standard form
\bn
 {\cal L}^{\Delta Q=0} &=& -
{g\over{2\cos\theta_w}}\left[ \bar \nu_L\gamma^\mu\nu_L
-\bar e_L\gamma^\mu e_L+2\sin^2\theta_w \bar e \gamma^\mu e\right] Z_\mu \nn
&=&-{g\over{\cos\theta_w}}  \bar l \gamma^\mu \left( T_{L3} -Q_l \sin^2\theta_w
\right) l Z_{\mu}     
\en
where the third components of weak isospin have the obvious values from (2.1),
$T_{L3}(\nu)=1/2,~T_{L3}(e)=-1/2 $ and $Q_l$ denotes the electric charges
of the leptons.
By combining the couplings to the photon and the above equation, we obtain
our final result in terms of generic fermion fields $\psi_i$
\be
{\cal L}^{\Delta Q=0} =-e\sum_i Q_i \bar \psi_i \gamma^\mu\psi_i A_\mu -
{g\over{\cos\theta_w}} \sum_i\bar \psi_i \gamma^\mu
\left( V_i+A_i\gamma^5 \right) \psi_i Z_\mu               
\ee
with the obvious identification of the vector and axial couplings
\be
V_i=T_{3i}-\sin^2\theta_w Q_i~~~~~~~~~~~A_i=T_{3i}         
\ee
The requirement of the gauge invariance fixes the interaction terms uniquely.
The charged current interactions are governed by (2.20) while the neutral
current weak interactions and the electromagnetic couplings are given by (2.31)
in the $SU(2)\times U(1)$ electroweak theory.
\bigskip\noindent
\section{ Goldstone bosons and Higgs mechanism }
\setcounter{equation}{0}
\bigskip
These two concepts will be introduced separately at first despite the
fact that they have to operate simultaneously in order to get a
well-defined electroweak theory with massive weak IVBs, yet massless photons.
Goldstone's theorem crudely states that, in a field theory
of scalars ($\phi_i$)
where the potential $V(\phi)$ and the {\bf Lagrangian} density
\be
{\cal L}(\phi) = {1\over 2} \partial_\mu \phi \partial^\mu \phi
-V(\phi)                                                  
\ee
are {\bf invariant} under some symmetry transformation but the
{\bf vacuum} state is {\bf not}, there must exist massless bosons, aptly called
Goldstone bosons. We reproduce below an elegant proof of this statement.

Let us assume that $\delta {\cal L}=0$ under some global gauge transformation
\be
\delta \phi_k =i \Lambda_a(T^a)_{kl} \phi_l                 
\ee
then the potential must also be invariant under (3.2)
\be
\delta V= {\delta V\over{\delta\phi_k}} \delta\phi_k =
i {\delta V\over{\delta\phi_k}} \Lambda_a(T^a)_{kl}\phi_l =0 .
\ee
Differentiating once more with respect to $\phi_m$ we obtain
\be
{{\delta^2 V}\over{\delta\phi_m\delta\phi_k}} (T^a)_{kl} \phi_l +
{{\delta V}\over {\delta\phi_k}}(T^a)_{km}=0                 
\ee
for arbitrary $\Lambda_a$. Suppose now that the vacuum is non-trivial, i.e. that
$<\phi>=v \neq 0$ minimizes $V$. Then, at this minimum, only the first term
of (3.4) is non-zero and we recognize the second derivative,
which measures the curvature of $V$ around the minimum, as being the
usual mass-squared matrix for a scalar field. Equation (3.4) then simply states
that
\be
M^2_{mk}(T^a)_{kl} v_l=0.
\ee
If $T^a$ is the generator of an unbroken subgroup, then trivially $T^a v=0$.
However when the vacuum breaks the symmetry for some generators,
then $T^a v \neq 0$ and it must be an eigenvector of $M^2$ with zero
eigenvalue, our massless Goldstone boson! There are as many massless
excitations as there are generators that do not leave the vacuum invariant.
An example is the chiral symmetry breaking of the $\sigma$-model
in which the pion emerges as a Goldstone boson.

The Higgs mechanism, which was originally invented to evade the appearance
of the massless Goldstone bosons by coupling the scalar fields to gauge fields,
turns out to be a convenient way
to generate mass terms for the vector gauge fields
when the scalar fields of the type (2.11) undergo spontaneous
symmetry breaking (SSB) as described above. In addition to preserving
gauge-invariance, this
process also preserves renormalizability as we will endeavor to show later on.
We start with complex scalar fields (hereafter called Higgs fields)
for which the gauge invariant form of (3.1) becomes
\be
{\cal L}(\phi) =  (D_\mu \phi)^\dagger (D^\mu \phi)-V(\phi)   
\ee
upon introducing the covariant derivative
$D_\mu=\partial_\mu+ig\vec A_\mu \cdot \vec T +ig'{Y\over 2} B_\mu$ but with
$\vec T$ and $Y$ acting on scalar fields $\phi$.
This generates the fermion-scalar couplings or Yukawa interaction
for one generic family generation.
\be
{\cal L}_{f\phi} =-f_e \left(\bar e_R\phi^\dagger l_L+\bar l_L\phi e_R \right)
- f_d \left(\bar d_R \phi^\dagger q_L+\bar q_L\phi d_R \right)
- f_u \left(\bar u_R \tilde \phi^\dagger q_L+\bar q_L \tilde \phi u_R \right)
\ee
where $\tilde\phi$ is the charge-conjugate of $\phi$
\bn
\tilde\phi=i \tau_2 \phi^* =\left(\begin{array}{c}
          \phi^{0*}\\
            \phi^-\end{array} \right),~~~~~~~~\qquad
\phi^-=-\phi^{+*}.
\en
Note that $\phi^\dagger l_L$ is an $SU(2)$ singlet with $Y=-2=Y(e_R^-)$
so that $\bar e_R \phi^\dagger l_l$ is
$SU(2)\times U(1)$ gauge-invariant. Similar uses
of (2.9) and (2.10) yield $Y(\phi^\dagger q_L)=-2/3=Y(d_R)$ and
$Y(\tilde \phi^\dagger q_L)=4/3=Y(u_R)$ so that all the terms in (3.7)
are gauge-invariant (the other terms are just Hermitean conjugates of the
ones for which we checked gauge invariance).

The crucial step is to write the Higgs potential as
\be
V(\phi) =m_0^2\phi^\dagger \phi +\lambda_0 \left(\phi^\dagger\phi \right)^2
\ee
with {\bf negative} bare mass $m_0^2<0$ such that (3.9) can be rewritten as
\be
V(\phi) = \lambda_0\left( \phi^\dagger\phi-{{v^2}\over 2}\right)^2 -
{{\lambda_0v^4}\over 4 }                                
\ee
adopting the famous ``mexican-hat" shape with a double minimum at $\vert \phi
\vert =\pm v/\sqrt 2$ and the identification $v^2\lambda=-m_0^2$.
Let us postulate that
the vacuum chooses spontaneously the state $<\phi^+>=0, ~<\phi^0>=v/\sqrt 2$.
Out of the four real scalar fields, only one denoted as $\eta$ remains since
\bn
\phi = \left( \begin{array}{c}
            \phi^+ \\
             \phi^0 \end{array} \right)
     = \left(\begin{array}{c}
            \phi^+_R + i \phi^+_I \\
            \phi^0_R+ i \phi^0_I \end{array}\right)~ \equiv
e^{i {{\vec\xi\cdot\vec\tau}\over{2v}}} 
       \left( \begin{array}{c}
           0 \\
          {{v+\eta}\over{\sqrt 2}} \end{array} \right)
\en
and we can choose the unitary gauge for which
$U=e^{-i {{\vec\xi\cdot\vec\tau}\over{2v}}}$, to ``gauge away" the $\vec \xi$
fields as follows
\bn
\phi \rightarrow \phi'=U\phi = \left( \begin{array}{c}
           0 \\
            {{v+\eta}\over{\sqrt 2}} \end{array} \right).
\en 
It can be shown, after some tedious manipulations, that ${\cal L}_\phi$ becomes
in terms of the remaining physical fields
\bn
{\cal L}_\phi &=& {1\over 2}\partial_\mu \eta \partial^\mu \eta
+{{v^2g^2}\over 8} \left( W_\mu^+W^{-\mu}+W_\mu^-W^{+\mu} \right) \nn
 & & + m_0^2\eta^2+
{{(v+\eta)^2}\over 8} \chi^\dagger \left[g\vec\tau\cdot\vec A_\mu+g'B_\mu
\right]^2\chi  +\cdots                                   
\en
with $\chi=\pmatrix {0\cr 1\cr}$ in isospin space.
This leads immediately to an identification for the masses squared
\be
{{M^2_{W^\pm}}\over 2} = {{v^2g^2}\over 8}
~~~~~~~~m_\eta^2=-2m_0^2=2\lambda_0 v^2 >0.
\ee
The three massless real scalar fields have been eaten up by the
gauge bosons which have become massive, thus conserving the number of
degrees of freedom of the theory: the 3 gauge bosons can now have longitudinal
as well as transverse components. In addition the remaining neutral Higgs
field ($\eta$) has seen a imaginary bare mass ($m_0^2<0$) transformed in a
real physical mass $m_\eta$.
For the two charged gauge bosons and the Higgs field, we have thus obtained
\be
M_{W^\pm}={{vg}\over 2}~~~~~~~~m_\eta= \sqrt{2\lambda_0}v
\ee
We can obtain further constraints on the parameters of our theory by focusing
on the charged current component of $\cal L$
\bn
{\cal L}^{\Delta Q \neq 0} &=& {g\over{\sqrt 2}}
\left(\bar \nu_L\gamma^\mu W^+_\mu e_L + h.c.\right)~~~~~~~~~~~~~~ \nn
&=& {g\over{2\sqrt 2}}
\left(\bar \nu_e \gamma^\mu (1+\gamma_5)W^+_\mu e + h.c.\right)=
{g\over{2\sqrt 2}} J^\mu (e_L) W^+_\mu + h.c.      
\en
Comparing this to the low-momentum transfer limit of the four-point Fermi
interaction
\be
{\cal L}^{\Delta Q \neq 0}={G_F\over{\sqrt 2}}
\left( J_\mu J^{\mu\dagger} +h.c.\right)={G_F\over{\sqrt 2}}
\left(\bar \nu_e\gamma^\mu (1+\gamma_5) e \right)
\left(\bar e\gamma^\mu (1+\gamma_5) \nu_e \right) + h.c.    
\ee
we conclude that the coupling strengths must obey the relationship
\be
 {{g^2}\over{8M_W^2}} = {{G_F}\over{\sqrt 2}}.              
\ee
Using the above expression for $M_W^2$ together
with the one of equation (3.14) we
conclude that, since $G_F= 1.16639(2)\times 10^{-5}$ GeV$^{-2}$, the
vacuum expectation value of our neutral Higgs has a value
\be
v={1\over{2^{1/4}\sqrt{G_F}}}= 246.22~ \mbox{GeV}
\ee
Here the Fermi constant $G_F$ is determined from the $\mu$-lifetime
$\tau_\mu=2.197035(40) \times 10^{-6}$ sec and
\be
\tau_{\mu}^{-1} =
{{G_F^2(M^2_W)m_\mu^5}\over{192\pi^3}} f \left({{m_e^2}\over {m^2_\mu}}\right)
\left( 1+{{3m_\mu^2}\over{5M_W^2}}\right) 
~\left[1+{{\alpha(m_\mu)}\over{2\pi}}
\left({{25}\over 4}-\pi^2 \right)\right]                      
\ee
where the three-body phase space factor takes the form
\be
 f(x)= 1-8x+8x^3-x^4-12x^2\ln x                           
\ee
and the electromagnetic fine structure constant at the muon's mass
$\alpha(m_\mu)$ obeys the equation
\be
\alpha^{-1}(m_\mu) = \alpha^{-1}
+{2\over{3\pi}}\ln{{m_\mu}\over{m_e}}+ {1\over{6\pi}}          
\ee
with $\alpha^{-1}=137.0359895(61)$, determined from the electron magnetic
moment anomaly $(g-2)$.
If we combine (3.14) and (2.26), we obtain a useful result for the $W$ mass
in GeV
\be
M_W={{ev}\over{2\sin\theta_w}} ={{37.2802}\over{\sin\theta_w}}  
\ee
where we have used the fact that, in our system of units, $e^2=4\pi\alpha$.
To obtain a similar relationship for the $Z$ mass,
we go back to (3.13) and find the term
\bn
 & &{{v^2}\over 8} \chi^\dagger \left(\begin{array}{cc}
      gA_{3\mu}+g'B_\mu & 0 \\
         0 & -gA_{3\mu}+g'B_\mu \end{array}\right)^2 ~\chi=
        {{v^2}\over 8} \left(g'B_\mu-gA_{3\mu}\right)^2 \nn
&=&{{v^2}\over 8}\left[ g'(\cos\theta_w A_\mu-\sin\theta_w Z_\mu)-
g(\sin\theta_w A_\mu-\cos\theta_w Z_\mu) \right]^2  \nn
&=&{{v^2}\over 8}
\left(g'\sin\theta_w+g\cos\theta_w\right)^2 Z_\mu Z^\mu   ~~~~~~  
\en
where we have used (2.26) to prove that the photon remains massless.
Using now (2.25) to eliminate $g'$ we find that (3.24) reduces to
\be
 {{g^2v^2}\over {8\cos^2\theta_w}} Z_\mu Z^\mu           
\ee
and we obtain the desired expression for the $Z$ mass in GeV
\be
M_Z={{gv}\over{2\cos\theta_w}} = {{M_W}\over{\cos\theta_w}}
= {{74.5604}\over{\sin 2\theta_w}}.
\ee
With the most recent Particle Data Group value for $\sin^2\theta_w=0.2259$
we obtain tree-level predictions for the gauge boson masses in GeV
\be
 M_W=78.44~~~~~~~~M_Z=89.15                              
\ee
which are already only a few percent off the experimental values
\be
 M_W=80.35\pm 0.13~ \mbox{GeV}~~~~M_Z=91.1863\pm 0.0020~ \mbox{GeV}  
\ee
the former being obtained indirectly from the new world average$^{[6]}$ of the
S$p\bar p$S and Fermilab Tevatron data and the latter
directly from $Z_0$ production at LEP I. One will have to wait for LEP II
to obtain comparable errors on the $W$ mass when the added center-of-mass
energy of the $e^+e^-$ Collider will finally allow $W^\pm$ pair production.
We should note that (3.26) is a direct consequence of our choice of a
Higgs doublet in the process of SSB. Since the Higgs sector is the one
with the least experimental constraints, model builders have considered many
variants: different numbers of doublets and more complex representations. Thus,
in a more general approach, we can define the ratio of the terms in (3.25)
as a parameter to be determined by experiment
\be
\rho= {{M_W^2}\over{M_Z^2 \cos^2\theta_w}}              
\ee
on the same footing as the weak-mixing angle.
We note that the charged fermions also gain masses from the same Higgs
mechanism. Namely after the SSB in the unitary gauge, the Yukawa couplings (3.7)
become
\be
{\cal L}_{f\phi} =-{{v+\eta}\over{\sqrt 2}}\left( f_e\bar e e+ f_d\bar d d
+ f_u \bar u u \right)                                         
\ee
so that
\be
 m_e={v\over{\sqrt 2}} f_e,~~~m_d={v\over{\sqrt 2}} f_d,~~~
m_u={v\over{\sqrt 2}} f_u.
\ee
Thus the kinetic terms in the Lagrangian density for the fermion fields
(2.13) can be combined with ${\cal L}_{f\phi}$ to give
\be
{\cal L}_{f-kinetic}= \sum_j \bar f_j\left( i\gamma^\mu \partial_\mu -m_j
-{{gm_j}\over{2M_W}}\right ) f_j +\sum_j \bar \nu_{Lj}\left(i\gamma^\mu
\partial_\mu\right) \nu_{Lj}.
\ee
These tree-level mass predictions will of course receive radiative
corrections of a few percent if and only if the renormalizability was not
lost through the mechanism of SSB. In other words, has the renormalizability
of the massless theory been spoiled by the spontaneous symmetry breaking
that gave us realistic masses? Gerhard 't Hooft$^{[7]}$ showed that SSB did not
affect renormalizability in the early '70s but his configuration-space
proof is not as transparent as Faddeev and Popov$^{[8]}$ momentum-space
formulation that we shall soon turn to.
There is one more ingredient in the standard model that one should mention
at this time. The quarks have an additional {\bf exact} $SU_c(3)$
symmetry, one that is {\bf not} broken in the above way, hence one where
the gauge bosons remain massless. Although traditions change according
to geographic location, this hidden degree of freedom is called ``color" and
the fundamental colors are chosen to be red, yellow and blue.
In this scheme, only ``white" or colorless hadrons can exist, forcing
baryons to consist of three quarks of different colors and mesons to
consist of quarks and antiquarks of given color and anticolor.
In addition, the gauge bosons mediating this strong force must carry color
and are, by construction, forbidden to appear as free states, just as the
isolation of a colored quark is equally verboten. Since these gauge
bosons glue so strongly the charged quarks together against the electromagnetic
force (in a volume of radius $\approx 1$ fm),
they are ``colorfully" called gluons and the theory is Quantum ChromoDynamics
or QCD for short.

A final remark is in order at this point. In
Quantum Field Theories (QFTs) where the vacuum state
is a complex state with continual creation and annihilation of
particle-antiparticle pairs, there is an effect of screening (or anti-screening)
of bare charges that depends on the probed distance from the charge.
In momentum space this translates into the coupling constants of the
theory being momentum dependent, the so-called ``running coupling constant".
For example, the renormalization scale dependence of the QCD coupling
constant $g$ is determined, to the three loop order, by
 
\be
 \mu {\partial g \over \partial \mu} = - {g^3 \over 16 \pi^2} [\beta_0 +
\beta_1 {g^2 \over 16 \pi^2} + \beta_2 {g^4 \over (16 \pi^2)^2} + \cdots
]~~~~
\ee
where
${\beta_0} = 11 - (2/3){n_f},~{\beta_1} = 102 - (38/3){n_f},~~
{n_f}$ being the number of quark flavors below the scale ${\mu}$, and
${\beta_2}$ is
scheme dependent. For abelian group
such as $U(1)_{em}$ the coupling strength increases
with decreasing distance or increasing momenta, revealing more and more of
the bare charge as we probe closer in. Our usual Coulombic charge at large
distances is thus
expected to be renormalized to a smaller quantity than the bare charge.
For a non-abelian theory such as QCD, the reverse is true, i.e. the coupling
constant $\alpha_s(Q^2)$ becomes weaker as the momentum $Q$ increases!
The quarks
behave at high energies as if they were asymptotically free just as Feynman's
partons were postulated to be. The quark-parton model is said to
exhibit asymptotic freedom. At the lower end of the energy regime, the coupling
gets stronger, we have infrared slavery and the quarks are forever bound to
colorless hadrons. 

The scale dependence of a running quark mass $m_{q}(\mu)$ is
determined by the equation
\begin{eqnarray*}
D_{\mu} m_q(\mu) = -\gamma_{\alpha_s} m_q(\mu)
\end{eqnarray*}
where
\begin{eqnarray*}
D_{\mu} &=& \mu \frac{d}{d \mu}, \\
\gamma_{\alpha_s} &=& \gamma_0 \alpha_s +\gamma_1 \alpha_s^2 +
  O(\alpha_s^3), \\
\gamma_0 &=& 2, ~~~~ \gamma_1 = \frac{101}{12}-\frac{5}{18}n_q,
\end{eqnarray*}
so that $m_q(\mu)$ turns out to be$^{[9]}$
\be
m_q = \tilde{m}_q \left(\frac{1}{2}L\right)^{-2\gamma_0/\beta_0}
     \left[ 1-\frac{\beta_1\gamma_0}{\beta_0^3}\frac{\ln{L}+1}{L}
     +\frac{8\gamma_1}{\beta^2_0 L}
     +O(L^{-2}\ln^2 L) \right]. \nonumber
\ee
where $L=\ln (\mu^2/\Lambda^2)$ and $\tilde{m_q}$ is the renormalization
group invariant mass, which is independent of 
$\ln (\mu^2/\Lambda^2)$.

We now turn to the difficult question of renormalizability and to the related
one of the appropriate Feynman rules to be used for the propagators of the
electroweak theory.
\bigskip \noindent
\section{ Quark mass matrices and Kobayashi-Maskawa flavor mixing}
\setcounter{equation}{0}
\bigskip
Although the standard model, with appropriate radiative corrections,
is in perfect experimental agreement with the observed properties and masses
of all the gauge bosons, it can merely accommodate the observed fermionic
mass spectrum and cannot predict it. In fact, the mere presence of
exactly three generations of fermions (a well-established experimental fact
since the precise measurements of the $Z_0$ decay width), is still
a theoretical puzzle. If I. Rabi could be heard to mutter while at a chinese
dinner, upon hearing of the discovery of the muon, ``Who ordered that?",
then what about its two quark partners, the $\tau$ lepton and its
partners too? 
With the discovery of the top quark$^{[5]}$, the three family structure
of the fermion sector has completely been determined.
Nevertheless, the flavor mixing and fermion masses and their hierarchical
patterns remain to be one of the basic problems in particle physics.
The light quark masses are known to be:
$m_u = 5.1 \pm 0.9 ~\mbox{MeV}, m_d = 9.3 \pm 1.4 ~\mbox{MeV}$ and
$m_s=175 \pm 25 ~\mbox{MeV}$, and the heavy quark masses are$^{[10]}$
$m_c=1.35\pm 0.05 ~\mbox{MeV}, m_b=5.3 \pm 0.1 ~\mbox{GeV}^{[11]}$ and
$m_t(\mu = 1~\mbox{GeV}) \simeq 280 - 450 $ GeV.

Because of the presence of several fermion generations, the weak interaction
eigenstates are not necessarily the mass eigenstates, and mixing will
inevitably occur between the different flavor states, adding more
parameters to the standard model (the final count of arbitrary parameters
in the standard model is actually 17).
Such flavor-mixing does not lead to any observational
consequences in the neutral weak current interactions in the tree
approximation as the standard model prevents flavor-changing neutral
currents up to the $\alpha G_F$ order.

Once  the origin of flavor mixing is seen to
originate from SSB in the electroweak sector,
several possible approaches to the problem will naturally arise.
We could ambitiously try to achieve calculability of the mixing
angles in terms of the physical masses which would then suggest an allowed
form for the mass matrices from which one could induce an appropriate
Higgs coupling structure. A more pragmatic and more popular approach is to
make an ansatz for the mass matrices and fit to experiment:$^{[12]}$ such are
the Fritzsch and Stech ans\"atze. Finally, the experimental approach is
to fit the parameters of the mixing matrix in a chosen form
(Kobayashi-Maskawa, Maiani or Wolfenstein) to various data after including
radiative corrections and making reasonable model assumptions for
non-perturbative effects; the resulting matrix can be checked for unitarity
or conversely unitarity can be used to deduce missing experimental inputs.

In the standard model with three generations containing left and right-handed
fermions $\psi_{iL}$ and $\psi_{jR},~i,j=1,2,3$ and Higgs scalars $\phi_a$,
the gauge-invariant Yukawa interaction is
\be
{\cal L}_Y = \sum_{i,j} \bar \psi_{iL} \Gamma^a_{ij} \phi_a \psi_{jR} + h.c.
\ee
which, after SSB, becomes
\be
M=\sum_{i,j}\left(\bar u_{0L}\right)_i M^{(u)}_{ij} \left(u_{0R}\right)_j +
\sum_{i,j}\left(\bar d_{0L}\right)_i M^{(d)}_{ij}\left(d_{0R}\right)_j + h.c.
\ee
where
\be
M^{(u)}=\sum_a \Gamma^a_u<\phi_a>_u~~~~~~~M^{(d)}=\sum_a \Gamma^a_d<\phi_a>_d
\ee
and the subscript$~_0$ indicates the weak basis.
Here $<\phi_a>_u$ and $<\phi_a>_d$ are the vacuum expectation values (v.e.v.)
of the neutral Higgs scalar components that contribute to the up and down quark
mass matrices respectively and $\Gamma^a_u$ and $\Gamma^a_d$ are the
associated coupling matrices. The mass matrix $M^{(e)}$ for charged
leptons can be obtained similarly. Gauge-invariant bare mass terms
can be added to (4.1) and (4.2) if there are no global phase transformation
invariances such as flavor number and lepton number conservations.

For obvious phenomenological reasons, we shall assume that the mass matrices
are non-singular, non-degenerate and can therefore be brought to diagonal
form by an appropriate rotation-redefinition of the quark fields
\be
q^Q_{L(R)}=U^Q_{L(R)} q^Q_{L(R)}, ~~~~~~Q=u,d      
\ee
where $q^Q_{L(R)}$ denotes the column vector representation of the left-handed
(right-handed) quarks with charge of $Q$ so that for $n$ generations
the mass matrix undergoes, due to (4.4) a biunitary diagonalization
\bn
U^Q_L M^{(Q)}U_R^{Q\dagger} = M^Q_D = \left( \begin{array}{cccc}
        M_1 &0 & \ldots &0 \\
        0&M_2&\ldots&0 \\
         \vdots&\vdots&\ddots &\vdots\\
          0&0&\ldots &M_n\end{array} \right)
\en
Here all $M_i$ are positive and we can assume $M_{i-1}<M_i$ without loss
of generality. The redefined quark fields (4.4) are the physical quark fields
up to certain phase factors which will be specified later.
Upon reexpressing the Lagrangian density in terms of the physical fields,
the charged weak current becomes
\be
J_\mu = \bar u_{0L} \gamma_\mu d_{0L} + h.c.=\bar u_L \gamma_\mu V d_L + h.c.
\ee
where $V$ is the generalized flavor-mixing matrix given {\bf only} in terms
of the left-handed rotation matrices
\be
V=U^{(u)}_L U^{(d)\dagger}_L                     
\ee
Because $V$ is independent of right-handed rotations, we can use this
freedom to reduce the number of non-zero elements in $M^{(Q)}$. Since the matrix
$V$ must be unitary, it can be parametrized, apart from $(2n-1)$ trivial
phases, by $n(n-1)/2$ real angles and $(n-1)(n-2)/2$ phases.
For 2 generations, we recover the lone Cabibbo angle and no non-trivial phase
\bn
 V_c =\left( \begin{array}{cc}
        V_{ud} & V_{us} \\
         V_{cd} &V_{cs} \end{array}\right) 
=\left( \begin{array}{cc}
         \cos\theta_c &  \sin\theta_c\\
         -\sin\theta_c & \cos\theta_c \end{array} \right)
= e^{i\theta_c \tau_2}                               
\en
For our real world of 3 generations, we have 3 angles and one
CP-violating non-trivial phase for $V$, so several possible
representations are possible. Historically the first one
is due to Kobayashi and Maskawa
\bn
V_{KM} &=& \left( \begin{array}{ccc}
        c_1 &s_1c_3 & s_1s_3 \\
-s_1c_2 & c_1c_2c_3-s_2s_3 e^{i\delta} &c_1c_2c_3+s_2c_3 e^{i\delta} \\
-s_1s_2 & c_1s_2c_3+c_2s_3 e^{i\delta} &c_1s_2s_3-c_2c_3 e^{i\delta} 
 \end{array}\right) \nn [.2in]
&=& e^{{i\over{\sqrt 6}}(\delta+\pi)} e^{-i\theta_2\lambda_7}
e^{{i\over{\sqrt 3}} (\delta+\pi)\lambda_8} e^{i\theta_1\lambda_2}
e^{i\theta_3\lambda_7}                                    
\en
where the $\lambda_i$'s are the usual Gell-Mann matrices
and the shorthand notation $s_i=\sin\theta_i,~c_i=\cos\theta_i$
has been introduced.
The Particle Data Group  has settled instead on the Maiani
representation which has the form
\bn
V_M = \left(\begin{array}{ccc}
    c_\theta c_\beta  &s_\theta c_\beta & s_\beta e^{i\phi} \\
-s_\theta c_\gamma -s_\gamma s_\beta c_\theta e^{-i\phi}&
c_\theta c_\gamma -s_\gamma s_\beta s_\theta e^{-i\phi} & s_\gamma c_\beta \\
s_\theta s_\gamma -c_\gamma s_\beta c_\theta e^{-i\phi}&
-c_\theta c_\gamma -c_\gamma s_\beta s_\theta e^{-i\phi} & c_\gamma c_\beta 
  \end{array}\right)
\en
mostly because the CP violating phase is identified with the $V_{ub}$
element. Finally the Maiani form can reduce to the Wolfenstein
parametrization upon setting the hierarchical relations
\be
s_\theta=\lambda,~s_\gamma=A\lambda^2,~s_\beta=A\rho\lambda^3
\ee
since $\lambda$ is empirically a small number close to 0.22
as we shall see later
\bn
V_W= \left(\begin{array}{ccc}
    1-\lambda^2/2 & \lambda & A\rho\lambda^3 e^{i\phi} \\
-\lambda &1-\lambda^2/2 & A\lambda^2 \\
A\lambda^3 (1-\rho e^{-i\phi})& -A\lambda^2 &1 \end{array}\right)
 + {\cal O}(\lambda^4)
\en
We now turn to the experimental determination of the moduli of the
elements of the flavor-mixing matrix in either of the above three forms.
\medskip
The element $V_{ud}$ is determined from the ratio of the rate for
$\beta$-decay in nuclei ($0^+ \rightarrow 0^+$ superallowed
transitions) to that for $\mu$-decay.
\be
{{\Gamma(d \rightarrow u +e^- + \bar \nu_e)}\over
{\Gamma(\mu^- \rightarrow \nu_\mu +e^- + \bar \nu_e)}}
\approx \vert V_{ud}\vert^2 \vert \rho^{cc}\vert^2          
\ee
where $\rho^{cc}$ is the electroweak radiative correction factor to define
the running Fermi constant
\be
G_F(\mu^2)
\equiv G_F(M_W^2)\rho^{cc}                       
\ee
The use of the nuclear superallowed Fermi transition gets rid of axial form
factor contributions as well as the weak magnetism term. The
normalization to $\mu$-decay is chosen because of its accurate measurements
and theoretically well-understood rate, including radiative corrections,
as given by $\tau_\mu^{-1}=\Gamma(\mu^- \rightarrow \nu_\mu +e^- + \bar \nu_e)$
in equations (3.20)-(3.22).
From the muon lifetime measurement and 8 superallowed decays, Sirlin
obtains the final result
\be
\vert V_{ud}\vert = 0.9736\pm 0.0010                    
\ee
so that, for the Wolfenstein parametrization $V_{ud}=1-\lambda^2/2$ we have
\be
 \lambda=0.23 \pm 0.004                                
\ee
We should stress that the 3-4\% radiative corrections of $\rho^{cc}$ were
instrumental in our determination, otherwise $V_{ud}$ would be so large
as to violate the unitarity of the flavor-mixing matrix with just
the first two elements $\vert V_{ud}\vert^2+ \vert V_{us}\vert^2 >1$.
\medskip
The element $V_{us}$ is similarly determined by taking a ratio
of strangeness changing $\beta$-decays to our reference reaction
$\mu$-decay. This element is just the sine of the Cabibbo angle $\sin \theta_c$
introduced to restore the universality between semi-leptonic hadronic
$\beta$-decays and $\mu$-decay. By studying the 3 body decays
$K^+\rightarrow \pi^0l^- \bar\nu_l$ and $K^0_L\rightarrow \pi^- l^+\nu_l$
which involve the matrix element
\be
<\pi(p_2)\vert \bar u \gamma_\mu s\vert K(p_1)>=C\left[(p_1+p_2)_\mu f_+(q^2)
+(p_1-p_2)_\mu f_-(q^2) \right]                       
\ee
where $q_\mu=(p_1-p_2)_\mu$ and $C$ is a Clebsch-Gordan coefficient ($1/\sqrt 2$
and $1$ for $K^+$ and $K^0$ decays
respectively). Since $m_l << m_K$ only $f_+(q^2)$ needs to
be parametrized linearly away from $q^2=0$.
The net combined result, after theoretical estimates of the relevant $f_+(0)$,
is $\vert V_{us}\vert=0.2196\pm 0.0023$. If we combine also information
on hyperon decays, we obtain the final result
\be
\vert V_{us}\vert=0.2205\pm 0.0018 =\lambda            
\ee
which is in good agreement with (4.16) confirming the unitarity of $V$.
\medskip
The elements $V_{cd}$ and $V_{cs}$ come from di-muon production in
deep-inelastic scattering $\nu_\mu N \rightarrow \mu^+\mu^- X$ which are
supposed to correspond to charmed particle production and their
semileptonic decays. Both the valence $d$-quark and the sea $s$-quark
contribute comparably but with different fractional longitudinal momentum
(Feynman $x$) distributions.
Combining the CDHS Tevatron and CLEO results, the Particle Data Group
determined,
\bn
\vert V_{cd}\vert &=0.224\pm 0.016 \\
\vert V_{cs}\vert &=1.01\pm 0.18
\en
consistent with unitarity but much less precise than (4.15) and (4.18).
\medskip
The elements $V_{ub}$ and $V_{cb}$ are first determined by obtaining their
ratio as given in the ratio of rates
\be
R={{\Gamma (b\rightarrow u l^-\bar\nu_l)}\over
{\Gamma (b\rightarrow c l^-\bar\nu_l)}}               
\ee
with, according to the spectator model
\be
\Gamma(b\rightarrow q l^-\bar\nu_l)= \vert V_{qb}\vert^2
{{G_F^2m_b^5}\over{192\pi^3}} f\left({{m_q^2}\over{m_b^2}}\right)
\left[1-{2\over 3}{{\alpha_s(m_b^2)}\over{\pi}} F\left({{m_q}\over{m_b}}
\right) \right]                                       
\ee
for $q=u,~c$. Here the 3-body phase space factor is non-negligeable
for the charmed quark and the lowest order QCD correction term$^{[13]}$ $F$
is about 2.5 for the charmed quark and 3.61 for the up quark
in the minimal subtraction scheme with $\Lambda_{\overline{MS}}=0.15$ GeV.
The strong running coupling constant has value $\alpha_s(m_b^2) \approx 0.16$.
The two components of the total semileptonic decay can, in principle,
be extracted because of the different shapes of the inclusive lepton
momentum spectrum near the end point.
From the CLEO and ARGUS observation of $b\rightarrow u$ transition in 
semileptonic $B$-decays, the Particle Data Group concludes
\be
\vert V_{ub}\vert / \vert V_{cb}\vert = 0.08 \pm 0.02
\ee
where the error is the combined uncertainties of experimental and
theoretical origin.
From the $B$-lifetime and the semi-leptonic exclusive $B \rightarrow
\bar{D^{\ast}}l\nu_{l}$  decay, we can obtain separately $V_{cb}$ 
\be
 \vert V_{cb}\vert = 0.041\pm0.003,
\ee
and we have the Wolfenstein parameters
\be
A=0.74-0.95,~~~~~~~~0.074<\rho^2+\eta^2 <0.207          
\ee
The remaining elements are then limited by unitarity if we assume
only 3 generations. Different weighing of experiments and a few
technically involved procedures lead to the Particle Data Group's most recent
(June 1996) range for the magnitude of the flavor-matrix elements listed below,
at the 90\% Confidence Level (CL)
\bn
 V=\left(\begin{array}{ccc}
 0.9745-0.9757 & 0.219-0.224 & 0.002-0.005 \\
              0.219-0.224 & 0.9736-0.9750 & 0.036-0.046 \\
              0.004-0.014 & 0.034-0.046 & 0.9989-0.9993 \end{array}\right)
\en 
corresponding to $s_\theta=0.219-0.223,~ s_\gamma=0.036-0.046,
~s_\beta=0.002-0.005$ at the 90\% CL in the Maiani representation.
Comparison of the ``experimental" mixing matrix elements with quark mass
matrix ans\"atze have been made. Generally the consistency between experiments
and theoretical models based on the calculability of the flavor-mixing
matrix in terms of the quark masses can be achieved but $m_t \approx 110$
GeV.$^{[12]}$
\bigskip\noindent
\section{ Renormalizability and radiative corrections}
\setcounter{equation}{0}
\bigskip
We recall that it is the masslessness of the photon, i.e. the absence of a
longitudinal component to the propagator, which insures that QED is
a renormalizable theory and thus that we can compute, to any order of
Perturbation Theory (PT), radiative corrections to various electrodynamic
processes such as the anomalous magnetic moment of the electron, for example.
Mathematically stated, the photon propagator
\be
 P^{\mu\nu}(k) = {{-g^{\mu\nu}+k^\mu k^\nu/k^2}\over {k^2}}  
\ee
obeys $k_\mu P^{\mu\nu}=0$ as well as having a well-defined
($\propto k^{-2}$) high-momentum (ultra-violet) behaviour.
We have seen however that the original electro-weak renormalizable theory
has suffered SSB with, as a result, all IVBs acquiring mass, leaving only
the photon massless. We would therefore expect a propagator
\be
 P^{\mu\nu}(k) = {{-g^{\mu\nu}+k^\mu k^\nu/M^2}\over {k^2-M^2}}
\ee
which has a bad ultra-violet behaviour and for which longitudinal components
exist, i.e. $k_\mu P^{\mu\nu} \neq 0$.
Renormalizability is however preserved, although not obviously,
as we can see by first writing down a time-ordered product of fields
in the path integral formalism of any Quantum Field Theory (QFT):
\be
<0 \vert T(A_\mu\phi\cdots) \vert 0> =
{{\int [d A_\mu][d\phi]\cdots (A_\mu\phi\cdots) e^{iS(A_\mu,\phi,\cdots)}}
\over {\int [d A_\mu][d\phi]\cdots  e^{iS(A_\mu,\phi,\cdots)}}}
\ee
where the action integral is the usual four-dimensional integral
over the Lagrangian density
\be
S(A_\mu,\phi\cdots)=\int d^4x{\cal L}(A_\mu,\phi,\cdots)     
\ee
Under a unitary gauge transformation $UU^\dagger=1$ i.e. such that
$(\partial^\mu U)U^\dagger +U\partial^\mu U^\dagger=0$ we thus have
simultaneously
\bn
\phi & \rightarrow & U(\Lambda) \phi \equiv \phi_U \nn
A^\mu & \rightarrow & U A^\mu U^\dagger-{i\over g} (\partial^\mu U) U^\dagger
=U A^\mu U^\dagger+{i\over g} U \partial^\mu U^\dagger \equiv A^\mu_U
\en
Within each orbit (the set of $A^\mu$ for an $U$) the integrand will be
independent of $\Lambda$. Hence the functional integration will introduce
spurious infinities due to multiple-counting of fields actually related
simply by a gauge transformation. As is usual to adepts of quantum
mechanics, the solution lies in introducing a cleverly chosen factor
of {\bf one} and inverting the order of integration. The former practice occurs
in the introduction of a complete sets of states in PT and the latter in
Dirac $\delta$-function manipulations.

Faddeev and Popov$^{[8]}$ introduce
\be
1=\Delta_f(A^\mu,\phi,\cdots)\int[dU]f(A^\mu_U,\phi_U)    
\ee
where the first term is a determinant, $[dU]$ is a gauge-invariant measure
and $f$ a functional, not necessarily gauge invariant, often called
a gauge-fixing term, made up of our constraint on the choice of gauge,
as expressed in $\delta$-functions (it should be clear now that the
determinant is related to the Jacobian of the transformation between
the gauge fixing equations and the fields so as to make the RHS of (5.6)
become one).
The absorption of the determinant into the action integral forces us to
introduce anticommuting fields $\eta_\alpha, \bar \eta_\alpha$ referred to
as Faddeev-Popov ghost particles which add a new term to the action.
In the general $R_\xi$-gauge, the gauge fixing term takes the form
\be
\prod_{x,\alpha} \delta(\partial_\mu A^\mu_\alpha+ i\xi <\phi^T>
g_\alpha T_\alpha \phi-C_\alpha)                        
\ee
which when weighted with a Gaussian measure for the $C_\alpha$ yield
a contribution to the action which results in the massive gauge boson
propagator taking the final form
\bn
 P^{\mu\nu}(k)&=&{{-g^{\mu\nu}+(1-\xi)k^\mu k^\nu/(k^2-\xi M^2)}
\over {k^2-M^2}} \nn
&=& {{-g^{\mu\nu}+k^\mu k^\nu/ M^2}\over
{k^2-M^2}} -{{k^\mu k^\nu/M^2}\over{k^2-\xi M^2}} 
\en
The badly behaved first term (which is just equation (5.2)) is now
compensated for by the $\xi$-dependent second term.
Three familiar gauges appear as special cases:
\begin{itemize}
\item{1)} the Landau gauge has $\xi=0$
\item{2)} the 't Hooft-Feynman gauge has $\xi=1$
\item{3)} the unitary gauge corresponding to $\xi \rightarrow \infty$
\end{itemize}
The Faddeev-Popov method modifies the action integral (5.4) in the path
integral formalism to an effective form
\bn
S_{eff}(A_\mu,\phi,\bar \eta,\eta,\dots) = S(A_\mu,\phi,\cdots) -
{1\over 2}\int d^4x\sum_\alpha(f_\alpha)^2 \nn
~~~~~~~~~~~~~~~~- \int d^4x d^4y \bar \eta_\alpha(x)
M_{\alpha\beta}(x,y)\eta_\beta(y)                            
\en
where $Det~ M = \Delta_ f$ is given by
\be
\delta f_\alpha(x)=\int d^4y M_{\alpha\beta}(x,y)f_\beta(y)  
\ee
and $\eta_\alpha$ are the ghost fields. In the $SU(2)\times U(1)$ model,
the gauge-fixing term needed is
\be
{1\over\xi}\vert \partial^\mu W^+_\mu +i\xi M_W\phi^+\vert^2+
{1\over{2\alpha}}\left(\partial^\mu Z_\mu+\alpha M_Z \eta_I\right)^2
+{1\over{2\alpha}}(\partial^\mu A_\mu)^2                 
\ee
where $\eta_I$ is the imaginary part of the neutral component of the shifted
Higgs scalar $\eta$. We then have to introduce the corresponding
anticommuting ghost fields $\eta_\pm, \eta_0$ and $\eta_A$ to compensate
the cross-terms like $\partial^\mu \phi^+ W^-_\mu, ~\partial^\mu\phi^0 Z_\mu$
etc. Then after some tedious but straightforward manipulations, one finds the
following propagators for the particles in the model
\bn
  W^\pm :&==>& i\left[ -g^{\mu\nu}+(1-\xi)k^\mu k^\nu /(k^2-\xi M^2_W)
\right] / (k^2-M_W^2) \nn
Z :&==>& i\left[ -g^{\mu\nu}+(1-\alpha)k^\mu k^\nu /(k^2-\alpha M^2_Z)
\right] / (k^2-M_Z^2) \nn
A :&==>& i\left[ -g^{\mu\nu}+(1-\alpha)k^\mu k^\nu /k^2 \right] / k^2\nn
\eta_\pm :&==>& -i\left( k^2-\xi M_W^2\right)^{-1} \nn
\eta_0 :&==>& -i\left( k^2-\alpha M_Z^2\right)^{-1} \nn
\eta_A :&==>& -i(k^2)^{-1} \nn
\phi^\pm :&==>& i\left[ k^2-(\xi M_W^2+\mu^2)\right]^{-1} \nn
\eta=Re \phi^0 :&==>& i(k^2-\mu^2)^{-1} \nn
\eta_I=Im \phi^0 :&==>& i\left[k^2-(\alpha M_Z^2+\mu^2)\right]^{-1} \nn
\psi_i :&==>& -i(\gamma^\mu p_\mu -m)^{-1}                   
\en
These propagators and the interaction terms in (2.15), (2.18), (2.30)
and (3.32) determine the complete Feynman rules of the $SU(2)\times U(1)$
theory. We leave the interaction terms governing the Yukawa couplings among the
particles as an exercise to the reader.
\begin{figure}[htb]
\centerline{
\epsfig{figure=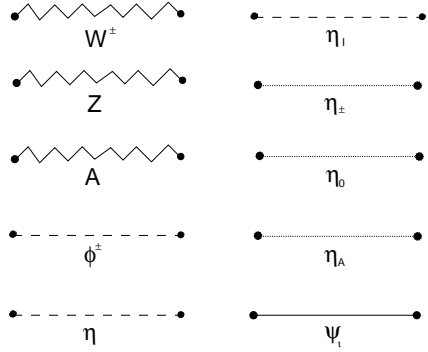, height=6cm, angle=0}}
\caption{propagators}
\end{figure}
\subsection{ Renormalization}
In perturbative theory the Lagrangian $\cal{L}$ has to be 
considered as the ``bare" Lagrangian of the electroweak theory with 
``bare" parameters which are related to the physical ones by
\begin{eqnarray}
                e_0 &=& e + \delta e \\
           M_W^{02} &=& M_W^2 + \delta M_W^2 \\
           M_Z^{02} &=& M_Z^2 + \delta M_Z^2
\end{eqnarray}
Then, the Lagrangian can be splitted into a ``renormalized" Lagrangian and a
counter term Lagrangian
\begin{equation}
        {\cal L}(\phi_0, g_0) = {\cal L}(Z_{\phi}^{1/2}\phi, Z_gg)
      ={\cal L}(\phi, g) + \delta {\cal L}(\phi, g, \delta Z_{\phi}, \delta g)
\end{equation}
which renders the results for all Green functions in a given order finite$^{[14]}$.

The simplest way to obtain a set of finite Green functions is the
`` minimal subtraction scheme"$^{[15]}$ where (in dimensional
regularization) the singular part of each divergent diagram is subtracted
and the parameters are defined at an arbitrary mass scale $\mu$.
This scheme, with slight modifications, has been applied in QCD where due
to the confinement of quarks and gluons there is no distinguished mass
scale in the renormalization procedure.

The situation is different in QED and in the electroweak theory.
There the classical Thomson scattering and the particle masses set
natural scales where the parameters can be defined.
In QED the favored renormalization scheme is the on-shell scheme
where $e = \sqrt{4\pi \alpha }$ and the electron, muon,... masses are used
as input parameters.
The finite parts of the counter terms are fixed by the renormalization 
conditions that the fermion propagators have poles
at their physical masses, and $e$ becomes the $ee\gamma$ coupling constant in
 the Thomson limit of Compton scattering.
The extraordinary meaning of the Thomson limit for the definition of the
renormalized coupling constants is elucidated by the theorem that the exact
Compton cross section at low energies becomes equal to the classical
Thomson cross section.
In particular this means that $e$ reps. $\alpha $ is free of infrared
corrections, and that its numerical value is independent of the order
od perturbation theory, only determined by the accuracy of the experiment.
This feature of $e$ is preserved in the electroweak theory.

{\bf (1)  Renormalization Conditions}\\
The on-shell subtraction of the self energies are satisfied with the 
conditions ;
\begin{equation}
       Re \tilde{\Sigma}^{WW}(M_W^2) 
     = Re \tilde{\Sigma}^{ZZ}(M_Z^2)
     = Re \tilde{\Sigma}^{f}(p=m_f) = 0
\end{equation}
where $~\tilde{}~$ denotes the renormalized self energies.
The generalization of the QED charge renormalization gives
the following relations ;
\begin{equation}
      \tilde{\Gamma }^{\gamma ee}_{\mu }(k^2=0, p=q=m_e)=ie\gamma_{\mu}
\end{equation}
In addition, there are conditions which should be satisfied by self-energies,
\begin{eqnarray}
       \tilde{\Sigma}^{\gamma Z}(0) = 0 \\
                \frac{\partial \tilde{\Sigma }^{\gamma \gamma}}
		     {\partial k^2}(0) = 0\\
                \lim_{k\rightarrow m_{f}}\frac{ \tilde{\Sigma}^{f}(k)u_{-}(k)}
                    {k-m_{f}} = 0
\end{eqnarray}
where $u_{-}$ is the $I_3 = -\frac{1}{2} $ fermion wave function. \\
{\bf (2) Mass renormalization}\\
Let us consider the gauge-boson propagators.
We restrict our discussion to the 't Hooft-Feynman gauge, i.e., to
the transverse parts $\sim g_{\mu \nu }$.
In the electroweak theory, differently from QED, the longitudinal
components $\sim q_{\mu}q_{\nu} $ of the vector boson propagators do not
give zero results in physical matrix elements.
But for light external fermions the contributions are suppressed by
$(m_f/M_Z)^2 $ and we are allowed to neglect them.
Writing the self-energies as
\begin{equation}
\Sigma^{W,Z}_{\mu \nu } = g_{\mu \nu } \Sigma^{W,Z} + \cdots 
\end{equation}
with scalar functions $\Sigma^{W,Z} (q^2) $ we have for the 1-loop 
propagators $(V=W,Z)$
\begin{eqnarray*}
               \frac{-ig^{\mu \sigma }}{k^2-M^{02}_V}
		    (-i\Sigma^{V}_{\sigma \rho }(k))
               \frac{-ig^{\rho \sigma }}{k^2-M^{02}_V}
 = \frac{-ig^{\mu \sigma }}{k^2-M^{02}_V}
		    \left(\frac{-\Sigma^{V}(k)}{k^2-M^{02}_V}\right)
\end{eqnarray*}
In the graphical representation, the self-energies for the vector
bosons denote the sum of all the diagrams with virtual fermions,
vector bosons, Higgs and ghost loops.
Resumming all self energy-insertions yields a geometrical series
for the dressed propagators:
\begin{eqnarray}
    & & \frac{-ig^{\mu \nu }}{k^2-M^{02}_V}
         \left[1+\left( \frac{-\Sigma^{V}}{k^2-M^{02}_V}\right) 
         +\left( \frac{-\Sigma^{V}}{k^2-M^{02}_V}\right)^2
         + \cdots \right] \nonumber \\
    &=& \frac{-ig^{\mu \nu }}{k^2-M^{02}_V+\Sigma ^{V}(k^2)} 
	       \nonumber \\
     &\equiv & -ig^{\mu \nu}D_V(k^2).
\end{eqnarray}
Since there are mixing of $\gamma $ and $Z$ at quantum level,
the propagator of the neutral boson has to be considered as a
$2\times 2$ matrix ;
\begin{eqnarray}
    \left( \begin{array}{cc}
	                 k^2+\Sigma^{\gamma \gamma }(k^2) & 
			     \Sigma^{\gamma Z}(k^2) \\
	                     \Sigma^{\gamma Z}(k^2) & k^2-M^{02}_Z 
			   + \Sigma ^{ZZ}(k^2)
	    \end{array} \right) .
\end{eqnarray}
Inverting this matrix gives the neutral gauge boson propagators as follows
\begin{eqnarray}
      D_{\gamma}(k^2) &=& \frac{1}{k^2+\Sigma ^{\gamma \gamma}(k^2)
                         -(\Sigma^{\gamma Z}(k^2))^2/(k^2-M_Z^{02}
			 +\Sigma^{ZZ}(k^2))}, \\
      D_{\gamma Z}(k^2) &=& \frac{\Sigma^{\gamma Z}(k^2)}
		                 {(k^2+\Sigma ^{\gamma \gamma}(k^2))
                            (k^2-M_Z^{02}+\Sigma^{ZZ}(k^2))
			    -(\Sigma^{\gamma Z}(k^2))^2} ,\\
      D_{Z Z}(k^2) &=& \frac{1}
			    {k^2-M_Z^{02}+\Sigma^{ZZ}(k^2)
			    -(\Sigma^{\gamma Z}(k^2))^2/
  	                     (k^2+\Sigma^{\gamma \gamma}(k^2))}.
\end{eqnarray}
Here, the last substracted term of the denominators are higer order
contributions and they can be ignored at one loop approximation.
In order to acquire the physical masses of the gauge bosons we use
the relation (5.14-5.15) and the definition of physical mass which is
identical to real part of the pole positions of corresponding 
propagators.
Upon requiring the renormalization condition (5.17) 
the mass counter terms $\delta M_{W,Z}^2$ get fixed as
\begin{eqnarray}
      \delta M^2_W &=& Re\Sigma^{WW}(M_W^2), \\
      \delta M^2_Z &=& Re\left(\Sigma^{ZZ}(M_Z^2) 
                    -\frac{(\Sigma^{\gamma Z}(M_Z^2))^2}
                          {M_Z^2+\Sigma^{\gamma \gamma }(M_Z^2)}\right).
\end{eqnarray}
{\bf (3) Charge renormalization}\\
Since the electroweak theory contains the electromagnetic charge $e$,
we have to maintain its definition as classical charge in Thomson
cross section $\sigma_{Thomson}=e^4/6\pi m_e^2$.
Accordingly, the Lagrangian carries the bare charge $e_o = e + \delta e $ with
the charge counter term $\delta e$ of 1-loop order.
The charge counter term $\delta e$ has to absorb the electroweak loop
contributions to the $e e \gamma $ vertex in the Thomson limit.
This charge renormalization condition is simplified by the validity
of a generalization of the QED Ward identity which implies that
those corrections related to the external particles cancel each other.

Then the bare $\gamma ee$ vertex is corrected to
\begin{equation}
      ie_0\gamma^{\mu }\rightarrow i\left[e_0-\frac{1}{2}e_0\Pi_{\gamma}(0)
                     + \frac{s_W}{c_W}
		       \frac{\Sigma^{\gamma Z}(0)}{M_Z^2}\right]\gamma^{\mu }
\end{equation}
where we have used the convention, 
       $\Pi^{\gamma}(k^2)=\Sigma ^{\gamma \gamma }(k^2)/k^2$
called  the ``vacuum
polarization" of the photon and $s_W$ and $c_W$ denoting
$\sin \theta_W$ and $\cos \theta_W$ respectively.
As an application of the renormalization condition (5.18), we obtain 
the following relation ;
\begin{equation}
      \frac{\delta e}{e} = \frac{1}{2}\Pi^{\gamma}(0)-\frac{s_W}{c_W}
                           \frac{\Sigma^{\gamma Z}(0)}{M_Z^2}.
\end{equation}
We note that the fermionic loop contributions to $\Sigma ^{\gamma Z}$ 
vanish at $q^2=0$; only the non-Abelian bosonic loops yield
$\Sigma^{\gamma Z}(0) \neq 0$.\\
\subsection{ Dimensional Regularization}
In general, the loop corrections involve the integrals with the UV
divergences as $p^2\rightarrow \infty $.
In order to remedy the difficulty, we need a regularization.
As usual, the dimensional regularization procedure is adopted
for gauge theories.
The main idea is to replace the space-time dimension 4 by a lower
dimension $d$ where the integrals become convergent ; 
\be
\int \frac{d^4p}{(2\pi )^4} \rightarrow \mu ^{4-d}\int 
\frac{d^dp}{(2\pi )^4}
\ee
where an arbitrary mass parameter $\mu $ is introduced in order to
keep the coupling constants in front of the integrals to be
dimensionless.
After calculations of physical quantities, we take the limit
$d\rightarrow 4$ and then results become finite.

Let us recall some algebraic relations; 
\begin{itemize}
\item {\bf metric ;} \\
\begin{eqnarray}
   g_{\mu \nu } = \left( \begin{array}{cccc}
		       1 &  0 &  0 &  0  \\
		       0 & -1 &  0 &  0  \\
		       0 &  0 & -1 &  0  \\
		       0 &  0 &  0 & -1   
		       \end{array}  \right),~~~~~~~~~~
   g_{\mu \nu }g^{\mu \nu } = Tr(1) = d
 \end{eqnarray}
\item {\bf Dirac algebra in $d$ dimensions ;} \\
 \begin{eqnarray}
         \{\gamma^{\mu }, \gamma^{\nu } \} &=& 2 g^{\mu \nu },~~~
  \{\gamma^{\mu }, \gamma_5 \} = 0 
  ,~~~ \gamma_5 = i\gamma^0\gamma^1\gamma^2\gamma^3 \nonumber \\
  \gamma_{\mu }\gamma^{\mu } &=& d , ~~~~~
  \gamma_{\rho }\gamma_{\mu }\gamma^{\rho } = (2-d)\gamma_{\mu } \nonumber \\
  \gamma_{\rho }\gamma_{\mu }\gamma_{\nu }\gamma^{\rho } &=& 4g_{\mu \nu }
    - (4-d)\gamma_{\mu }\gamma_{\nu }, \nonumber \\
  \gamma_{\rho }\gamma_{\mu }\gamma_{\nu }\gamma_{\sigma }\gamma^{\rho }
     &=& -2\gamma_{\sigma }\gamma_{\nu }\gamma_{\mu }+(4-d)
	 \gamma_{\mu }\gamma_{\nu }\gamma_{\sigma }
 \end{eqnarray}
 \item {\bf  the trace relations ;} \\
 \begin{eqnarray}
   Tr(\gamma^{\mu }\gamma^{\nu }) &=& 4g^{\mu \nu } \\
   Tr(\gamma^{\mu }\gamma^{\nu }\gamma^{\rho }\gamma^{\sigma }) &=&
   4\left[g^{\mu \rho }g^{\nu \sigma }-g^{\mu \nu }g^{\rho \sigma }\right.
     +\left. g^{\mu \sigma }g^{\nu \rho }\right] \\
   Tr(\gamma_5 \gamma^{\mu }\gamma^{\nu }\gamma^{\rho }\gamma^{\sigma})
     &=& 4i\epsilon^{\mu \nu \rho \sigma } \\
\vspace{12mm}
   Tr(\gamma_5\gamma^{\mu }\gamma^{\nu }) &=& 0
 \end{eqnarray}
 \end{itemize}
A consistent treatment of $\gamma_5 $ in $d$ dimensions is more subtle$^{[16]}$.
In theories which are anomaly free like the standard model we can use $\gamma_5$
as anticommuting with $\gamma_{\mu}$:
\begin{equation}
 \{\gamma_{\mu}, \gamma_{5} \} = 0.
\end{equation}
\subsection{ Calculation of loop-integrals} 
 For convenience, we define several types of integrals 
 \begin{eqnarray}
 \int Dp~ \frac{1}{[~~1~~]}
 &\equiv &\frac{i}{16\pi ^2}~\mbox{I}~(m_1), \\
 \int Dp~ \frac{1}{[~~1~~][~~2~~]}
 &\equiv &\frac{i}{16\pi ^2}~\mbox{II}_0(k^2,m_1,m_2), \\
 \int Dp~ \frac{p^{\mu }}{[~~1~~][~~2~~]}
 &\equiv &\frac{i}{16\pi ^2}~k^{\mu }~\mbox{II}_1(k^2,m_1,m_2), \\
 \int Dp~ \frac{p^{\mu }p^{\nu }}{[~~1~~][~~2~~]}
 &\equiv &\frac{i}{16\pi ^2}\left[~{k^{\mu }k^{\nu }
                ~\mbox{II}_{21}(k^2,m_1,m_2)
 -g^{\mu \nu }~\mbox{II}_{22}(k^2,m_1,m_2)}~\right].
 \end{eqnarray}
where $ \mu ^{\epsilon}\int \frac{d^dp}{(2\pi )^d}\equiv \int Dp,
~\epsilon = 4-d,~ p^2-m_1^2\equiv [~~1~~],$ and
 $(p^2+k^2)-m_2^2\equiv [~~2~~]$. \\
The 1-point integral $\mbox{I}$ in (5.40) 
can be transformed into a Euclidean integral:
\be
 \frac{i}{16\pi ^2}~\mbox{I}~(m) = -i\frac{\mu^{\epsilon}}{(2\pi)^d}
 \int \frac{d^dp_E}{p^2_E + m^2}. 
\ee
This $p_E$- integral is a special 
of the rotationally invariant integral in a $d$-dimensional Euclidean space,
\begin{eqnarray*}
 \int \frac{d^dp_E}{(p^2_E + m^2)^n}
\end{eqnarray*}
They can be evaluated in $d$-dimensional polar coordinates $(p_E^2 =P)$
\begin{eqnarray*}
 \int \frac{d^dp_E}{(p^2_E + m^2)^n} = \frac{1}{2}\int d\Omega_d
 \int^{\infty}_{0} dP P^{\frac{d}{2}-1} \frac{1}{(P+m^2)^n},
\end{eqnarray*}
yielding
\be
\frac{\mu^{\epsilon}}{(2\pi)^d} \int \frac{d^dp_E}{(p^2_E + m^2)^n}
=\frac{\mu^{\epsilon}}{(4\pi)^{d/2}}\frac{\Gamma (n-\frac{d}{2})}
 {\Gamma(n)}\cdot (m^2)^{-n+\frac{d}{2}}.
\ee
The singularities of the original 4-dimensional integrals are
now recovered as poles of the $\Gamma$-function for $d=4$ and
$n\leq 2$.

Although the LHS of Eq.(5.45) as a $d$- dimensional integral is
sensible only for integer values of $d$, the RHS has an
analytic continuation in the variable $d$: it is well defined for all
complex values $d$ with $n-\frac{d}{2} \neq 0, -1, -2, \cdots,$
in particular for $d=4-\epsilon $ with $\epsilon > 0$.
For physical reasons we are interested in the vicinity of $d=4$.
Hence we consider the limiting case $\epsilon \rightarrow  0$ and
perform an expansion around $d=4$ in powers of $\epsilon$.
For this task we need the following properties of the
$\Gamma$-function at $x\rightarrow 0$:
\begin{eqnarray}
\Gamma (x) =\frac{1}{x} - \gamma + O(x), \nonumber \\
\Gamma (-1+x) =-\frac{1}{x} + \gamma -1 + O(x)
\end{eqnarray}
with $\gamma = -\Gamma^{\prime}(1) =0.577 \cdots$ known as Euler's
constant.
Combining (5.44) and (5.45) we obtain the scalar 1-point integral 
for $d=4-\epsilon$:
\begin{eqnarray}
\mbox{I}(m) &=& -\frac{\mu^{\epsilon}}{(4\pi)^{-\epsilon/2}}
          \frac{\Gamma(-1+\frac{\epsilon}{2})}{\Gamma(1)}
           (m^2)^{1-\epsilon/2} \nonumber \\
    &=& m^2 \left( \frac{2}{\epsilon} - \gamma
        +\ln 4\pi - \ln \frac{m^2}{\mu^2} + 1 \right) +O(\epsilon) \nonumber \\
  &\equiv & m^2 \left( \Delta - \ln \frac{m^2}{\mu^2} +1 \right)
      +O(\epsilon).
\end{eqnarray}
Here we have introduced the abbreviation for the  singular part
\begin{equation}
\Delta = \frac{2}{\epsilon} - \gamma + \ln 4\pi .
\end{equation}
\bigskip
\\
With help of the Feynman parameterization
\begin{eqnarray*}
\frac{1}{AB} = \int^1_0 dx \frac{1}{[Ax + B(1-x)]^2}
\end{eqnarray*}
and after a shift in the $p-$variable, the two point function
$\mbox{II}_0$ can be written in the form
\be
 \frac{i}{16\pi^2}\mbox{II}_0(k^2,m_1,m_2) =  \int^{1}_{0}dx
 \frac{\mu^{\epsilon}}{(2\pi)^d} \int \frac{d^dp}
 {[p^2-x^2k^2 +x(k^2+m_1^2-m_2^2)-m_1^2]^2}.
\ee
The advantage of this parameterization is a simpler $p-$integration
where the integrand is only a function of $p^2=(p^0)^2-\vec{p}^2$.
In order to transform it into a Euclidean integral we
perform the substitution
$p^0 = ip^0_E,~ \vec{p}=\vec{p}_E,~ d^dp =i d^dp_E $
where the new integration momentum $p_E$ has a definite metric:
$p^2 = -p_E^2,~ p_E^2 = (p^0_E)^2+\cdots +(p_E^{d-1})^2 $.
This leads us to a Euclidean integral over $p_E$:
\be
 \frac{i}{16\pi^2}\mbox{II}_0 =  i\int^{1}_{0}dx
 \frac{\mu^{\epsilon}}{(2\pi)^d} \int \frac{d^dp_E}
 {(p_E^2 + K)^2}
\ee
where $K=x^2k^2-x(k^2+m_1^2-m_2^2)+m_1^2-i\epsilon $ is a constant with
respect to the $p_E-$integration.
Using Eq. (5.45) with $n=2$, we obtain
\begin{eqnarray}
 \mbox{II}_0(k^2,m_1,m_2) &=& \Delta - \int^{1}_{0}dx \ln {
 \left[~\frac{x^2k^2-xk^2-x(m_1^2-m_2^2)+m^2_1-i\epsilon}{\mu ^2}~\right]}\nn
 &=& \Delta - \ln \frac{m_1m_2}{\mu^2} + C(k^2, m_1, m_2).
\end{eqnarray}
where $C(k^2, m_1, m_2)$ is essentially the same integral as in the first
line except that $\mu^2$ is replaced by $m_1m_2$ for convenience.
The remaining integrals in (5.42) and (5.43) can be related to
$\mbox{I}(m)$ and $\mbox{II}_0$:
 \begin{eqnarray}
 \mbox{II}_1(k^2,m_1,m_2) &=& \frac{1}{2k^2}
 \left[~ \mbox{I}~(m_1)-\mbox{I}~(m_2)-(m_1^2-m^2_2+k^2)~\mbox{II}_0
      (k^2,m_1^2,m_2^2)~\right], \\
 \mbox{II}_{21}(k^2,m_1,m_2) &=& \frac{1}{3k^2}
        \left[~ \mbox{I}~(m_2)-2(m_1^2-m^2_2+k^2)
        ~\mbox{II}_1(k^2,m^2_1,m^2_2)\right. \nn
        & &\left. -m^2_1 ~\mbox{II}_0(k^2,m_1^2,m_2^2) 
            -\frac{1}{2}(m_1^2+m_2^2-\frac{1}{3}k^2)~\right] ,\\
 \mbox{II}_{22}(k^2,m_1,m_2) &=& \frac{1}{6}~
      \left[~ -\mbox{I}_1(m_2)-2m_1^2~\mbox{II}_0(k^2,m_1,m_2)
          -(m_1^2-m_2^2+k^2)\right. \nn
       & & \left. \times ~\mbox{II}_1(k^2,m_1,m_2)-(m^2_1+m^2_2
           -\frac{1}{3}k^2)~\right].
 \end{eqnarray}

 In particular, in cases of equal masses $m_1=m_2=m$,
$\Pi_0$ and $\Pi_1$ become
 \begin{eqnarray}
 \mbox{II}_0(k^2,m,m) &=& \Delta - \ln {\frac{m^2}{\mu ^2}}+C(k^2,m,m), \\
 \mbox{II}_1(k^2,m,m) &=& -\frac{1}{2}~\mbox{II}_0(k^2,m,m), 
 \end{eqnarray}
 where
 \begin{eqnarray}
 C(k^2,m,m) &=& -\int ^{1}_{0}dx \ln {\left[(x^2-x)\frac{k^2}{m^2}+1\right]}\nn
           \nn
             &=& \left\{ \begin{array}{l}
                2\left[~1-\sqrt{4(m^2/k^2)-1}\arcsin\left( 
		\frac{1}{\sqrt{4m^2/k^2}}\right)~\right] ,~~~4(m^2/k^2) > 1 \\
           \\
		2\left[~1-\sqrt{1-4(m^2/k^2)}\ln \left(
		\frac{1+\sqrt{1-4(m^2/k^2)}}{\sqrt{4m^2/k^2}}\right)~\right] 
		,~~~4(m^2/k^2) < 1 
		  \end{array}
		  \right. \\
 C(0,m,m)   &=& 0, \\
 C(k^2,m,m) &=& \left\{ \begin{array}{l}
                  2-\ln {\frac{k^2}{m^2}} + i\pi ~~~~~\mbox{for }~~~k^2 >>m^2 \\
                  \frac{1}{6}\left(\frac{k^2}{m^2}\right)
                  +\frac{1}{60}\left(\frac{k^2}{m^2}\right)^2
                   ~~~~~\mbox{for }~~~k^2 <<m^2  \end{array} \right.
\end{eqnarray}
Note that the second term for $k^2 <<m^2$ contributes to $C$ less than
$3\%$ when $k^2 \leq M_Z^2$ and $m=m_t$.
\subsection{ One-loop calculation of gauge boson self energies}
 {\bf (1) Fermion loop contribution to gauge boson self energies.} \\
 Fermion loop contribution to gauge boson self energies is given by Fig.2,
which we will denote by [{\bf FL}]

\begin{figure}[htb]
\centerline{
\epsfig{figure=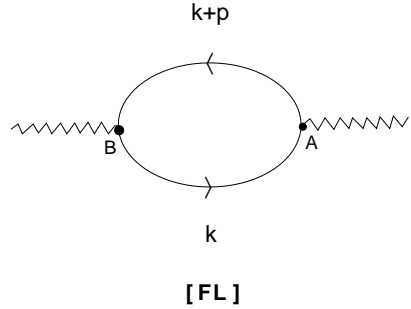, height=6cm, angle=0}}
\caption{fermion loop contribution to the gauge boson self energy.}
\end{figure}
 \begin{eqnarray}
 [\mbox{{\bf FL}}] &=& -i\Sigma^{\mu \nu }(k)  \nn
		        &=& -(AB)\int Dp~
	Tr\left\{~\frac{i(p+m_1)}{[~~1~~]}~\gamma^{\mu }(a_B+b_B\gamma^5)~
  \frac{i(p+k+m_2)}{[~~2~~]}~\gamma^{\nu }(a_A+b_A\gamma^5)~\right\}
 \end{eqnarray}
 where $A,B, a's$ and $b's$  are coupling constants at each vertices 
in $SU(2)_L \times
U(1)$ theory as given in Table 1.

 Now, let us write $\Sigma^{\mu \nu }(k)$ in the form
 \begin{equation}
 \Sigma^{\mu \nu }(k)=\left(g^{\mu \nu}-\frac{k^{\mu }k^{\nu }}{k^2}\right)
 \Sigma_1(k^2) + \frac{k^{\mu }k^{\nu }}{k^2}
     \left( k^2\Sigma _2(k^2)+\Sigma _1(k^2)\right).
 \end{equation}
 Note that only the transverse amplitude $\Sigma_1(k^2)$ contributes
 to S-matrix elements when contracted with
 a polarization vector. 
 Using the integrals of the previous section, we obtain
 \begin{eqnarray}
 \Sigma_1(k^2) &=& -\frac{(AB)}{4\pi^2}~
 \left  \{ ~\frac{1}{3}(a_Aa_B+b_Ab_B) ~\left[~\left(m_1^2+m_2^2
  -\frac{1}{3}k^2\right)
 -(~\mbox{I}~(m_1)+\mbox{I}~(m_2)~)\right. \right.\nn
 & & -\frac{1}{2k^2}(m_1^2-m_2^2)~(~\mbox{I}~(m_1)-\mbox{I}~(m_2)~)
  -\left(\frac{m_1^2+m_2^2}{2}-k^2
  +\frac{(m_1^2-m_2^2)^2}{2k^2}\right) \nonumber \\
 & & \left. \left. \times~ \mbox{II}_0(k^2,m_1,m_2)~\right] 
   + (a_Aa_B-b_Ab_B)~m_1m_2~\mbox{II}_0(k^2,m_1,m_2)~\right\}.
\end{eqnarray}
The fermion contributions to gauge 
boson self energies, $(\gamma \gamma),(\gamma Z),(Z Z)$ and $(W W)$,
can readily be read off from (5.62) by using the appropriate couplings
$A,B,a_i$ and $b_i$ for each cases from Table 1:
\begin{table}
\begin{center}
\begin{tabular}{|c||c|c|c|c|c|c|c|}\hline \hline
self energy type & $m_1$ & $m_2$ & $a_A$ & $a_B$ & $b_A$ & $b_B$ & $AB$ 
\\ \hline
($\gamma , \gamma $) & $m_f$ & $m_f$ & 1 & 1 & 0 & 0 & $-e^2Q_f^2$ \\
($\gamma , Z $) & $m_f$ & $m_f$ & $1$ & $v_f$ & 0 & $-a_f$ & $e^2Q_f$ \\
($Z , Z $) & $m_f$ & $m_f$ & $v_f$ & $v_f$ & $-a_f$ & $-a_f$ & $-e^2$ \\
($W , W $) & $m_1$ & $m_2$ & 1 & 1 & -1 & -1 & $-e^2/8s_W^2$ \\ \hline
\end{tabular}
\caption{Classification of parameters in Eq. (5.60)}
\end{center}
\end{table}
\begin{eqnarray}
\Sigma ^{\gamma \gamma }(k^2) &=& \sum_{f}\frac{e^2Q^2_f}{4\pi^2}~
 \frac{N_c^f}{3}~\left[~ 2m_f^2-\frac{1}{3}k^2-2\mbox{I}~(m_f)~+
 (k^2+2m_f^2)~\mbox{II}_0(k^2,m_f,m_f)~\right], \nonumber \\
 &=& \frac{\alpha }{3\pi }\sum_{f}Q_f^2k^2N_c^f
 \left[~ \Delta_f+P_f(k^2) \right], \\
\Sigma ^{\gamma Z }(k^2) &=& -\sum_{f}\frac{e^2Q_f}{4\pi^2}
 \frac{N_c^f}{3}v_f
 \left[~2m_f^2-\frac{1}{3}k^2-2\mbox{I}~(m_f)+(k^2+2m_f^2)
 ~\mbox{II}_0(k^2,m_f,m_f)~\right], 
  \nonumber \\
 &=& -\frac{\alpha }{3\pi }\sum_{f}Q_fv_fN_c^fk^2
 \left[~ \Delta_f+P_f(k^2) \right], \\
\Sigma ^{Z Z }(k^2) &=& \frac{\alpha }{3\pi}\sum_{f}N_c^f\left[(v_f^2+a_f^2)k^2
 \left(~ \Delta_f+P_f(k^2) \right)-\frac{3m_f^2}{8s_W^2c_W^2}
 \left(\Delta_f+C(k^2,m_f,m_f)\right) \right], \\
\Sigma ^{W W}(k^2) &=& \frac{\alpha }{4\pi s_W^2}
  \frac{1}{3}\sum_{f=(1,2)}N_c^f \left[ m_1^2+m_2^2-\frac{1}{3}k^2-
  (~\mbox{I}~(m_1)+\mbox{I}~(m_2)~) +\frac{1}{2k^2}~(m_1^2-m_2^2) \right. \nn
& &\left. \times~(\mbox{I}~(m_1)-\mbox{I}~(m_2)) 
   +~\left(k^2-\frac{(m_1^2+m_2^2)}{2}-\frac{(m_1^2-m_2^2)^2}{2k^2}\right)~
     \mbox{II}_0(k^2,m_1,m_2)\right], 
\end{eqnarray}
where 
\begin{eqnarray}
N_c^f = \left\{ \begin{array}{c}
          1 ~~~~~~~~\mbox{for~~leptons} \nn
          3 ~~~~~~~~\mbox{for~~quarks} \nonumber
           \end{array} \right\}
\end{eqnarray}
and $\Delta_f=\Delta -\ln \frac{m_f^2}{\mu^2},~~
v_f=(I_{3L}^f-2Q_fs_W^2)/2s_Wc_W, ~~a_f=I_{3L}^f/2s_Wc_W$ and 
\begin{eqnarray*}
P_f(k^2) &=& \left(1+\frac{2m_f^2}{k^2}\right)~C(k^2,m_f,m_f)-\frac{1}{3} \\
       &=& \left\{ \begin{array}{c}
		-\ln \frac{|k^2|}{m_f^2}+\frac{5}{3}+i\pi \theta(k^2),
                ~~~\mbox{for}~~|k^2| >> m_f^2 \\
           \\
              ~~~~ \frac{k^2}{5m_f^2},
                ~~~~~~~~~~~~~~~~~~\mbox{for}~~|k^2| << m_f^2 
		  \end{array}
		  \right.
\end{eqnarray*}
\bigskip
\\
  Now let us discuss the light and heavy fermion contributions separately.
\begin{table}
\begin{center}
\begin{tabular}{|c|c|c|}\hline \hline
fermions & $V_f$ & $A_f$ \\ \hline
 neutrino & $\frac{1}{2}$ & $\frac{1}{2}$   \\
$ e, \mu , \tau $ & $-\frac{1}{2}+2 s_W^2$ & $-\frac{1}{2}$   \\
$ u, c, t $ & $\frac{1}{2}-\frac{4}{3} s_W^2$ & $\frac{1}{2}$  \\
$ d, s, b $ & $-\frac{1}{2}+\frac{2}{3} s_W^2$ & $-\frac{1}{2}$  \\
 \hline \hline
\end{tabular}
\caption{we denote $V_f\equiv (2 s_W c_W) v_f, A_f\equiv 
(2 s_W c_W) a_f$}
\end{center}
\end{table}
  In particular, we represent the self energies at $k^2=0$ and
  $k^2=M_Z^2(~\mbox{or}~ M_W^2)$. \\
\bigskip
For heavy fermions (t,b), we get from Eqs.(5.63-5.66)\\
\begin{eqnarray}
  \Sigma ^{ \gamma \gamma}(0) &=&  0,	 \\
  \Sigma ^{ \gamma \gamma}(M_Z^2) &=&  \frac{\alpha }{9\pi}M_Z^2
      \left[5\Delta_t + \frac{4}{5}\frac{M_Z^2}{m_t^2}
       -\ln \frac{M_Z^2}{m_t^2}+\frac{5}{3}+i\pi\right], \\
  \Sigma ^{ \gamma Z}(0) &=&  0, \\
  \Sigma ^{ \gamma Z}(M_Z^2) &=& -\frac{\alpha }{\pi}M_Z^2
      \left[Q_tv_f\left(\Delta_t+\frac{M_Z^2}{5m_t^2}\right)+
        Q_bv_b\left(\Delta_Z+\frac{5}{3}+i\pi \right)\right], \\
  \Sigma ^{ ZZ}(0) &=& -\frac{\alpha }{\pi }\frac{3}{8s_W^2c_W^2}
     \left[m_t^2\Delta_t+m_b^2\Delta_b \right]  , \\
  \Sigma ^{ ZZ}(M_Z^2) &=&  \frac{\alpha }{\pi}M_Z^2
      \left[(v_t^2+a_t^2)\left(\Delta_t+\frac{M_Z^2}{5m_t^2}\right)+
      (v_b^2+a_b^2)\left(\Delta_Z+\frac{5}{3}+i\pi \right) \right. \nonumber \\
   & &\left. -\frac{3}{8s_W^2c_W^2}
     \frac{m_t^2}{M_Z^2}\left(\Delta_t+\frac{M_Z^2}{6m_t^2}
     +\frac{1}{60}\frac{M_Z^4}{m_t^4}\right)
      \right],  \\
\frac{d\Sigma^{ZZ}}{dk^2}(M_Z^2)&=& \frac{\alpha }{\pi}
\left[(v_t^2+a_t^2) \left(\Delta_t+\frac{M_Z^2}{6m_t^2}\right)
+ (v_b^2+a_b^2)\left(\Delta_Z + \frac{5}{3}+i\pi \right) \right.\nn
& & -\left.\frac{3}{8s_W^2c_W^2}\left(\frac{1}{6}+\frac{1}{30}
      \frac{M_Z^2}{m_t^2}\right)\right], \\
  \Sigma ^{ WW}(0) &=& -\frac{\alpha }{4\pi s_W^2}
  \left[\frac{3}{2}m_t^2\Delta_t+\frac{3}{4}m_t^2\right], \\
  \Sigma ^{ WW}(M_W^2) &=& \frac{\alpha }{4\pi s_W^2}
  \left[\left(M_W^2-\frac{3}{2}m_t^2\right)\Delta_t - 
  \frac{3}{4}m_t^2 + \frac{1}{3}M_W^2 \right].
  \end{eqnarray}
\bigskip
  The contributions of the light fermions $(m^2_f << M_Z^2)$ are,
  \begin{eqnarray}
  \Sigma ^{\gamma \gamma}(0) &=& 0, \\
  \Sigma ^{\gamma \gamma }(M_Z^2) 
   &\simeq & \sum_{f=light}\frac{N_c^f\alpha }{3\pi }Q_f^2M_Z^2
   \left(\Delta_Z+\frac{5}{3}+i\pi \right), \\
  \Sigma ^{ \gamma Z}(0) &=& 0, \\
  \Sigma ^{ \gamma Z}(M_Z^2) &=& -\frac{\alpha }{3\pi}\sum_{f}
    N_c^fQ_fv_fM_Z^2
   \left(\Delta_Z+\frac{5}{3}+i\pi \right), \\
  \Sigma ^{ ZZ}(0) &=& -\frac{\alpha }{3\pi}\sum_{f}
  \frac{3N_c^f}{8c_W^2s_W^2}m_f^2\Delta _f, \\
  \Sigma ^{ ZZ}(M_Z^2) &=& 
              \frac{\alpha }{3\pi}\sum_{f}N_c^fM_Z^2\left[(v_f^2+a_f^2)
    \left(\Delta_Z+\frac{5}{3}+i\pi \right)
   - \frac{3m_f^2}{8c_W^2s_W^2M_Z^2}
   \left(\Delta_Z+2 + i\pi \right)\right], \\
\frac{d\Sigma^{ZZ}}{dk^2}(M_Z^2) &=& \frac{\alpha }{3\pi}
  \sum_{f}N_c^f (v_f^2+a_f^2)\left[\Delta_Z+\frac{5}{3}
       +i\pi  \right], \\
  \Sigma^{WW}(0) &=& -\frac{\alpha }{16\pi s_W^2}\sum_{f=(1,2)}N_c^f
  \left[2(m_1^2\Delta_1+m_2^2\Delta_2)+(m_1^2+m_2^2)
    -\frac{2m_1^2m_2^2}{m_1^2 - m_2^2} \ln \frac{m_1^2}{m_2^2} \right], \\
  \Sigma ^{ WW}(M_W^2) &=& -\frac{\alpha }{12\pi s_W^2}\sum_{f=(1,2)}
  N_c^f M_W^2\left[\frac{\Delta_1+\Delta_2}{2}+C(M_W^2,m_1,m_2)-\frac{1}{3}
     \right].
\end{eqnarray}
\bigskip
{\bf (2) Vector and Scalar boson contributions} \\
The vector and scalar boson contributions to the photon and $Z$ boson self
energies are given by the diagrams in Fig.3\\
\begin{figure}[htb]
\centerline{
\epsfig{figure=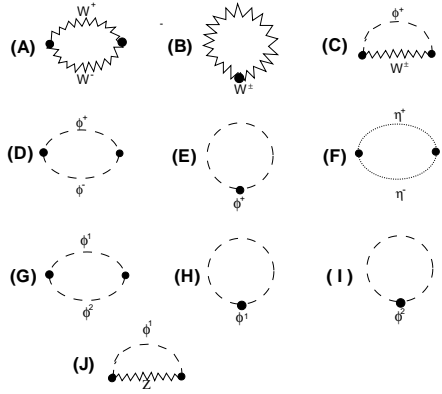, height=6cm, angle=0}}
\caption{Vector and scalar boson contributions to the gauge boson self
energies. We have omitted the external $\gamma $ and $Z$ boson lines for
simplicity which are obviously attached to the dots, except those in
({\bf B}), ({\bf E}) and ({\bf H}) to which two external gauge boson lines
are attached.}
\end{figure}

Let us define the integrals relevant to Fig. 3 by the following formulae ;
\begin{eqnarray}
(\mbox{{\bf A}}) &=& -i\int Dp~
		  \frac{E_{\mu \nu }}{[~~1~~][~~2~~]} \equiv 
           -i(k_{\mu }k_{\nu }~\mbox{I}(m^{V}_1)-g_{\mu \nu }~
            \mbox{I}_{V_2}), \\
(\mbox{{\bf B}}) &=& \int Dp~
		\frac{H_{\mu \nu }}{[~~1~~]} \equiv 
            (6-2\epsilon)~\mbox{I}(m^V_1), \\
(\mbox{{\bf C}}) &=&  i\int Dp~
		  \frac{g_{\mu \nu }}{[~~1~~][~~2~~]} \equiv 
            ig_{\mu \nu }~\mbox{II}_0, \\
(\mbox{{\bf D}}) &=& -i\int Dp~
		  \frac{(2p+k)^{\mu }(2p+k)^{\nu }}{[~~1~~][~~2~~]} \equiv 
	   -i(k_{\mu }k_{\nu }~\mbox{I}(m^{\phi}_1)-g_{\mu \nu }
             ~\mbox{I}_{\phi _2}), \\
(\mbox{{\bf E}}) &=& -\int Dp~
		 \frac{g_{\mu \nu }}{[~~1~~]}\equiv  
            -g_{\mu \nu }~\mbox{I}(m^{\phi}_1), \\
(\mbox{{\bf F}}) &=& i\int Dp~
		 \frac{p^\mu (p+k)^{\nu }}{[~~1~~][~~2~~]}\equiv 
	   -i(k_{\mu }k_{\nu }~\mbox{I}(m^{\eta}_1)-g_{\mu \nu }
               ~\mbox{I}_{\eta^2}) ,	
\end{eqnarray}
where the couplings at each verteices are left out but the Lorentz factors
\begin{eqnarray}
      E_{\mu \nu } &=&   k_{\mu }k_{\nu }(d-6)
		        +(k_{\mu }p_{\nu }+p_{\mu }k_{\nu })
                         (2d-3) + p_{\mu }p_{\nu }(4d-6) \nonumber \\
                   & &  +g_{\mu \nu }[(2k+p)^2+(k-p)^2], \\
      H_{\mu \nu } &=&   g_{\alpha \beta }[2g_{\mu \nu }g_{\alpha \beta}
                        -g_{\alpha \nu}g_{\mu \beta }
			+g_{\mu \alpha}g_{\beta \nu}] \nonumber \\
	           &=& (6-2\epsilon)g_{\mu \nu}, 
\end{eqnarray}
and
\begin{eqnarray}
      \mbox{I}_{V_2}  &=& 10~\mbox{II}_{22}-
      [~\mbox{I}~(m_1)+\mbox{I}~(m_2)+(m_1^2+m_2^2+4p^2)~\mbox{I}_0~]
           	       -4\epsilon ~\mbox{II}_{22}, \\
      \mbox{I}_{\phi_2}   &=& 4~\mbox{II}_{22}, \\
      \mbox{I}_{\eta^2}      &=& -\mbox{II}_{22}. 
 \end{eqnarray}
Note that the three diagrams ({\bf E}), ({\bf H}) and ({\bf  I }) have
the same form of the integral as (5.89), the two diagrams
({\bf C}) and ({\bf J}) have the integral (5.87), and the
the two diagrams ({\bf D}) and ({\bf G}) have the integral (5.88).
 In terms of the integrals defined above,  we can obtain the bosons contributions to
 the gauge self energies,
 \begin{eqnarray}
  \Sigma^{\gamma \gamma }(p^2) &=& \frac{i}{16\pi^2}
 \left[ (-e^2) (\mbox{{\bf A}}) + (-ie^2) (\mbox{{\bf B}})+ (-2e^2c_W^2M_Z^2) 
  (\mbox{{\bf C}})
 + (-e^2) (\mbox{{\bf D}}) + (2ie^2) (\mbox{{\bf E}}) + (-2e^2) (\mbox{{\bf F}})
    \right] \nn
  \Sigma^{\gamma \gamma }(0)  &=& 0, \\
  \Sigma^{\gamma \gamma}(M_Z^2) &=&- \frac{\alpha }{4\pi }M_Z^2
                 [~3\Delta_W +(3+4c_W^2)C(M_Z^2,M_W,M_W)~], \\
\nn
  \Sigma^{\gamma Z}(p^2) &=& \left(\frac{c_W}{s_W}\right)
        \Sigma^{\gamma \gamma }(p^2) + \frac{i}{16\pi^2}
 \left(\frac{e^2}{c_Ws_W}\right)\left[2M_W^2 (\mbox{{\bf C}})+ 
 \frac{1}{2} (\mbox{{\bf D}})+ (-i) (\mbox{{\bf E}}) \right],\nn
  \Sigma^{\gamma Z}(0) &=& \frac{\alpha c_W}{2\pi s_W}M_Z^2\Delta_W ,\\
  \Sigma^{\gamma Z}(M_Z^2) &=& \frac{\alpha }{4\pi s_Wc_W}M_Z^2
             \left[~ \left(5c_W^2+\frac{1}{6}\right)\Delta_W
       + \left(\frac{1}{6}+\frac{13}{3}c_W^2+4c_W^4\right)C(M_Z^2,M_W,M_W)
    +\frac{1}{9}~\right], \\
\nn
  \Sigma^{ZZ}(p^2) &=& \left(\frac{c_W^2}{s_W^2}\right)
        \Sigma^{\gamma \gamma }(p^2) + \frac{i}{16\pi^2}
 \left(\frac{e^2}{s_W^2}\right)\left[(4c_W^2-2)M_Z^2 (\mbox{{\bf C}})+ 
 \left(1-\frac{1}{4c_W^2}\right) (\mbox{{\bf D}}) \right. \nn 
& &\left. ~~~ +i\left(\frac{1}{2c_W^2}-2\right) (\mbox{{\bf E}}) 
  -\left(\frac{1}{4c_W^2}\right) (\mbox{{\bf G}})
+\left(\frac{i}{4c_W^2}\right) [(\mbox{{\bf H}})+(\mbox{{\bf I}})]
+\left(-\frac{M_Z^2}{c_W^2}\right) (\mbox{{\bf J}}) \right], \nn
  \Sigma^{ZZ}(0) &=& \frac{\alpha }{4\pi }M_Z^2\left[
 \left(4+\frac{1}{c_W^2}-\frac{1}{s_W^2}\right)\Delta_W+
  \frac{1}{12 c_W^2s_W^2}\left(
  \frac{21}{2}-\frac{3}{2}h-\frac{9h}{h-1}\ln h+12\ln c_W^2\right)
  \right], \\
 \Sigma^{ZZ}(M_Z^2) &=& \frac{\alpha }{4\pi}
\left(7+\frac{7}{6c_W^2}-\frac{25}{6s_W^2}\right)M_Z^2\Delta_W \nonumber \\
& &+\left(\frac{\alpha }{48\pi c_W^2s_W^2}M_Z^2\right)\left\{
[-c_W^4(40+80c_W^2)+(c_W^2-s_W^2)^2(8c_W^2+1)+12c_W^2]\right. \nonumber \\
  & &\times C(M_Z^2,M_W,M_W)
 +(12-4h+h^2)C(M_Z^2,M_H,M_Z)-(h-1)^2+\frac{4}{3}(1-2c_W^2)  \nn
& &\left.+\left(-6-h+\frac{h^2}{2}\right)\ln h + 13 \ln c_W^2\right\}, \\
\frac{d\Sigma^{ZZ}}{dk^2}(M_Z^2)&=&\frac{\alpha }{4\pi}
\left(3-\frac{19}{6s_W^2}+\frac{1}{6c_W^2}\right)\Delta_W \nn
& &+\frac{\alpha }{48\pi c_W^2s_W^2}\left\{[-40c_W^4+(c_W^2-s_W^2)^2]
  C(M_Z^2,M_W,M_W) \right. \nn
& &+[-c_W^4(40+80c_W^2)+(c_W^2-s_W^2)^2(8c_W^2+1)+12c_W^2]
       \tilde{C}(M_Z^2,M_W,M_W) \nn
& & +\left(2h-h^2\right)C(M_Z^2,M_H,M_Z)+ [12-4h+h^2 ] 
      \tilde{C}(M_Z^2,M_H,M_Z)  \nn
& &  \left. +\left(1-\frac{h+1}{2(h-1)} \ln h
   -\frac{1}{2}\ln \frac{h}{c_W^4}\right)+\frac{4}{3}
     (1-2c_W^2)\right\}.
\end{eqnarray}
where ${\tilde C}(k^2,m_1,m_2) =dC/dk^2$ and $h=M_H^2/M_Z^2$. \\
For $\Sigma^{WW} $, the vector and scalar boson contributions are from the
diagrams in Fig. 4 as following,

\begin{figure}[htb]
\centerline{
\epsfig{figure=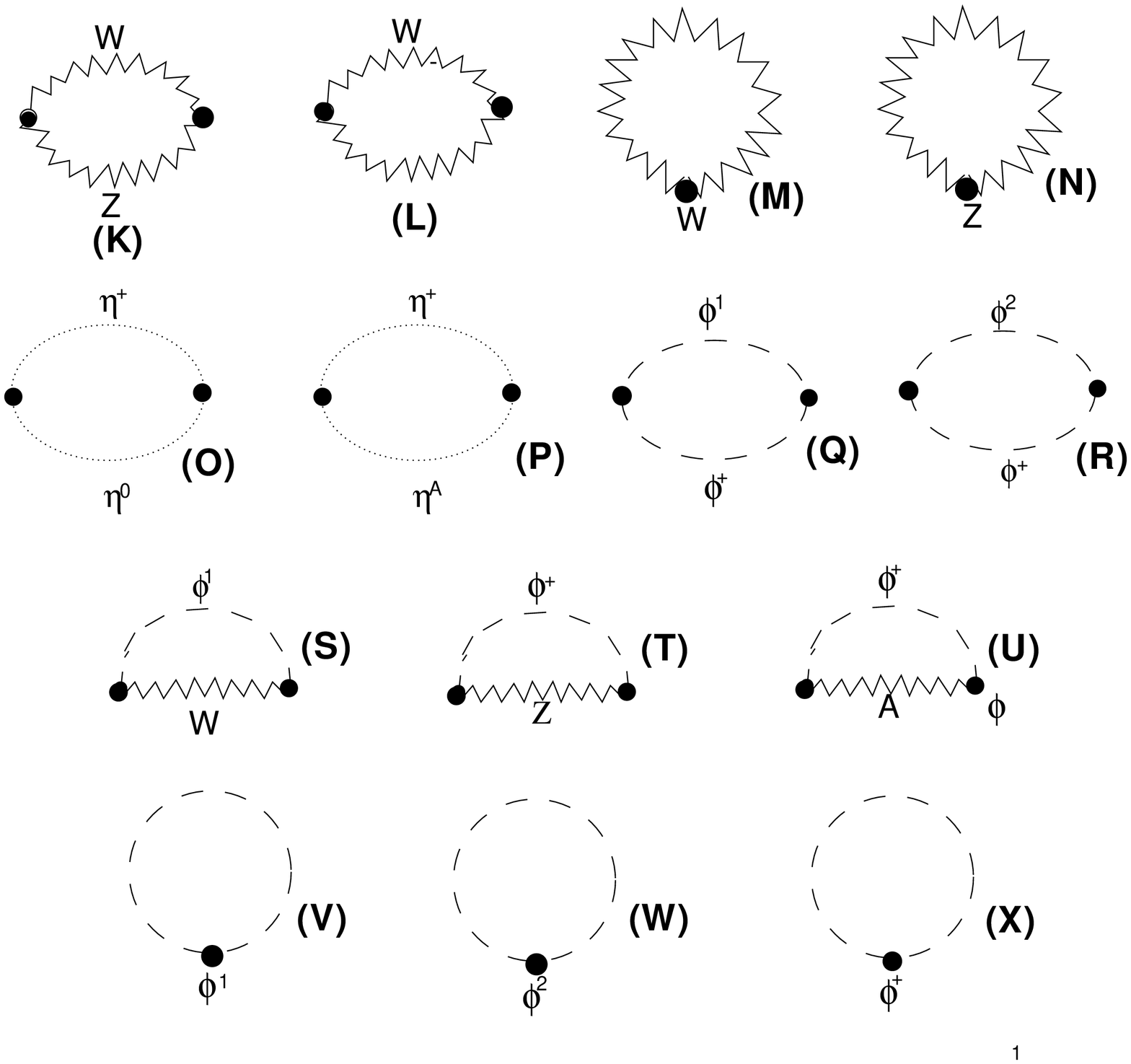, height=6cm, angle=0}}
\caption{Vector and scalar boson contributions to the $W$ boson self energy.
External $W$ lines are attached to the dots except those in
({\bf M}), ({\bf N}), ({\bf V}), ({\bf W}) and
({\bf X}) to which two $W$ lines are attached.}
\end{figure}
\begin{eqnarray}
\Sigma^{WW}(p)&=& \frac{i}{16\pi^2 }
\left[(-g^2c_W^2)(\mbox{{\bf K}}) + (-e^2) (\mbox{{\bf L}})
+\frac{1}{2}ig^2[2g_{\mu \nu}g_{\lambda \rho }
-g_{\mu \lambda}g_{\nu \rho }
-g_{\mu \rho}g_{\nu \lambda }] (\mbox{{\bf M}}) \right. \nn
& & ~~~~+\frac{1}{2}ig^2c_W^2 (\mbox{{\bf N}})
    +(-2g^2c_W^2) (\mbox{{\bf O}})
    +(-2e^2) (\mbox{{\bf P}})
    +(-\frac{g^2}{4}) (\mbox{{\bf Q}}) \nn
& & ~~~~+(-\frac{g^2}{4}) (\mbox{{\bf R}})
    +(-g^2M_W^2) (\mbox{{\bf S}})
    +(-g^2M_Z^2s_W^4) (\mbox{{\bf T}})
    +(-e^2M_W^2) (\mbox{{\bf U}}) \nonumber \\
& & ~~~~\left.  +(\frac{ig^2}{4}) (\mbox{{\bf V}})
  +(\frac{ig^2}{4}) (\mbox{{\bf W}})
  +(\frac{ig^2}{2}) (\mbox{{\bf X}}) \right] \nn
\Sigma^{WW}(0)&=& \frac{\alpha}{4\pi s_W^2}M_W^2
\left[\left(\frac{s_W^2}{c_W^2}-1\right)\Delta_W+
\frac{3}{4(1-c_W^2/h)}\ln \frac{c_W^2}{h}
 -\frac{h}{8c_W^2}\right. \nonumber \\
& &+\left.  \left(s_W^2+\frac{s_W^4}{c_W^2}-\frac{1}{8c_W^2}
-\frac{39}{12}+\left(\frac{s_W^2}{c_W^2}+3
-\frac{17}{4s_W^2}\right)\ln c_W^2\right)\right],\\
 \Sigma ^{WW}(M_W^2) &=& \frac{\alpha }{4\pi }\frac{1}{s_W^2}M_W^2
\left[-\left(\frac{25}{6}-\frac{s_W^2}{c_W^2}\right)
       \Delta_W\right. \nonumber \\
   & &  + \left(\frac{1}{12c_W^4}+\frac{4}{3c_W^2}
     -\frac{17}{3}-4c_W^2\right)(C(M_W^2,M_Z,M_W)-\ln c_W)
         \nonumber \\
   & &  \left(1-\frac{h}{3c_W^2}+\frac{1}{12}\frac{h^2}{c_W^4}\right)
\left(C(M_W^2,M_H, M_W)-\frac{1}{2}\ln \frac{c_W^2}{h}\right)\nn
   & & -\frac{h}{h-c_W^2}\ln \frac{h}{c_W^2}
\left(\frac{7}{4}-\frac{h}{3c_W^2}+\frac{h^2}{12c_W^4}\right) \nn
& &+\left[\frac{1}{12c_W^4}+\frac{7}{3c_W^2}-\frac{107}{12}-10c_W^2\right]
\frac{1}{1-c_W^2}\ln c_W^2 \nn
& &\left.  +\frac{1}{12c_W^4} +\frac{13}{6c_W^2} -\frac{233}{18}
 -\frac{h}{2c_W^2} +\frac{h^2}{12c_W^4}-4c_W^2\right].
\end{eqnarray}
\subsection{ Expressions of $\Delta \alpha, \Delta \rho $ and $\Delta r$}
{\bf (1) $\Delta \alpha $}\\
$\Delta \alpha $ is defined by the subtracted photon vacuum polarization
and is composed of four contributions,
\begin{eqnarray}
      \Delta \alpha &=& -Re\Pi^{\gamma }(M_Z^2)
			+\Pi^{\gamma}(0) \nonumber \\
                    &=& \Delta \alpha_{lept} + \Delta \alpha_{had}^{(5)} +
                        \Delta \alpha_{top} + \Delta \alpha_{W}.
\end{eqnarray}
The leptonic subtracted part, $\Delta \alpha _{lept}$, is given by
\begin{eqnarray}
      \Delta \alpha_{lept} &=& -Re\Pi^{\gamma}_{lept}(M_Z^2) 
                              + \Pi^{\gamma}_{lept}(0) \nonumber \\
                           &=& \sum_{l=e,\mu, \tau}\frac{\alpha}{3\pi}
                       \left(\ln{\frac{M_Z^2}{m_l^2}}-\frac{5}{3}\right).
\end{eqnarray}
We note that the 5 flavor contributions to $\Delta \alpha^{(5)}_{had}
=0.02804(65)$
can be derived from experimental data with the help of a dispersion 
relation$^{[17]}$
\begin{equation}
      \Delta \alpha ^{(5)}_{had}
	     =Re\left(-\frac{\alpha}{3\pi}M_Z^2\right.
		      \int^{\infty}_{4m_{\pi}^2}ds^{\prime}
                \left.\frac{R^{\gamma}(s^{\prime})}{s^{\prime}
                          (s^{\prime}-M_Z^2-i\epsilon)}\right),
\end{equation}
with
\begin{equation}
       R^{\gamma}(s)=\frac{\sigma (e^+e^-\rightarrow
				   \gamma^{*}\rightarrow hadrons) }
			  {\sigma (e^+e^-\rightarrow
				   \gamma^{*}\rightarrow \mu^{+}\mu^{-})},
\end{equation}
as an experimental quantity up to a scale $s_1$ and applying perturbative
QCD for the tail region above $s_1$. 
The contribution of a heavy top quark is
\begin{equation}
       \Delta \alpha_{top}
            \simeq  -\frac{\alpha }{\pi}Q_t^2\frac{M_Z^2}{5m_t^2} 
\end{equation}
It can be seen that $W$ boson contribution to $\Delta \alpha $ can be
evaluated by using eq.(5.97),
\begin{equation}
      \Delta \alpha_{W} = \frac{\alpha }{4\pi}
                          \left[(3+4c_W^2)C(M_Z^2,M_W,M_W)-\frac{2}{3}\right],
\end{equation}
which is negligible.
Then, we may take $\Delta \alpha $ as,
\begin{eqnarray}
      \Delta \alpha &=&
                       \frac{\alpha }{3\pi}\sum_{l}Q_l^2
	               \left( \ln{\frac{M_Z^2}{m_l^2}}-\frac{5}{3}\right)
             	      +\Delta \alpha ^{(5)}_{had} \nn
      &=& 0.05940(65).
\end{eqnarray}
{\bf (2) $\Delta \rho$ }\\
The fermionic contribution to $\Delta \rho $ is calculated from
the formulae
\begin{equation}
\Delta \rho = \frac{\Sigma^{ZZ} (0)}{M^2_Z}-\frac{\Sigma^{WW} (0)}{M^2_W}.
\end{equation}
For a doublet of fermions with masses $(m_1,m_2)$, it is in general given by
\begin{equation}
      (\Delta \rho)_{ferm} = 
		      N_c\frac{\alpha }{16\pi s_W^2c_W^2M_Z^2}
                      \left( m_1^2+m_2^2-\frac{2m_1^2m_2^2}{m_1^2-m_2^2}\right.
  		      \left. \ln\frac{m_1^2}{m_2^2}\right).
\end{equation}
We can see that their singular parts cancel and that 
the contributions of the quarks except top quark are very small.
Then a finite term which is quadratic in $m_t$ remains
\begin{equation}
      (\Delta \rho )_{ferm}=\frac{3 \alpha }{16\pi s_W^2c_W^2}
  		               \frac{m_t^2}{M_Z^2}.
 \end{equation}
The Higgs mass dependence in $\Delta \rho$, up to one-loop order,
is given by
\begin{eqnarray}
       (\Delta \rho )_H &=& \frac{3\alpha }{16\pi s_W^2}
                       \left\{\frac{1}{c_W^2}\left(\Delta -
                            \ln \frac{M_Z^2}{\mu^2}\right)
                  -\left(\Delta -\ln \frac{W_Z^2}{\mu^2}\right)
			    +\frac{5}{6}\frac{s_W^2}{c_W^2}\right. \nonumber \\
                    & &~~~~~\left. +\frac{h}{c_W^2(1-h)}
			    \ln h 
                            -\frac{h}{c_W^2-h}
			     \ln {\frac{h}{c_W^2}} \right\}.
\end{eqnarray}
{\bf (3) $\Delta r$}\\
To the one-loop order, the contributions to the $\mu- $ decay amplitude are
obtained by adding the diagrams, \\
\begin{figure}[htb]
\centerline{
\epsfig{figure=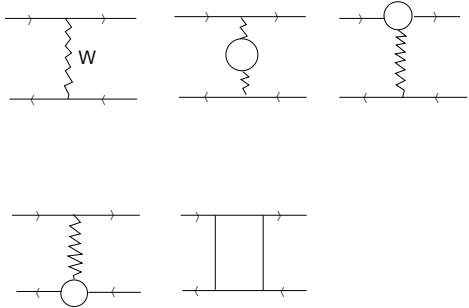, height=6cm, angle=0}}
\caption{$\mu-$ decay amplitude}
\end{figure}

The vertex corrections and box diagrams in the $\mu- $ decay amplitude are 
shown in Fig.5.
\begin{figure}[htb]
\centerline{
\epsfig{figure=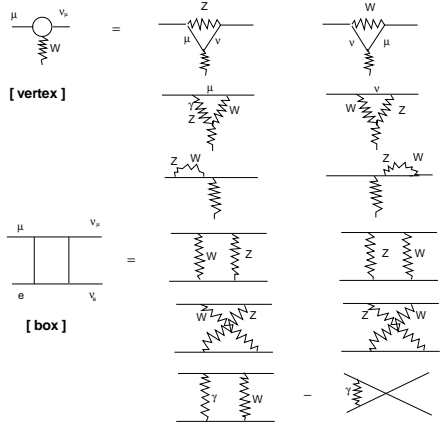, height=6cm, angle=0}}
\caption{vertex and box contributions}
\end{figure}

Then, using the bare parameter relations (5.13-5.15), $G_{\mu }/\sqrt{2}$ 
can be written as
\begin{eqnarray}
      \frac{G_{\mu }}{\sqrt{2}} &=& 
	      \frac{e^2_0}{8s_W^{02}M_W^{02}}
              \left[ 1+\frac{\Sigma^{WW}(0)}{M_W^2} + [\mbox{{\bf vertex}}]+
                [\mbox{{\bf box}}]\right]
	      \nonumber \\
          &=& \frac{e^2}{8s_W^{2}M_W^{2}}
              \left[ 1+2\frac{\delta e}{e}-\frac{c_W^2}{s_W^2}\right.
              \left(\frac{\delta M_Z^2}{M_Z^2}\right.
	     -\left.\frac{\delta M_W^2}{M_W^2}\right) 
             -\frac{\delta M_W^2}{M_W^2}
              \left.  +\frac{\Sigma^{WW}(0)}{M_W^2} + [\mbox{{\bf vertex}}]+
              [\mbox{{\bf box}}]\right] 
	      \nonumber \\
    &\equiv & \frac{\pi \alpha }{2s_W^2M_W^2}[1+\Delta r],
\end{eqnarray}
where we have used the relations
$e_0^2=e^2\left(1+2\frac{\delta e}{e}\right)$, 
$M_W^{02}=M_W^2 + \delta M_W^2$ and
\begin{equation}
       \sin^2\theta_W^0 =  \sin^2\theta_W
                              +\cos^2\theta_W
                               \left(\frac{\delta M_Z^2}{M_Z^2}\right.
			       -\left. \frac{\delta M_W^2}{M_W^2}\right),
\end{equation}
and 
\begin{eqnarray}
      [\mbox{{\bf vertex}}]+[\mbox{{\bf box}}] &=& \frac{\alpha }{\pi s_W^2}
                        \Delta_W +\frac{\alpha }{4\pi s_W^2}
                        \left(6+\frac{7-4s_W^2}{2s_W^2} \ln{c_W^2}\right) 
			\nonumber \\
                    &=& \frac{2}{c_Ws_W}\frac{\Sigma^{\gamma Z}(0)}{M_Z^2}
                        +\frac{\alpha }{4\pi s_W^2}
                         \left(6+\frac{7-4s_W^2}{2s_W^2} \ln{c_W^2}\right) ,
\end{eqnarray}
where we have used the result eq.(5.98).
Then,
\begin{eqnarray}
\Delta r &=&
            \Pi^{\gamma}(0) -\frac{c_W^2}{s_W^2}
            \left(\frac{\delta M_Z^2}{M_Z^2}\right.
	    -\left.\frac{\delta M_W^2}{M_W^2}\right) 
            -\frac{\delta M_W^2-\Sigma^{WW}(0)}{M_W^2}  \nonumber \\
         & & +2\frac{c_W^2}{s_W^2}\frac{\Sigma^{\gamma Z}(0)}{M_Z^2}
            + \frac{\alpha}{4\pi s_W^2}
	      \left(6+\frac{7-4s_W^2}{2s_W^2}\ln{c_W^2}\right)\nonumber \\
         &=& \Pi^{\gamma}(0) - Re\Pi^{\gamma}(M_Z^2) 
	     -\frac{c_W^2}{s_W^2}
            \left(\frac{\Sigma^{ZZ}(0)}{M_Z^2}\right.
	    -\left. \frac{\Sigma^{WW}(0)}{M_W^2}\right) \nonumber \\
         & & + Re\Pi^{\gamma}(M_Z^2)+2\frac{c_W}{s_W}
     	      \frac{\Sigma^{\gamma Z}(0)}{M_Z^2} 
             -\frac{Re\Sigma^{WW}(M_W^2)-\Sigma^{WW}(0)}{M_W^2} + \nonumber \\
         & & -\frac{c_W^2}{s_W^2}
              \left[\frac{Re\Sigma^{ZZ}(M_Z^2)-\Sigma^{ZZ}(0)}{M_Z^2}\right. 
             -\left.\frac{Re\Sigma^{WW}(M_W^2)-\Sigma^{WW}(0)}{M_W^2}\right]
	      +\cdots \nonumber \\
         &=& \Delta \alpha -\frac{c_W^2}{s_W^2}\Delta \rho 
	     + (\Delta r)_{rem}.
\end{eqnarray}
It should be noted that the remainder term also contains a logarithmic term 
in the top mass,
\begin{equation}
      (\Delta r)^{top}_{rem} = \frac{\alpha }{4\pi s_W^2}
       \left(\frac{c_W^2}{s_W^2}-\frac{1}{3}\right)\ln {\frac{m_t}{M_Z}}.
\end{equation}
Also the Higgs boson contribution is part of the remainder 
and is given by
\begin{eqnarray}
    (\Delta r)^{Higgs}_{rem} 
       &=& \frac{\alpha}{4\pi s^4_W}
           \left\{\left(\frac{h}{3}-1
          -\frac{h^2}{12}\right) \right.
           \left(C(M_Z^2,M_H,M_Z)-1+\frac{h+1}{2(h-1)}\ln h\right) \nonumber \\
       & & +\left(2-\frac{1}{c_W^2}\right)
           \left(c_W^2-\frac{h}{3}
          +\frac{h^2}{12c_W^2}\right)\left(C(M_W^2,M_H,M_W) 
           -1 + \frac{h+c_W^2}{2(h-c_W^2)}\ln \frac{h}{c_W^2}\right) \nn
       & & +\left(\frac{11}{24}-\frac{h}{12}\right)
            \frac{h+1}{h-1}\ln h
           +\left(\frac{11}{24}+\frac{h}{12}-\frac{3}{4}
            \frac{c_W^2h}{h-c_W^2}\right) \ln \frac{h}{c_W^2}
           \nonumber \\
       & & \left.  -\left(\frac{3}{8}-\frac{h}{12}\right)
            \ln c_W^2
	   +s_W^2\left(\frac{1}{24}\frac{h}{c_W^2}
          -\frac{59}{72}\right)\right\},
\end{eqnarray}
where we choose the renormalization scale as $\mu ^2 = M_Z^2$.
For large $M_H$, it increases only logarithmically,
\begin{equation}
      (\Delta r)^{Higgs}_{rem} \simeq \frac{\alpha }{16\pi s_W^2}
       \frac{11}{3}\left(\ln {\frac{M_H^2}{M_W^2}}-\frac{5}{6}\right).
\end{equation}
The explicit form of $\Delta r$ is written as
\begin{eqnarray}
      \Delta r &=& (\Delta r)_{ferm} + (\Delta r)_{boson}, \\
      (\Delta r)_{ferm} &=&
		 (\Delta \alpha)_{ferm}-\frac{\alpha }{4\pi s_W^4}
        \left[\left(\frac{1}{2}-\frac{2}{3}s_W^2\right)\ln \frac{M_Z^2}{m_t^2}
       -\left(\frac{5}{6}-\frac{10}{9}s_W^2+\frac{20}{27}s_W^4\right)\right. \nn
    & & \left. +\left(\frac{16}{45}s_W^4+\frac{4}{3}s_W^2-\frac{19}{40}\right)
         \frac{M_Z^2}{m_t^2}\right]
      +\sum_{f\neq t} \frac{\alpha}{4\pi s_W^2}
        \left(1-\frac{c_W^2}{s_W^2}\right) \frac{N_c^f}{6}\ln c_W^2+\cdots, \\
     (\Delta r)_{boson} &=&
		  (\Delta \alpha)_W+\frac{\alpha }{4\pi s_W^4}
       \left\{\left(s-\frac{1}{c_W^2}\right)
        \left(\frac{1}{12c_W^2}+\frac{4}{3}-\frac{17}{3}c_W^2-4c_W^4\right)
        C(M_W^2, M_Z, M_W) \right. \nn
  & & + \left(\frac{32}{3}c_W^4 + \frac{2}{3}c_W^2-\frac{37}{12}\right)
         C(M_Z^2, M_W, M_W) + \left(\frac{1}{3}h-1-\frac{h^2}{12}\right)
        C(M_Z^2, M_H, M_Z) \nn
 & & +\left(2-\frac{1}{c_W^2}\right)
       \left(c_W^2-\frac{1}{3}h+\frac{h^2}{12c_W^2}\right)C(M_W^2, M_H, M_W)\nn
 & & +\ln \frac{h}{c_W^2}(c_W^2-s_W^2)
   \left[\frac{1}{2}-\frac{1}{6c_W^2}+\frac{h^2}{24c_W^4}
  -\frac{h}{h-c_W^2}\left(\frac{7}{4}-\frac{h}{3c_W^2}
  +\frac{h^2}{12c_W^4}-\frac{3}{4}\frac{s_W^2}{c_W^2-s_W^2}\right)\right] \nn
 & & +\left(\frac{1}{2}-\frac{h}{12}-\frac{h^2}{24}\right)\ln h  
  +\left(\frac{11}{6}-\frac{3}{4c_W^2}-\frac{141}{6}c_W^2-24c_W^4
     +\frac{1}{12s_W^4}-\frac{1}{6s_W^2}\right)\ln c_W^2 \nn
 & & +\frac{h}{24}\left(2h+\frac{9}{c_W^2}-13+\frac{4h}{c_W^2}-\frac{2h}{c_W^4}
        \right) 
  +\frac{211}{24}+\frac{23}{8c_W^2}+\frac{13}{12c_W^4}
     -\frac{233}{12}c_W^2 \nn
 & & \left. +6s_W^2 + \frac{7-4s_W^2}{2}\ln c_W^2 \right\}.
\end{eqnarray}
\subsection{ Higher order corrections}
There are large logarithms of the form $\alpha ^{n}\ln^m(M_Z^2/m_f^2)$
in the context of the effective electromagnetic charge.
These corrections are effectively incorporated when the one-loop
correction $1+\Delta \alpha $ is replaced by  $\frac{1}{1-\Delta \alpha}$
according to renormalization group arguments$^{[18]}$.

Due to large mass of top quark, the next-leading large contributions to
$\Delta \rho $ have to be considered, which are $O(G_{\mu}^2m_t^4)^{[19]},
O(\alpha_sG_{\mu}m_t^2)^{[20,21]}$ and $O(\alpha_s^2G_{\mu}m_t^2)^{[22]}$.
Including the first two corrections, we can write the parameter
$\Delta \rho $ as
\begin{equation}
\Delta \rho = 3x_t\left[1+x_t\rho^{(2)}(\frac{M_H}{m_t}) \right.
     \left. -2\alpha_s(m_t)\frac{\pi^2+3}{9\pi}\right].
\end{equation}
where $x_t=G_{\mu}m_t^2/8\pi^2\sqrt{2}$ and $\rho^{(2)}$ is strongly
dependent on $M_H^{[19]}$.
For light Higgs boson $M_H<<m_t$, the function $\rho^{(2)}$ is
$19-2\pi^2$ and for heavy Higgs $M_H>>m_t$, it is given by the
asymptotic expression with $r=M_H/m_t$,
\begin{equation}
\rho^{(2)}=6\ln^2 r-27 \ln r +\frac{49}{4}\pi^2+O(\ln^2 r/r^2)
\end{equation}
The third term is originated from $t\bar{t}$ threshold, which is discussed
elsewhere$^{[19]}$.
Though the QCD corrections in $\Delta \rho $ are sufficient for very large 
top mass $m_t>>M_W$, one can take into account the subleading corrections
$O(\alpha_s G_{\mu}M_W^2)^{[19]}$  for realistic values of $m_t$.
They can amount to 20 \% of the perturbative QCD correction to $\Delta r$
at $m_t=150 ~\mbox{GeV}$.
We will include above corrections except $t\bar{t}$ threshold in the ZFITTER,
as well as the dominant two-loop corrections$^{[19]}$  of 
$O(\alpha^2m_t^2)$ to $\Delta \rho = 1-1/\rho $ and the QCD corrections
to the leading electroweak one-loop term$^{[23]}$ of 
$O(\alpha \alpha_s m_t^2)$ and $O(\alpha \alpha_s^2 m_t^2)$.
These new theoretical advances in radiative corrections coupled with the
experimental developments in the electroweak data, $m_t$ and $M_W$ are
additional reasons to update the precision tests of the SM.

Thanks to the higher order corrections in $\Delta \alpha $ and $\Delta \rho $, 
Eq. (5.116) can be written as
\begin{equation}
\sqrt{2}G_{\mu}=\frac{\pi \alpha }{M_W^2\sin^2\theta_W}
\left[\frac{1}{(1-\Delta \alpha)\cdot
(1+\frac{\cos^2\theta_W}{\sin^2\theta_W}\Delta \rho )}
+\Delta r_{remainder}\right].
\end{equation}
The proper incorporation of the non-leading high order terms containing mass
singularities of the type $\alpha^2\ln (M_Z/m_f)$ enable us to rewrite
eq.(5.128) as
\begin{eqnarray}
\sqrt{2}G_{\mu} &=& \frac{\pi \alpha }{M_W^2\sin^2\theta_W}
\left[~\frac{1}{(1-\Delta \alpha)\cdot
(1+\frac{\cos^2\theta_W}{\sin^2\theta_W}\Delta \rho )
-\Delta r_{remainder}}~\right] \nonumber \\
 &\equiv & \frac{\pi \alpha }{M_W^2\sin^2\theta_W}\frac{1}{1-\Delta r}.
\end{eqnarray}

\subsection{ $Z$ physics in $e^+e^-\rightarrow f\bar{f}$}
Following the general principles discussed above we attach multiplicative
renormalization constants to each free parameter and each symmetry
multiplet of fields in the symmetric Lagrangian:
\begin{eqnarray}
      W_{\mu }^a &\rightarrow &(Z_2^W)^{1/2}W_{\mu }^a \nn
      B_{\mu } &\rightarrow & (Z_2^B)^{1/2}B_{\mu } \nn
      \psi^L_j &\rightarrow & (Z^j_L)^{1/2}\psi^L_j \nn
      \psi^R_{j\sigma} &\rightarrow & (Z^{j\sigma}_R)^{1/2}
      \psi^R_{j\sigma} \nn
       \Phi &\rightarrow & (Z^{\Phi})^{1/2}\Phi \nn
      g_2 &\rightarrow & Z_1^W(Z_2^W)^{-3/2}g_2 \nn
      g_1 &\rightarrow&  Z_1^B(Z_2^B)^{-3/2}g_1 \nn
       v &\rightarrow & (Z^{\Phi})^{1/2}(v-\delta v) \nn
       g_{j\sigma} &\rightarrow & (Z^{\Phi})^{-1/2}
       Z_1^{j\sigma } g_{j\sigma} \nn
       \mu^2 &\rightarrow & (Z^{\Phi})^{-1}(\mu^2-\delta \mu^2) \nn
       \lambda &\rightarrow & (Z^{\Phi})^{-2}Z_{\lambda} \lambda  
\end{eqnarray}
where the RHS represent the bare  fields and parameters, 
the quantities without
the $Z$-factors are the corresponding renormalized fields and parameteres
and the variation of the renormalization constants is given by $Z_i
=1+\delta Z_i $,
and
\begin{equation}
 \left( \begin{array}{c}
              \delta Z_i^{\gamma} \\
              \delta Z_i^{Z} 
         \end{array}
 \right) =
 \left( \begin{array}{cc}
              s_W^2 & c_W^2  \\
              c_W^2 & s_W^2 
        \end{array} 
 \right)
 \left( \begin{array}{c}
              \delta Z_i^{W}  \\
              \delta Z_i^{B} .
        \end{array}
 \right).
\end{equation}

Field renormalization ensures that we end up with finite Green functions.
The field renormalization in (5.130) is performed in a way that it respects the
gauge symmetry by introducing the minimal number of field renormalization
constants. Therefore also the counter term Lagrangian and the renormalized
Green functions reflect the gauge symmetry. The price for this, however, is
that not all residues of the propagators can be normalized to unity. As a 
consequence, any calculation with the renormalized Lagrangian will have to 
include finite multiplicative wave function renormalization factors for some of
the external lines in S matrix elements.

It is of course possible to perform the renormalization in such a way that 
these finite corrections do not appear$^{[24,25,26,26]}$. 
But then the Lagrangian will
contain many constants which have to be calculated in terms of the few 
fundamental parameters.

The independent renormalization of the Higgs vacuum expectation value $v$  
absorbs the linear term in the Higgs potential, which is induced by the 
appearance of
tadpole diagrams in one-loop order, in such a way that the relation
$v=\frac{2\mu}{\sqrt{\lambda}}$
remains valid for the renormalized parameters with $v$  being the minimum of the
Higgs potential at the one-loop level. As a practical consequence of this 
tadpole renormalization, all tadpole graphs can be omitted in the renormalized 
amplitudes and Green functions. They are, however, necessary to make the mass 
counter terms gauge independent.

The systematic way for obtaining results for physical amplitudes in one-loop
order is scheduled as follows: The expansion (5.16) yields the renormalized 
Lagrangian $\cal{L}$  which can now be re-parametrized in terms of the physical
parameters $e, M_W, M_Z, M_H, m_f$  and the physical fields 
$A_{\mu}, Z_{mu}, W_{\mu}^{\pm}, H $ 
 (also the unphysical Higgs field 
components $\phi^{\pm}, \eta $ and the ghost fields $\eta_{\pm}, \eta_0,
\eta_A $ are present in the 
$R_{\xi}$ gauge), and the
counter term Lagrangian $\delta \cal{L}$. From $\delta \cal{L}$ the counter 
term Feynman rules are derived.
After rewriting them in terms of $e, M_W, M_Z, M_H, m_f$ the counter term graphs have to be added 
to the 1-loop vertex functions calculated from $\cal{L}$. 
The renormalization constants in
(5.16) are fixed afterwards by imposing the appropriate renormalization 
conditions.
The results are finite Green functions in terms of the above physical parameter
set from which the S matrix elements for all processes of interest can be 
obtained. 
 Then the renormalized gauge boson self energies ($\tilde{\Sigma }$)
can be expressed in terms of the non-renormalized ones ($\Sigma $),
\begin{eqnarray}
      \tilde{\Sigma}^{\gamma \gamma }(k^2) &=& 
			  \Sigma^{\gamma \gamma }(k^2)-k^2
                          \Pi^{\gamma  }(0) \\
      \tilde{\Sigma}^{\gamma Z}(k^2) &=& \Sigma^{\gamma Z}(k^2)-
                          \Sigma^{\gamma Z}(0) + k^2
                          \left[2\frac{\Sigma^{\gamma Z}(0)}{M_Z^2}\right. 
                         -\frac{c_W}{s_W}
			  \left(\frac{\delta M_Z^2}{M_Z^2} \right.-
                          \left.\left.\frac{\delta M_W^2}{M_W^2}\right)\right]\\
      \tilde{\Sigma}^{ZZ}(k^2) &=& \Sigma^{ZZ}(k^2)
			  -\delta M_Z^2 + \delta Z_2^Z(k^2-M_Z^2) \\
      \tilde{\Sigma}^{WW}(k^2) &=& \Sigma^{WW}(k^2)-
			   \delta M_W^2 +  \delta Z_2^W(k^2-M_W^2) 
\end{eqnarray}
with
\begin{eqnarray}
\delta Z_2^Z &=& -\Pi^{\gamma}(0)-2\frac{c_W^2-s_W^2}{s_Wc_W}
                  \frac{\Sigma^{\gamma Z}(0)}{M_Z^2}
                 +\frac{c_W^2-s_W^2}{s_W^2}
                  \left(\frac{\delta M_Z^2}{M_Z^2}\right.
		 -\left. \frac{\delta M_W^2}{M_W^2}\right) \\
\delta Z_2^W &=& -\Pi^{\gamma}(0)-2\frac{c_W}{s_W}
                  \frac{\Sigma^{\gamma Z}(0)}{M_Z^2}
                 +\frac{c_W^2}{s_W^2}
                  \left(\frac{\delta M_Z^2}{M_Z^2}\right.
		 -\left. \frac{\delta M_W^2}{M_W^2}\right) 
\end{eqnarray}
Replacing those renormalized self energies and renormalized masses with
non-renormalized ones in eq.(5.25-5.27), one can obtain the renormalized
gauge boson propagators.
The two constant $Z_W$ and $Z_B$ are sufficient to make the self
energies (resp. the propagators) and the vertex corrections finite.
These additional two field renormalization constants allow to fullfil
the further two renormalization conditions:
vanishing of the $\gamma Z$ mixing propagator for real photons
$(k^2=0)$; residue$=1$ for the photon propagator (in analogy to 
pure QED).
The residues of the $W$ and $Z$ propagators, however, are different from unity.

{\bf (1) Effective neutral current couplings}\\
The weak dressed $Z$ exchange amplitude can be written as follows
\begin{figure}[htb]
\centerline{
\epsfig{figure=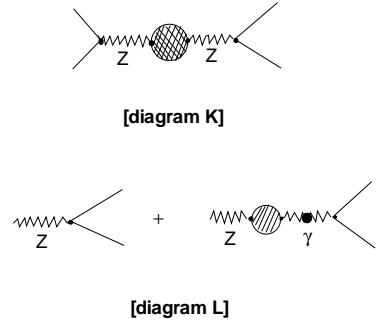, height=6cm, angle=0}}
\caption{The $Z-$exchange amplitude and effective 
neutral current couplings}
\end{figure}
\begin{eqnarray}
 [\mbox{{\bf diagram~ K}}] &=& \frac{e^2}{4s_W^2c_W^2}J^{(e)}\cdot
        \tilde{D}_Z(k^2) \cdot J^{(f)} \nonumber \\
            &=& \frac{e^2}{4s_W^2c_W^2}\frac{J^{(e)}\cdot J^{(f)}}
	        {[k^2-M_Z^2+\Sigma^{ZZ}(k^2)]} \nonumber \\
            &=& \frac{e^2}{4s_W^2c_W^2}\frac{1}{[1+\Pi^Z(k^2)]}
	        \frac{J^{(e)}\cdot J^{(f)}}
	        {[k^2-M_Z^2+iIm\Sigma^{ZZ}(k^2)/[1+\Pi^Z(k^2)]]}
\end{eqnarray}
where,
\begin{eqnarray}
\Sigma^{ZZ}(k^2) &=& \tilde{\Sigma}^{ZZ}(k^2)-
       \frac{\tilde{\Sigma}^{\gamma Z}(k^2)}{k^2+
	    \hat{\Sigma}^{\gamma \gamma}(k^2)}, \\
J^{(e)} &=& \bar{v}_{e}[\gamma_{\mu }(I_{3L}^e-2Q_es_W^2)-
	    \gamma_{\mu }\gamma_5I_{3L}^e]u_e, \\
J^{(f)} &=& \bar{u}_{f}[\gamma_{\mu }(I_{3L}^f-2Q_fs_W^2)-
	    \gamma_{\mu }\gamma_5I_{3L}^f]v_f, \\
\Pi^{Z}(k^2) &=& \frac{Re\Sigma^{ZZ}(k^2)-\delta M_Z^2}{k^2-M_Z^2}
	        -\Pi^{\gamma}(0) +\frac{c_W^2-s_W^2}{s_W^2}
                 \left(\frac{\delta M_Z^2}{M_Z^2}\right.
     	        -\frac{\delta M_W^2}{M_W^2}-2\frac{s_W}{c_W}
                 \left.\frac{\Sigma^{\gamma Z}(0)}{M_Z^2}\right), \\
Re \Sigma^{ZZ}(k^2) &=& (k^2-M_Z^2)\Pi^Z(k^2).
\end{eqnarray}
Then we can interpret the real part in the denominator of eq.(5.138) 
as the modified effective neutral current couplings.
From the relation (5.129), we can see that
\begin{eqnarray}
      \frac{\pi \alpha }{s_W^2c_W^2}\frac{1}{1+\Pi^Z} &=&
      \sqrt{2}G_{\mu }M_Z^2\frac{1-\Delta r}{1+\Pi^Z} \nonumber \\
      &=&\sqrt{2}G_{\mu }M_Z^2[1+\Delta \rho +\cdots ] \nonumber \\
      &=&\sqrt{2}G_{\mu }M_Z^2\rho_f .
\end{eqnarray}
The form factor $\rho_f$ can be expressed in terms of the self energy and
vertex contributions which explicitly depend on the type of the
external fermions,
\begin{equation}
  \rho_f=1+\Delta \rho_{se} + \Delta \rho_{f, vertex}.
\end{equation}
Here the self energy contributions are given by
\begin{equation}
  \Delta \rho_{se} = \Delta \rho + \Delta \rho_{se,rem},	
\end{equation}
where
\begin{eqnarray}
\Delta \rho &=& \frac{\Sigma^{ZZ}(0)}{M_Z^2}-\frac{\Sigma^{WW}(0)}{M_W^2}
	       -2\frac{\sin \theta_W}{\cos \theta_W}
	       \frac{\Sigma^{\gamma Z}(0)}{M_Z^2}, \\
\Delta \rho_{se,rem} &=& \frac{\Sigma^{ZZ}(M_Z^2)}{M_Z^2}
			-\frac{\Sigma^{ZZ}(0)}{M_Z^2}
	                -\left(\frac{d\Sigma^{ZZ}}{d k^2}\right)(M_Z^2).
\end{eqnarray}
For the vertex contributions we get
\begin{eqnarray}
\Delta \rho_{f,vertex} &=& \frac{\alpha}{16\pi s^2_W}
\left[2(3v_f^2+a_f^2)\Lambda_2(k^2,M_Z)-4c^2_W(1-2(1-|Q_f|)s^2_W)
    \Lambda_2(k^2,M_W) \right. \nonumber \\
  & & \left. +24c_W^4\Lambda_3(k^2,M_W)\right]-\Delta r_{vertex+box},
\end{eqnarray}
where $\Delta r_{vertex+box}$ is given by eq.(5.118) and
\begin{eqnarray}
\Lambda_2(k^2,M) &=& -\frac{7}{2}-2\frac{M^2}{k^2}
           -\left(2\frac{M^2}{k^2}+3\right)\ln \frac{M^2}{k^2} \nonumber \\
       & & +2\left(1+\frac{M^2}{k^2}\right)^2
          \left[\ln {\left(\frac{M^2}{k^2}\right)}
          \ln{\left(\frac{1+M^2/k^2}{M^2/k^2}\right)}
         +\int^1_0 \frac{dt}{t}\ln {\left(1+\frac{k^2}{M^2}t\right)}\right],\\
 \Lambda_3(k^2,M) &=& \frac{5}{6}-\frac{2M^2}{3k^2}
         +\frac{2}{3}\left(2\frac{M^2}{k^2}+1\right)
          \sqrt{4\frac{M^2}{k^2}-1} \arctan{
          \left[\frac{1}{\sqrt{4\frac{M^2}{k^2}-1}}\right]
							    } \nonumber \\
       & & -\frac{3}{8}\frac{M^2}{k^2}\left(\frac{M^2}{k^2}+2\right)
            \left(\arctan{\frac{1}{\sqrt{4\frac{M^2}{k^2}-1}}}\right)^2 
  ~~~~~(for ~~~k^2 < 4M^2).
\end{eqnarray}
In particular, we note that for $f=b$ the vertex corrections have a strong
dependence on $m_t$.
On can find the additional terms resulting from the heavy top quarks up
to two-loop order,
\begin{eqnarray}
\Delta \rho^{add}_{b, vertex} &=& -\frac{\alpha }{8\pi s_W^2}
\left[\frac{m_t^2}{2M_W^2}+\left(\frac{8}{3}+\frac{1}{6c_W^2}\right)
\ln \frac{m_t^2}{M_W^2} \right.\nonumber \\
 & &+\left.\frac{\alpha}{64\pi s_W^2}
\left(\frac{m_t^2}{M_W^2}\right)^2
(\tau^{(2)}-1)-\frac{\alpha_s}{2\pi}\frac{\pi^2-3}{3}\right],
\end{eqnarray}
where $\tau^{(2)} $ is the two-loop coefficient which depends on Higgs mass
and is asymptotically given by 
$ \tau ^{(2)} = 9 - \frac{\pi}{3} $ for small Higgs masses ($M_H << m_t$)
and $\tau^{(2)}=\frac{1}{144}[311+24\pi^2+282\ln r + 90 \ln^2 r -4 r]$
 for $M_H > 2m_t $ ($r=(m_t/M_H)^2)$.
The explicit form of $\rho_f$ is written by
\begin{eqnarray}
\rho_f&=&1+\frac{3\alpha m_t^2}{16\pi s_W^2 M_W^2}+
       \frac{\alpha}{\pi}(v_t^2+a_t^2)\frac{M_Z^2}{30m_t^2}
         +O(\frac{m^2_{f=light}}{M_W^2}) \nn
& &  +\frac{\alpha}{4\pi s_W^2}\left\{
     \frac{1}{12c_W^2}\left[\left(-80c_W^6+8(c_W^2-s_W^2)^2c_W^2+12c_W^2\right)
      C(M_Z^2, M_W, M_W) \right. \right. \nn
& & +\left(40c_W^4(1+2c_W^2)-(c_W^2-s_W^2)^2(1+8c_W^2)-12c_W^2\right)
     {\tilde C}(M_Z^2, M_W, M_W) \nn
& & +(12-6h +2h^2)C(M_Z^2, M_H, M_Z)-(12-4h+h^2){\tilde C}(M_Z^2, M_H, M_Z)\nn
& & \left. +(2h-h^2-2)+\left(-6-h+\frac{h^2}{2}+\frac{(h+1)}{2(h-1)}
     \right)\ln h \right] \nn
& & +\left(\frac{1}{24c_W^2}+\frac{3h}{4(h-c_W^2)}\right)\ln \frac{h}{c_W^2}
     +\frac{h}{8c_W^2}+\left(\frac{1}{24c_W^2}-2+\frac{17}{4s_W^2}\right)
       \ln c_W^2 \nn
& & \left. -\left(s_W^2+\frac{s_W^4}{c_W^2}-\frac{1}{8c_W^2}-\frac{39}{12}
    \right)\right\}
%
\end{eqnarray}

Also, the appearance of the $\gamma Z$ mixing beyond the tree level may 
be viewed as a redefinition of the neutral current vector coupling 
constants $v_f, a_f$.
This can be seen from
\begin{eqnarray}
[\mbox{{\bf diagram~ L}}]
   &=& \gamma_{\mu }(v_f-a_f\gamma_5)+\gamma_{\mu }Q_f
       \frac{\Pi^{\gamma Z}(k^2)}{1+\Pi^{\gamma}(k^2)} \nonumber \\
   &=& \gamma_{\mu }
       \left[\frac{I_3^f-2Q_fs_W^2}{2s_Wc_W}+Q_f\right.
       \left.\frac{\Pi^{\gamma Z}(k^2)}{1+\Pi^{\gamma}(k^2)}\right]
      -\frac{I_3^f}{2s_Wc_W}\gamma_{\mu }\gamma_5 \nonumber \\
   &=& \frac{1}{2s_Wc_W}\left\{\gamma_{\mu }\left[I_3^f-2Q_f 
       \left(s_W^2-s_Wc_W \frac{\Pi^{\gamma Z}(k^2)}
        {1+\Pi^{\gamma }(k^2)}\right)\right]
     - I_3^f\gamma_{\mu }\gamma_5 \right\}
\end{eqnarray}
where $\Pi^{\gamma }(k^2)=\tilde{\Sigma}^{\gamma \gamma }(k^2)/k^2$
and $\Pi^{\gamma Z}(k^2)=\tilde{\Sigma}^{\gamma Z}(k^2)/k^2.$
The last line of eq.(5.154) allows a redefinition of an effective mixing angle
$\bar{s}_W^2$ as
\begin{eqnarray}
      \sin^2{\theta_W^{eff}}\equiv \bar{s}_W^2 &=& s_W^2-s_Wc_W
                     Re\frac{\Pi^{\gamma Z}(k^2)}
			  {1+\Pi^{\gamma \gamma}(k^2)} \nonumber \\
                &\equiv &\kappa_f s_W^2
\end{eqnarray}
where $\kappa_f = 1-\frac{c_W}{s_W}\frac{\Pi^{\gamma Z}(k^2)}
                  {1+\Pi^{\gamma}(k^2)}$.
Similarly to the form factor $\rho_f$ we can express $\kappa_f$
as
\begin{equation}
\kappa_f=1+\Delta\kappa_{se}+\Delta \kappa_{f, vertex},
\end{equation}
where
\begin{eqnarray}
 \Delta \kappa_{se} &=& \frac{c_W^2}{s_W^2}\Delta \rho
		      + \Delta \kappa_{se,rem} \nonumber \\
   &=&\frac{\cos^2\theta_W}{\sin^2\theta_W}\left[\frac{\Sigma^{ZZ}(M_Z^2)}
     {M_Z^2}-\frac{\Sigma^{WW}(M_W^2)}{M_W^2}
     -\frac{\sin \theta_W}{\cos \theta_W}\frac{\Sigma^{\gamma Z}(M_Z^2)
     +\Sigma^{\gamma Z}(0)}{M_Z^2}\right].
\end{eqnarray}
The vertex contributions of the light quarks are given by
\begin{eqnarray}
\Delta \kappa_{f,vertex} &=& 
            \frac{\alpha }{16\pi s_W^2}
            \left\{(1-4|Q_f|s_W^2)(1-2|Q_F|s_W^2)\right.
            \Lambda_2(M_Z^2,M_Z) \nonumber \\
       & &-\frac{2}{s_W^2}(1-2(1-|Q_f|)s_W^2)\Lambda_2(M_Z^2,M_W)
           +\left. 12 \frac{c_W^2}{s_W^2}\Lambda_3(M_Z^2,M_W)\right\} .
\end{eqnarray}
For b quark, there are additional terms,
\begin{eqnarray}
\Delta \kappa^{add}_{b,vertex}&=& -\frac{1}{2}\Delta \rho^{add}_{b,vertex}+
	    6\left(\frac{\alpha m_t}{8\pi M_W^2s^2_W}\right)^2.
\end{eqnarray}
The explicit form of $\kappa_f$ turns out to be
\begin{eqnarray}
\kappa_f &=& 1 -\frac{\alpha }{4\pi s_W^4} 
             \left\{ -\frac{1}{2}\frac{m_t^2}{M_Z^2}
            +\left(\frac{1}{6}-\frac{2}{3}c_W^2+ 
	     \frac{11}{6}\frac{m_t^2}{M_Z^2}-\frac{4}{3}c_W^2
    \frac{m_t^2}{M_Z^2}\right)C(M_Z^2,m_t,m_t)\right.  \nonumber \\
        & &+\left(c_W^2-\frac{1}{2}\frac{m_t^2}{M_Z^2}-\frac{1}{2} 
             \frac{m_t^4}{c_W^2M_Z^4}\right)C(M_W^2,m_t,0) 
        +\left(\frac{1}{6}+\frac{1}{3}c_W^2\right)\ln \frac{M_Z^2}{m_t^2} \nn
        & & +\left(\frac{1}{12}+\frac{17}{6}c_W^2 
           + \frac{16}{3}c_W^4\right)
	     C(M_Z^2,M_W,M_W)  \nonumber \\
         & &   +\left(\frac{1}{12c_W^2}+\frac{4}{3} 
	   - \frac{17}{3}c_W^2-4c_W^4\right)
             C(M_W^2,M_Z,M_W) \nonumber \\
        & &+\left(-1+\frac{h}{3} - \frac{h^2}{12}\right)
              C(M_Z^2,M_H,M_Z) 
        +\left( c_W^2-\frac{h}{3} +  \frac{h^2}{12c_W^2}\right)
              C(M_W^2,M_H,M_W) \nonumber \\
        & & -\frac{h^2-1}{24}\ln h +\ln \frac{h}{c_W^2}
            \left[\left(\frac{11}{24}+\frac{h}{4}-\frac{c_W^2}{2}
           -\frac{h^2}{24c_W^2}\right) \right. \nn
     & & \left. +\frac{h}{h-c_W^2}\left(\frac{1}{4}c_W^2-\frac{h}{3}
         +\frac{h^2}{12c_W^2}\right)\right]
          -\ln c_W^2 \left[ -\frac{h}{12}+\frac{33}{4(1-c_W^2)}
       -\frac{69}{8}-3c_W^2 \right. \nn
 & &  \left.\left. +\frac{1+c_W^2}{1-c_W^2}\left(\frac{1}{24c_W^2}+\frac{2}{3}
        -\frac{17}{6}c_W^2-2c_W^4\right) \right]
        -\frac{3}{4}-\frac{43}{6}c_W^2+8c_W^4-\frac{1}{12c_W^2}
        -\frac{h^2s_W^2}{12c_W^2} \right\}
\end{eqnarray}
{\bf (2) $Z$ decay width}\\
With the help of the form factors, $\kappa_f$ and $\rho_f$, 
we can calculate the $Z$ width $\Gamma _Z$.
It can be expressed as the sum over the fermionic partial decay widths
\begin{equation}
\Gamma_Z = \sum_f \Gamma (Z\rightarrow f\bar{f}) 
\end{equation}
to a good approximation because other rare decay modes contribute to
$\Gamma_Z$ less than $0.1\%$.
In terms of the effective coupling constants $\bar{V}^2_f$ and $\bar{A}_f^2$,
\begin{eqnarray}
\bar{V}_f^2 &=& \sqrt{\rho_f}(I_{3L}^f - 2Q_f\kappa_f \sin^2\theta_W )\\
\bar{A}_f^2 &=& \sqrt{\rho_f}I_{3L}^f,
\end{eqnarray}
the partial widths, $\Gamma(Z\rightarrow f\bar{f})$, can be written as
\begin{equation}
\Gamma(Z\rightarrow f\bar{f}) = N_c^f\frac{M_Z^3}{12\pi}\sqrt{2}G_{\mu}
~[\bar{V}^{2}_f+\bar{A}_f^2]~R_{QED}R_{QCD}.
\end{equation}
The additional photonic QED and
the gluonic corrections to the hadronic final state,
$R_{QED}$ and $R_{QCD}$, are given by
\begin{eqnarray}
R_{QED} &=& 1+\frac{3Q_f\alpha }{4\pi } \\
R_{QCD} &=& N_c \left[1+\frac{\alpha_s(M_Z^2)}{\pi}
+1.409\left(\frac{\alpha_s(M_Z^2)}{\pi}\right)^2
-12.77\left(\frac{\alpha_s(M_Z^2)}{\pi}\right)^3 \right]
\end{eqnarray}
where $N_c=1$ for leptons while $N_c=3$ for quarks.
Note that the $Z\rightarrow f\bar{f}$ widths contain a number of additional
 corrections such as fermion mass effects and further QCD corrections 
proportional to the running quark mass $m_q^2(M_Z^2)$.
These contributions are very small for light quarks except $b$ quark.
For $b$ quark, the additional corrections$^{[28]}$ are
$\Delta R_{QCD}=c_1(m_b^2)\frac{\alpha_s(M_Z^2)}{\pi}
+c_2(m_b^2,m_t^2)\left(\frac{\alpha_s(M_Z^2)}{\pi}\right)^2
+c_3(m_b^2)\left(\frac{\alpha_s(M_Z^2)}{\pi}\right)^3~~ $.
\section{Precision Test of the Standard Model}
\setcounter{equation}{0}
Recent LEP measurements have improved so precise that
LEP's sensitivity can even detect the passing of TGV train,
while the long-awaited top quark has now been measured.
Those experimental advances should enable us to examine some 
important questions concerning the higher order radiative corrections
and the existence of the Higgs boson for which
we do not have direct evidence.
In this paper, we will report the results of our precision tests of the
standard electroweak model  including the radiative corrections developed
in section 5 with the 1995 electroweak precision data$^{[29]}$
and the experimental $m_t^{[5]}$  and $M_W$.
We will see that the precision test is sensitive not only to the
choice of the data set, i.e. how many data points, but also to
accuracy of the data used.
The experimental data sets are from the measurements at LEP, SLD and Fermilab.
In particular, the parameters $R_b$, $R_c$ from LEP and 
$\sin^2\theta^{lept}_{eff}$ from SLD 
should carefully be treated since the measurements of those parameters
deviate significantly from the SM predictions.
We will see how sensitive the indirect bounds of the Higgs boson mass
are to those parameters.
The Higgs boson mass range depends on whether or not the top quark mass
$m_t$ is treated as constrained by the experimental mass range.
Also we will show the uncertainty of the electroweak quantities in the
precision tests to the choice of and  the uncertainties
in the input data as well as in the parameters  $\alpha_s(M_Z)$, $\alpha(M_Z)$ 
and $m_t$.  
The new CDF result combined with the new D0 (as of spring 1996) reduced
the uncertainty in $m_t$ significantly, i.e., $m_t=176\pm 13 $ GeV.
Also the preliminary D0 measurement (as of spring 1996) of the W boson
mass revises the world average value, i.e., $M_W=80.26\pm 0.16 $ GeV.
In the present work we use these values\footnote{Note added in proof: 
These values have been further improved
as of 1996 summer: $m_t=175\pm 6 $ GeV,
$M_W=80.356 \pm 0.125 $ GeV and $M_Z=91.1863\pm 0.0020$ GeV.
See Ref.$^{[6]}$.}
of $m_t$ and $M_W$ along with
the QCD and QED couplings $\alpha_s(M_Z)$ and $\alpha(M_Z)$ at the $Z$
mass scale as given by $\alpha_s(M_Z)=0.123 \pm 0.006 $ corresponding to
the value deduced from the event-shape measurements at LEP$^{[30]}$ and
$\alpha^{-1}(M_Z)=128.89\pm 0.09^{[17]} $.

The electroweak sector of the standard model (SM) contains, besides the 
masses of the fermions and the Higgs bosons, three independent parameters,
$g, g^{\prime}$ and $v$ which are the $SU(2)_L$, $U(1)_Y$ couplings and
the vacuum expectation value of neutral component of the Higgs field.
One can instead choose $g, M_W$ and $M_Z$ to be the three independent
parameters.
On the other hand, there are three fundamental parameters measured with
high precision, which poses as the obvious choice of the three experimental
inputs; the hyperfine structure constant $\alpha=1/137.0359895(61)$,
the Fermi coupling constant from the muon decay $G_{\mu}=
1.16639(2) \times 10^{-5}~\mbox{GeV}^{-2}$ and the $Z$ boson mass
\footnote{See the Footnote 3.}
$M_Z=91.1884(22)$ GeV.
What enters in the electroweak observable quantities is however the effective
QED coupling constant $\alpha(M_Z)$ at the $M_Z$ scale, which is known
only within $0.1\%$ due to the uncertainty mainly in the hadronic
contribution to the running of QED coupling from low energy to the $M_Z$
scale. With the choice of $\alpha(M_Z), G_{\mu}$, and $M_Z$ as the
fundamental three input parameters, one can predict $M_W$ and the on-shell 
weak mixing angle $\sin^2{\theta_W}$.

The $W$-mass relation from the charge-current Fermi coupling constant
depends on the higher order radiative correction $\Delta r$ which depends
on the masses of the fermions including $m_t$, of the Higgs boson
$m_H$ and the gauge bosons $M_W$ and $M_Z$ along with $\alpha_s(M_Z)$ and
$\alpha(M_Z)$.
Thus the $W$-boson mass determination should have a self-consistency, i.e.,
the $M_W$ entering in calculating the radiative correction $\Delta r$
should be the same as the final output $M_W$ from the $W$-mass relation.
It is obvious that the $W$-mass relation can result only a correlation
between $M_W$ and $m_t$ for a given $m_H$ or $M_W$ {\it vs} $m_H$ for 
given $m_t$ at the moment.

What distinguishes our work$^{[31]}$ from other works of the precision test
is the imposition of the consistency of $M_W$ in the minimal $\chi^2$-fit
to the electroweak data. 
The $Z$-decay parameters are numerically calculated for the self-consistent
sets of ($m_t, m_H, M_W$) and the minimal $\chi^2$-fit solution is searched
by fitting them to the experimentally observed values.
We present the results$^{[31]}$ of the fits to the 1995 LEP and SLD
 data$^{[29]}$
obtained by using the ZFITTER program$^{[32]}$ that
includes the dominant two-loop corrections$^{[18]}$  of 
$O(\alpha^2m_t^2)$ to $\Delta \rho = 1-1/\rho $ and the QCD corrections
to the leading electroweak one-loop term$^{[22]}$ of 
$O(\alpha \alpha_s m_t^2)$ and $O(\alpha \alpha_s^2 m_t^2)$.
These new theoretical advances in radiative corrections coupled with the
experimental developments in the electroweak data, $m_t$ and $M_W$ are
additional reasons to update the precision tests of the SM.

\subsection {Global fits to the LEP and SLD data }
In Table 3, we give various sets of $(m_t,m_H)$ that give the best $\chi^2$
fits to the 1995 data of the eleven observables measured at LEP when
$m_t$ is restricted to the range $176\pm 13 $ GeV.
Each set of $(m_t, m_H)$ is correlated self-consistently to the $W$-boson 
mass given in the first row in each case, i.e., these masses
along with given lighter fermion masses and  $\alpha(M_Z)$ and
$ \alpha_s(M_Z)$ give the
radiative correction factor $\Delta r$ which in turn reproduces the $M_W$
from the $W$-mass relation.
The errors in the electroweak parameters due to the uncertainty in QED and 
QCD coupling constants at
$M_Z$ scale are also given in the parenthesis.

Numerical results in Table 3 show in general a good agreement with the SM
predictions except for $R_b=R(\Gamma_{b\bar{b}}/\Gamma_{had})$ and
$R_c=R(\Gamma_{c\bar{c}}/\Gamma_{had})$, which are about $3.5\sigma $ and
$2.3 \sigma $ away from the SM predictions, and are the major contributors to
the $\chi^2$-value in the fits to the 1995 LEP data.
Note that the deficiency in the predicted $R_b$ and the excess in $R_c$
(compared to the measured values) tend to decrease and thus the $\chi^2$
gets improved as $m_t$ is decreased toward the lower limit of the measurements.
It is obvious therefore that the absolute minimal $\chi^2$-fit solution of the 
global fit to the 1995 LEP data would be reached for $m_t$ well below the 
experimental
value. In fact the global minimal fit to the 1995 LEP data occurs when
$m_t=145$ GeV and $m_H=42$ GeV with $\chi^2/dof=19.0/11$,
which is to be compared to $m_t=176 \pm 13 $ GeV and $m_H>66$ GeV from
the LEP search.
On the other hand, if we choose to ignore $R_b$ and $R_c$ from the input data
to fit, the global minimal solution occurs for $m_t=161$ GeV and
$m_H=106$ GeV with significantly improved $\chi^2/dof=2.99/9$
and predicts $R_b=0.2162$ and $R_c=0.1722$.
A similar result,( $m_t=160$ GeV, $m_H=93$ GeV ), follows with 
$\chi^2/dof=8.84/10 $
when only $R_b$ is ignored from the input data set, which means that the global
minimal solution is sensitive to $R_b$ data
and favors to ignore the $R_b$ data\footnote{Note added in proof: 
New ALEPH result of five different tags
gives $R_b=0.2158(9)(11)$ and the combined LEP result is $R_b=0.2179
\pm 0.0012 $ bringing much closer agreement with the SM prediction.
 See Ref.$^{[6,33]}$.}.
As shown in Table 3, the uncertainty in the predicted $m_H$ due to
$\Delta \alpha^{-1}(M_Z)=0.09$ and $\Delta \alpha_s (M_Z)=0.006 $ is
about 100 GeV each around $m_t=175$ GeV.
There is however a clear evidence of the electroweak radiative corrections
in each of the electroweak $Z$-parameters.
The best fit solutions to the 1995 LEP data with and without $R_b$
give a stable output $M_W=80.331(24)$ GeV for $m_t=176 \pm 13 $ GeV.
$M_W$ can be shifted by another 13 MeV and 4 MeV due to uncertainties in 
$\alpha^{-1}(M_Z)$ and $\alpha_s(M_Z)$.

If we set $\alpha_s(M_Z)$ free, the global minimal fit solution occurs at
$\alpha_s(M_Z)=0.123$ with the values of ($m_t, m_H)$ given above 
irrespectively of $R_b$ and $R_c$ in the data set.
However the global minimal fit solution is achieved at
$\alpha_s(M_Z)=0.122$, for the combined set of LEP and  SLD data as given
in Table 4, with $( m_t=153 ~\mbox{GeV}, m_H=27 ~\mbox{GeV},
\chi^2/dof=28.3/14)$ for all data and with 
($m_t=159 ~\mbox{GeV}, m_H=36 ~\mbox{GeV}, \chi^2/dof=11.7/12)$
if $R_b$ and $R_c$ are excluded from the fit.
More or less the same result as the latter follows if only $R_b$ is excluded,
suggesting that the effect of $R_b$ and $R_c$ are not so influential 
compared to the case of LEP data alone.
The two $\alpha_s(M_Z)$ values are well within the input value 
$\alpha_s(M_Z)=0.123(6)$ used in our calculations.
In particular, we
note that inclusion of the SLD data makes  the output $m_H$ to
shift to even lower value, i.e., to $m_H=36$ GeV from $m_H=106$ GeV
in the case of the LEP+SLD data set excluding $R_b$ and $R_c$.

Of the three data from SLD, $\sin^2 \theta_{eff}^{lept}$ or $A_{LR}^0$  
deviates 
the most from the SM prediction$^{[6]}$ and therefore influences the most
the minimal $\chi^2$-fit solution.
To see this, we searched for the minimal $\chi^2$-solutions with the data sets 
including all three SLD parameters and only 
$\sin^2 \theta_{eff}^{lept}$ with or without
excluding $R_b$ and/or $R_c$ and found that they give the same results, 
for instance, $m_t=158$ GeV and $m_H=30$ GeV in the case of the LEP+SLD
data set with $R_b$ excluded and with either all three or one SLD parameter 
$\sin^2 \theta_{eff}^{lept}$ included.
The $\chi^2/dof$ is $11.8/12$ and $8.15/10$ respectively.
Thus it seems that either the SLD 
$\sin^2 \theta_{eff}^{lept}$, if supported by further 
measurements, implies new physics beyond the SM or is not supported
by the global precision test.
Note also that the inclusion of the SLD data predicts from the minimal 
$\chi^2$-fit solutions a stable $W$-boson
mass, $M_W=80.377(23)$ GeV where the error is due to
$\Delta m_t=\pm 13$ GeV around 176 GeV.
This is to be compared to the world average value 
\footnote{See Footnote 3.} $M_W=80.26\pm 0.16 $ GeV.
As before, there can be another shift of $13$ MeV and 4 MeV in $M_W$ due to
$\Delta \alpha^{-1}=0.09$ and $\Delta \alpha_s=0.006$ respectively.


\begin{table}
\begin{center}
\begin{tabular}{|c||c||l|l|l|} \hline \hline
& (Experiment) & & & \\
$m_t$~(GeV) & $176 \pm 13 $  & 
              $189$ & 
              $176$ & 
              $163$ \\
 & & & &  \\
$ m_H$~(GeV) & 60 $\leq m_H \leq 1000$ & 
              $578^{(213)(256)}_{(158)(168)}$ &
              $276^{(110)(122)}_{(85)(82)}$ & 
              $123^{(61)(59)}_{(56)(34)}$ \\
 & & & &  \\
 \hline
 & & & &  \\
$M_W$~(GeV) & $80.26\pm 0.16$ & 
              $80.355(13)(4)$ &
              $80.332(13)(4)$ &
              $80.307(13)(4)$ \\
 & & & & \\
$ \Gamma_Z $~(MeV) & $2496.3\pm 3.2 $ &
      $2497.2(7)(32)$ &
      $2497.0(7)(32)$ &
      $2496.5(7)(33)$ \\
 & & & & \\
$ \sigma_{h}^P(nb)$ & $41.488\pm 0.078 $ & 
       $41.462(2)(32)$ &
       $41.452(2)(32)$ & 
       $41.443(1)(32)$ \\
 & & & &  \\
$R(\Gamma_{had}/\Gamma_{l\bar{l}})$ & $20.788 \pm 0.032$ &
          $ 20.757(5)(40)$&
          $ 20.769(5)(40)$ & 
          $ 20.781(5)(40)$ \\
 & & & &  \\
$ A^{0,l}_{FB}$ & $ 0.0172\pm 0.0012 $& 
          $0.0156(5)$ & 
          $0.0156(4)$ &
          $0.0156(5)$ \\
 & & & &  \\
$ A_{\tau} $ & $ 0.1418\pm 0.0075$& 
          $0.1442(20)$ &
          $0.1440(20)$ & 
          $0.1441(20)$ \\
 & & & &  \\
$ A_{e} $ & $ 0.1390\pm 0.0089$& 
          $0.1442(20)$ &
          $0.1440(20)$ & 
          $0.1441(20)$ \\
 & & & &  \\
$R(\Gamma_{b\bar{b}}/\Gamma_{had})$ & $0.2219 \pm 0.0017$ &
          $0.2152$ &
          $0.2157$ & 
          $0.2161$ \\
 & & & &  \\
$R(\Gamma_{c\bar{c}}/\Gamma_{had})$ & $0.1543 \pm 0.0074$ & 
          $0.1725$ & 
          $0.1723$ & 
          $0.1722$ \\
 & & & & \\
$ A^{0.b}_{FB}$ & $  0.0999\pm 0.0031$& 
          $0.1010(14)$ &
          $0.1009(14)$ & 
          $0.1011(14)$ \\
 & & & &  \\
$ A^{0.c}_{FB}$ & $  0.0725\pm 0.0058$& 
          $0.0721(11)$ &
          $0.0720(11)$ & 
          $0.0721(11)$ \\
 & & & &  \\
$ \sin^2\theta^{lept}_{eff}$ & $0.2325\pm 0.0013 $ &
          $0.2319(2)$ &
          $0.2319(3)$ & 
          $0.2319(3)$ \\
 & & & & \\
\hline
$\chi^2 $ & & 25.3& 22.6 & 20.5 \\
\hline \hline
\end{tabular}
\caption{Numerical results including full EWRC for
eleven experimental parameters (the second column) of the Z-decay and $M_W$. 
Each pair of $m_t$ and $m_H$ represents the case of the best $\chi ^2$-
fit to the 1995 LEP data and $M_W = 80.26(16)$ GeV for $\alpha_s(M_Z)
=0.123(6)$ and $\alpha^{-1}(M_Z)=128.89(9)$. 
The numbers in the first and second () represent the explicit errors due to
$\Delta \alpha^{-1}(M_Z)=\pm 0.09$ and  $\Delta \alpha_s(M_Z) = 
\pm 0.006$ respectively.}
\end{center}
\end{table}
%
\begin{table}
\begin{center}
\begin{tabular}{|c||c||l|l|l|} \hline \hline
& (Experiment) & & & \\
$m_t$~(GeV) & $176 \pm 13 $  & 
              $189$ & 
              $176$ & 
              $163$ \\
 & & & & \\
$ m_H$~(GeV) & 60 $\leq m_H \leq 1000$ & 
              $338^{(142)(102)}_{(106)(77)}$ &
              $143^{(77)(48)}_{(57)(33)}$ & 
              $51^{(39)(12)}_{(24)(5)}$ \\
 & & & & \\
 \hline
 & & & & \\
$M_W$~(GeV) & $80.26\pm 0.16$ & 
              $80.400(13)(4)$ &
              $80.379(13)(4)$ & 
              $80.354(13)(4)$ \\
 & & & & \\
$ \Gamma_Z $~(MeV) & $2496.3\pm 3.2 $ &
      $2499.4(7)(32)$ &
      $2499.2(7)(32)$ &
      $2497.8(7)(32)$ \\
 & & & & \\
$ \sigma_{h}^P(nb)$ & $41.488\pm 0.078 $ & 
       $41.460(1)(32)$ & 
       $41.451(1)(32)$ & 
       $41.441(1)(32)$ \\
 & & & & \\
$R(\Gamma_{had}/\Gamma_{l\bar{l}})$ & $20.788 \pm 0.032$ &
          $ 20.762(4)(40)$ & 
          $ 20.775(4)(40)$ & 
          $ 20.788(5)(40)$ \\
 & & & & \\
$ A^{0,l}_{FB}$ & $ 0.0172\pm 0.0012 $& 
          $0.0161(4)$ & 
          $0.0162(4)$ &
          $0.0163(5)$ \\
 & & & & \\
$ A_{\tau} $ & $ 0.1418\pm 0.0075$& 
          $0.1466(19)$ & 
          $0.1469(20)$ & 
          $0.1476(20)$ \\
 & & & & \\
$ A_{e} $ & $ 0.1390\pm 0.0089$& 
          $0.1466(19)$ & 
          $0.1469(20)$ & 
          $0.1476(20)$ \\
 & & & & \\
$R(\Gamma_{b\bar{b}}/\Gamma_{had})$ & $0.2219 \pm 0.0017$ &
          $0.2152$ &
          $0.2157$ & 
          $0.2161$ \\
 & & & & \\
$R(\Gamma_{c\bar{c}}/\Gamma_{had})$ & $0.1543 \pm 0.0074$ & 
          $0.1725$ &
          $0.1724$ & 
          $0.1722$ \\
 & & & & \\
$ A^{0.b}_{FB}$ & $  0.0999\pm 0.0031$& 
          $0.1027(14)$ & 
          $0.1029(14)$ & 
          $0.1035(14)$ \\
 & & & & \\
$ A^{0.c}_{FB}$ & $  0.0725\pm 0.0058$& 
          $0.0734(11)$ & 
          $0.0736(11)$ & 
          $0.0739(11)$ \\
 & & & & \\
$ \sin^2\theta^{lept}_{eff}$ & $0.2325\pm 0.0013 $ &
          $0.2316(3)$ & 
          $0.2315(3)$ & 
          $0.2315(3)$ \\
 & & & & \\
\hline
(SLD) & & & &\\
$ \sin^2\theta^{lept}_{eff}$ & $0.23049\pm 0.00050 $ &
          $0.2316(3)$ & 
          $0.2315(3)$ & 
          $0.2315(3)$ \\
 & & & & \\
$ A_b $ & $0.841\pm 0.053 $ &
          $0.934$ & 
          $0.935$ & 
          $0.935$ \\
 & & & & \\
$ A_c $ & $0.606\pm 0.090 $ &
          $0.668(1)$ & 
          $0.668(1)$ & 
          $0.668(1)$ \\
 & & & & \\
\hline
$\chi^2 $ & & 36.8& 33.2 & 32.2 \\
\hline \hline
\end{tabular}
\caption{Numerical results including full EWRC for
fourteen experimental parameters (the second column) of the Z-decay and $M_W$. 
Each pair of $m_t$ and $m_H$ represents the case of the best $\chi ^2$-
fit to the 1995 LEP and SLD data and $M_W = 80.26(16)$ GeV for $\alpha_s(M_Z)
=0.123(6)$ and $\alpha^{-1}(M_Z)=128.89(9)$. 
The numbers in the first and second () represent the explicit  errors due to
$\Delta \alpha^{-1}(M_Z)=\pm 0.09$ and  $\Delta \alpha_s(M_Z) = 
\pm 0.006$ respectively.}
\end{center}
\end{table}

We presented the updated results of precision tests of the SM with the 1995 LEP
and SLD data within the framework where $G_{\mu}, \alpha$ and $M_Z$ are
given as input.
The $W$-boson mass $M_W$ has been treated self-consistently throughout
the calculation.
The results show that there is a good agreement with the SM predictions
with radiative corrections except $R_b$ and possibly $R_c$ and SLD 
$\sin^2\theta^{lept}_{eff}$.
The numerical fits for arbitrary $m_t$ and
$m_H$ show that the global minimal fit solution prefers to ignore the
$R_b$ data from the LEP data set, in order to achieve a better output
$m_t$ and $m_H$ and with an improved $\chi^2/dof$.
Inclusion of the SLD data, in particular the $\sin^2\theta^{lept}_{eff}$
parameter, has
the effect to shift $m_H$ to even a lower value below the experimental
lower bound in the global fits with arbitrary $m_t$ and $m_H$.
Either the $\sin^2\theta^{lept}_{eff}$
parameter or equivalently $A_{LR}^0$
needs to  be remeasured to reconcile the difference between LEP and SLD.
However the global minimal solutions tend to give $m_t$ lower than the
experimental measurements.

For $m_t$ fixed in the experimental range, the output $m_H$ is insensitive 
to the $R_b$ parameter in the data set.
In general, a minimal $\chi^2$-fit solution with a similar $\chi^2/dof$
is obtained for all $m_t$ in the experimental range but with a wide
range of $m_H$.
Inclusion of the SLD data again has the effect to lower $m_H$ from that for 
the LEP data alone, i.e., $m_H=88^{+63}_{-41}$ GeV compared to
$m_H=187^{+122}_{-81}$ GeV at $m_t=170$ GeV and $143^{+83}_{-56}$ GeV
compared to $m_H=276^{+160}_{-106}$ GeV at $m_t=176$ GeV.


Finally we want to add a remark on the $R_b$ parameter which caused a lot of
theoretical activities with the hope of discovering a new physics effect,
such as the idea$^{[33]}$ to associate the possible new physics effects
of $R_b$  and the low-energy determination of $\alpha_s$; an idea to 
invoke the supersymmetry$^{[35]}$ or  
the extended technicolor$^{[36]}$; and  to fine-tune the additional
contribution to the $Zb\bar{b}$ vertex from $Z-Z^{\prime}$ mixing model$^{[37]}$
.
None of the ideas  appear to be natural in spite of numerous efforts.

\section{ Bounds of the Higgs Mass }
\setcounter{equation}{0}
The vacuum stability problem is related to the negative sign of the running
Higgs quartic self-coupling $\lambda (\mu )$.
For a negative $\lambda (\mu)$, the Higgs potential is unbounded from 
below, and the vacuum is destabilized.
One way to remedy the problem is to introduce an embedding 
scale $\Lambda $ beyond
which the validity of the SM breaks down.
From the requirement of positive $\lambda (\mu )$, one can obtain a lower 
bound on \HS which, of course, depends on the scale $\Lambda $.

The minimum of the radiatively corrected Higgs potential lies outside
the validity region of the perturbative calculation as the higher order 
terms contain higher powers of
$\lambda \ln(\phi^2/M^2)$ which is large for $\phi \equiv <0|H^0|0>$ where
$M^2$ is a renormalization scheme dependent mass scale.
Thus in general the vacuum instability is expected when $\phi $ is much
larger than all mass scales of the theory and one should
make use of the renormalization group (RG) improved
form of the Higgs potential, which greatly extends the region of validity
of the perturbative calculation$^{[38,39]}$.
Since \HS is related to $\lambda (\mu )$, one can calculate the former
by solving the RG equations.
Particularly, since the $\beta $ function for $\lambda $ is strongly 
correlated with the top-Yukawa coupling constant,
\HS is given as a function of $m_t$ as well as 
the scale $\Lambda $.

Recently, Altarelli and Isidori$^{[40]}$ have reanalyzed the lower bound
on \HS from the requirement of the SM vacuum stability at the two loop level.
The resulting lower bound on \HS , for $m_t=140 - 210 $ GeV and 
$\Lambda = 10 ^{19}$ GeV, was given by the fitted formulae,
\begin{equation}
  m_h^{SM} > 135 + 2.1[m_t-174] - 4.5 \left[\frac{\alpha_s(M_Z)-0.118}
	{0.006}\right], 
\end{equation}
where \HS  and $m_t$ are expressed in GeV.
Also, independently, Casas et. al.$^{[41]}$ have given the corresponding
bounds on \HS at the two loop level as,
\begin{equation}
  m_h^{SM} > 127.9 + 1.92[m_t-174] - 4.25 \left[\frac{\alpha_s(M_Z)-0.124}
	{0.006}\right],
\end{equation}
which is  valid for $m_t= 150 -  200$ GeV and $\Lambda = 10^{19}$ GeV.
Both results are consistent within a few GeV.
We note that the lower bound on \HS does not deviate much
as the cut-off $\Lambda $ is increased beyond $\sim 10^{10}$ GeV.

\begin{figure}[htb]
\centerline{
\epsfig{figure=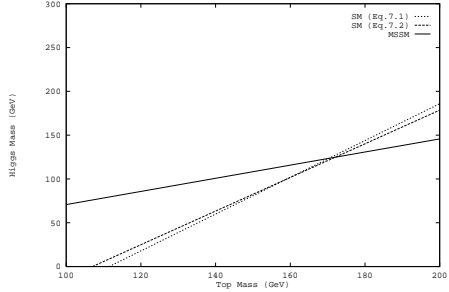, height=6cm, angle=270}}
\caption{Plots of the lower bound on the SM Higgs boson mass (dotted and
 dashed lines) and the upper bound from MSSM (solid line)
 as a function of the top quark mass.}
\end{figure}
In Fig.8, we show the results of the bound on \HS :
the dotted line is from Eq.(7.1) and the dashed one from Eq.(7.2)
for a fixed $\alpha_s(M_Z)=0.123$. \\

In the minimal supersymmetric extension of the standard model$^{[42]}$,
the Higgs sector is consisted of two CP-conserving Higgs-doublets 
with opposite hypercharge.
After the Higgs mechanism is imposed, there remain five physical Higgs particles :
two CP-even scalars, one CP-odd scalar and a pair of charged Higgs bosons.
One of the interesting phenomenological consequences of the MSSM Higgs sector
is that the tree-level bound on the lightest CP-even Higgs mass,
$m_h^{MSSM} \leq M_Z $, which could be the most significant implication for 
future experiments
at LEP 200 since this type of  Higgs boson, if exists, should be found
there.
However, the tree-level bound is spoiled when radiative corrections are 
incorporated.
Several groups$^{[43]}$ have computed the radiatively corrected upper 
bounds on $m_h^{MSSM} $
by assuming that the effective low energy theory
below the supersymmetry breaking scale is the SM with a single Higgs doublet.
Accordingly,
when the masses of superpartners of the SM particles 
and of the Higgs sector
but the lightest Higgs are taken to be large compared to $M_Z$, say,
of the order of a TeV, the lightest MSSM Higgs boson behaves much like
 the SM Higgs boson in
its production channels and decay modes$^{[44]}$.

\subsection{ Indirect determination of the Higgs mass }
Recent LEP data  has become so accurate that 
the prediction of the Higgs mass deserves to be seriously considered.
Several groups $^{[45]}$ have studied the Higgs boson mass range 
also with the LEP precision data, and confronted their analysis with
the perturbative lower bounds on the $m_H^{SM}$ and the theoretical
upper bound on the $m^{MSSM}_H$.
They predicted two Higgs boson mass ranges, one from the fit to the LEP
precision data alone and another one from the fit that combines 
the CDF/D0 $m_t$ measurement,
and indicated that the light Higgs boson of the MSSM type would be
surprisingly more consistent with the data than that of the SM type,
``though not significantly so".

The results based on the earlier data are reported in$^{[31]}$ in which
the importance of determining $M_W$ in a self-consistent manner from
the $W-$mass relation with radiative corrections is emphasized to test
 the genuine electroweak radiative effects$^{[46]}$.
In the recent work$^{[31]}$, we studied the $m_t-M_W$ correlations for
the CDF $m_t$ range and the errors to the predicted values of the 
fit resulting from the uncertainties in $\alpha_s$ 
and $\alpha $ with the LEP data  as of the Glasgow meeting.
In principle, the mass of Higgs boson can be determined from this within
the context of the SM if $M_W$ and $m_t$ become known to be sufficiently
precise.
We will examine the necessary degree of the accuracy of $M_W$ and $m_t$ 
in order to differentiate $m_H$ by better than $100$ GeV in future experiments.

The global minimal fit solution obtained from the LEP data alone as input
gives the $1\sigma $ range $m_t=145^{+14}_{-10}$ GeV and
$m_H=42^{+86}_{-24}$ GeV, and even lower range of $m_H$ for the 
LEP+SLD combined data set.
When compared to the bounds$^{[41]}$ of the Higgs bosons from the SM and 
the minimal supersymmetric standard model (MSSM), a light Higgs boson of
the MSSM type appears to be slightly more probable for this $1\sigma $ range
but $m_t$ is just barely consistent with the experimental value at the 
$1\sigma $ level.
This is because the upper limit of $m_H$ obtained from the boundary 
condition for the Higgs quartic self-coupling $\lambda $ at the renormalization
scale in the MSSM and the lower limit of $m_H$ resulting from the vacuum
stability requirement at the two-loop level in the SM interest around
$m_t=173$ GeV and $m_H=125$ GeV.
In view of possible uncertainty of $4$ GeV in $m_t$ due to the uncertainty
$\Delta \alpha_s(M_Z)=0.006$, there is a region in which both the SM and
MSSM types can be compatible for $m_t<169$ GeV.
There is slightly larger region of ($m_t, m_H$) from the minimal $\chi^2$-fit
solution at the $1\sigma $ level that coincide with the MSSM type.

However because $m_t$ obtained from the global minimal $\chi^2$-fit is
somewhat smaller than the experiment, it may be interesting to fix $m_t$
in the experimental range.
If we limit the solutions to have $m_H$ safely larger than the experimental 
lower bound $66$ GeV at the $1\sigma $ level, we find that for the LEP data
alone the solutions for $165 < m_t < 180$ GeV have more or less the same
$\chi^2/dof$, while for the LEP+SLD data those for $175 < m_t < 185 $ GeV 
have a similar $\chi^2/dof$.
In particular for $m_t=165$ GeV, the $1\sigma $ range of the Higgs boson 
mass is $m_H=140^{+96}_{-59} $ GeV with $\chi^2/dof=20.8/11$ for the LEP data
alone and for $m_t=176$ GeV, $m_H=143^{+83}_{-56}$ GeV 
with $\chi^2/dof=33.2/14$ for the LEP+SLD data.
In addition if we choose to ignore $R_b$ from the input data, we get
$m_H=138^{+93}_{-61}$ GeV with $\chi^2/dof=8.89/10$ in the former case while
$m_H=141^{+80}_{-59}$ GeV with $\chi^2/dof=18.7/13 $ in the latter case.
While $m_H$ from the best-fit solution is insensitive to the $R_b$ parameter
as soon as $m_t$ is fixed,
the inclusion of the SLD data can achieve the minimal $\chi^2$-fit solutions in 
general for larger fixed $m_t$ and a smaller output $m_H$ than in the case
of LEP data alone, i.e., $m_H \simeq 143$ GeV is achieved for fixed
$m_t=176$ GeV in the case of LEP+SLD input data,
while the minimal fit solution for fixed $m_t=176 $ GeV in the case LEP
data alone gives $m_H\simeq 276$ GeV.
Since these solutions fall in mostly where $m_t<177$ GeV, it is not
possible to distinguish from the minimal fit solutions the type or
origin of $m_H$ between the SM and the MSSM. \\

Finally, we noticed that the uncertainties in $\alpha^{-1}(M_Z)$ 
and $\alpha_s(M_Z)$ 
cause an uncertainty of 70-100 GeV and 50-100 GeV each in $m_H$.
Thus experimental determination of $m_H$ with an error smaller than 100 GeV
will mean that the measurement surpass  theoretically intrinsic uncertainty 
attainable in the precision test.
Also an uncertainty of 5 GeV 
\footnote{Consult the $M_W-m_t$ correlations in Fig. 2 of the first
reference in Ref.$^{[31]}$.}
in $m_t$ around 175 GeV can cause a shift in
$m_H$ by 60-80 GeV depending on the data set, which in turn implies a
shift in $M_W$ by about 30-40 MeV from the $M_W-m_t$ correlation obtained
from the $W$-mass relation.
Thus we can say that if $M_W$ is determined to within 40 MeV uncertainty, 
$\Delta r$ will be tightly constrained to distinguish the radiative corrections
and the minimal $\chi^2$-fit can discriminate the mass range of $m_t$ and $m_H$
within 5 GeV and 80 GeV respectively, given the current accuracy in the
input parameters and electroweak data set.

\section{ Concluding  Remarks}
\bigskip
We have tried to construct the standard model's weak sector as the simplest
gauge theory containing the known charged weak current coupling to charged
IVBs which are the $I_3=\pm 1$ members of the adjoint representation of
$SU_L(2)$. The $I_3=0$ neutral gauge boson could then
mix with the $U(1)$ resulting in a physical $Z_0$ gauge boson and the photon.
A doublet of complex Higgs bosons, by the mechanism of SSB,
then gave masses to the three IVBs
(swallowing up three of their four degrees of freedom) leaving one
observable neutral Higgs and the massless photon.
The standard model is very successful with experiments and the renormalization
correction is only a few percent from, say, the on-shell definition
\be
M_W^2= \left( {{\pi \alpha}\over {\sqrt 2 G_F}} \right) /
\left[\sin^2\theta_w(1-\Delta r)\right]                  
\ee
which gives $\Delta r \approx 0.04$ for $m_t \approx 170$ GeV and $sin^2
\theta_w=0.224$. The Higgs mass has a lower bound given by Eq. (7.1) and (7.2)
which follows from the quantum corrections to the Higgs potential and an
upper bound
\be
m_H \leq \left( {{4\pi\sqrt 2}\over{G_F}}\right)^{1/2} \approx 1.2~ TeV
\ee
from the unitarity limit of $WW \rightarrow WW$ via $\gamma, Z$ and Higgs
exchanges. If $m_H$ turns out to be heavier than 1.2 TeV, it simply would
mean that the perturbation calculation is no longer valid and the weak
interaction behaves like a strong interaction in such a case. Non-standard SSB
can of course add complexity to the Higgs sector without perturbing low-energy
phenomenology. The unbroken strong sector consisted of eight massless gluons
belonging to $SU_c(3)$ and QCD formed a copy of the incredibly successful QED.

Quarks were seen to gain masses from the Higgs mechanism and mixing resulted
from the weak eigenstates differing from the mass eigenstates.
Several parametrizations for this mixing matrix were surveyed and the
magnitudes of the matrix elements were deduced from a variety of
experimentally measured decays. Neutrinos gained their tiny mass by a quite
different method: the seesaw mechanism and as a result could also mix
and oscillate into each other. If these oscillations were amplified in the
sun's interior, one could resolve the solar neutrino problem.
For the seesaw mechanism to work a heavy right-handed neutrino is needed
and, since $SO(10)$ is the smallest GUT group providing one naturally,
a specific model was studied more deeply as a useful example of the
principles espoused above.
What have we intentionally left out for lack of time?

It seems that experimentally, the running coupling constants
do not meet at a single Grand Unified energy, suggesting the
existence of a more complex SSB scheme at the GUT scale. Amongst the many
popular extensions of the standard model that can
offer such a new scale are those inspired from supersymmetry.
As a bonus, they offer fermionic partners (``-inos") to every boson in 
the theory
(and vice versa) that are the best candidates for the missing invisible (dark)
mass needed to just close our Universe after the Big Bang. Given sufficient
abundance and mass these generic neutralinos (photinos, Higgsinos, gluinos,
Zinos..) form what is called cold dark matter because they only interact
gravitationally at the present epoch. Of course the zoo of yet unobserved
particles (and even strings!) does not stop there.
Thus one may have liked to supplement with more discussions of GUTs and
supersymmetric GUTs and their implications for the cosmological problems such
as dark matter and baryogenesis.

Another subject that we left out is to search for a replacement of SSB with
elementary Higgs scalars, i.e. Dynamical Symmetry Breaking (DSB)
and compositeness
of quarks, leptons and Higgs-like scalars. It would be extremely interesting
to explore the possibility of understanding the flavor problem of fermions,
including the hierarchy of their masses, in some framework of supersymmetric
GUTs of more fundamental preons. While the standard model seems to be able to
explain the observed experiments, it is not the fundamental theory as it
contains many arbitrary parameters including the fermion masses.
Supersymmetric GUTs (and the related superstring theory) and the DSB
scheme are just two possible ways to go beyond the standard model.

We discussed in great detail the on-shell renormalization scheme of
the standard model as a field theory.
One-loop corrections are calculated by the dimensional regularization.
Because of the experimental advances in precision measurements
the higher order radiative corrections must be made to the S- matrix.
We have made a precision test of the standard model including 
one-loop and most of the dominant two-loop corrections against the
electroweak data from LEP and SLD.
In view of the elusive nature of the Higgs boson,
the precision test can reduce to only a correction between $m_t-M_W$
for given $m_H$ or a correlation between $m_t-m_H$ for given $M_W$.
The accuracy of $m_t$ or $M_W$ required to corner $m_H$ and the
uncertainty in $m_H$ and in $Z-$decay parameters due to the experimental
uncertainties in $m_t, M_W$ and QED and QCD couplings are discussed
with the possibility to discover new physics beyond the standard model
from such precision test with $Z-$decay parameters.
The bounds on the Higgs boson mass deduced from the precision test are
compared with the lower bound due to vacuum stability of the Higgs
sector in the standard model and the upper bound from the MSSM.

\section*{Acknowledgements}
This work is supported in part by the U.S.DOE Contract DE-FG-02-91ER
40688-Task A and also in part by the Brown-SNU CTP Exchange Program.
One of us (KK) would like to thank Professor Choonkyu Lee for
the invitation to lecture at the 15th Symposium on Theoretical Physics
and other colleagues at SNU CTP for the warm hospitality.
 
\vfill\eject

\end{document}